# Continental Diversity of *Chenopodium album* Seedling Recruitment

Andujar, J.G.[1], D.L. Benoit[2], A. Davis[3], J. Dekker[4], F. Graziani[5], A. Grundy[6], L. Karlsson[7], A. Mead[6], P. Milberg[7], P. Neve[6], I.A. Rasmussen[8], J. Salonen[9], B. Sera[10], E. Sousa[11], F. Tei[5], K.S. Torresen[12], J.M. Urbano[13].

[1] I.A.S.-CSIC, Auda. Meueuder Pidal, Cordoba, Spain; [2] Agriculture and Agri-Food Canada, Horticulture Research and Development Centre, Saint-Jean-sur-Richelieu, Quebec, Canada; [3] USDA-ARS, Photosynthesis and Global Change Research Unit, Urbana, IL, USA, [4] corresponding author, Weed Biology Laboratory, Agronomy Dept., Iowa State University, USA; [5] Dept. Agricultural and Environmental Sciences, University of Perugia, Italy; [6] Warwick HRI, University of Warwick, Wellesbourne, Warwickshire, UK; [7] IFM Biology, Division of Ecology Linköping University, Sweden; [8] University of Aarhus, Department of Integrated Pest Management, Research Centre Flakkebjerg, Slagelse, Denmark; [9] MTT Agrifood Research Finland, Plant Protection, Jokioinen, Finland; [10] Institute of Systems Biology and Ecology, Academy of Sciences of the Czech Republic, Ceske Budejovice, Czech Republic; [11] Depto. Proteccao das Plantes e Eitoecologia, Instituto Superior de Agronomia, Lisboa, Portugal; [12] Norwegian Institute for Agricultural and Environmental Research (Bioforsk), Plant Health and Plant Protection Division, Ås, Norway; [13] Dpto. Ciencias Agroforestales, University of Sevilla, Sevilla, Spain.

**ABSTRACT**.

*Chenopodium album* seedling emergence studies were conducted at nine European and two North American locations comparing local populations with a common population from Denmark (DEN-COM).

**Mortality risk**. Weedy plant life history conforms to the time of mortality risk from either the environment (e.g. frozen soil in the winter) or cropping systems practices (e.g. herbicide use). *C. album* life histories are constrained by the environmental and cropping system practices of a locality. It is hypothesized that *C. album* seedling recruitment timing and magnitude have adapted to these local conditions and are expressed in emergence behavior. Recruitment periodicity and quantity were found to be closely related to predictable cropping system disturbances, experimentally induced disturbances, and seasonally-available plant resources and conditions. Recruitment patterns were the consequence of seasonal limitations in the habitat (filter 1), predictable cropping system disturbances (filter 2) and experimentally induced disturbances to simulate early season tillage. As a consequence of these filters *C. album* seized locally available seedling emergence opportunity.

**Filter 1: Habitat.** Limitations in the habitat (filter 1) are reflected in local *C. album* population recruitment season length: Julian weeks (JW) 2-37, duration 12-37 weeks. Generally, the duration of seedling recruitment (undisturbed, experimentally disturbed) of both populations (local; DEN-COM) increased with decreasing latitude, north-to-south. In general, compared to the local population, DEN-COM recruitment at locations north of Denmark was longer and south of Denmark was shorter, and ended sooner. The early season onset of recruitment was more consistent than that in the late season, which had smaller numbers and greater variability over longer times. Autumnal recruitment in the burial year (2005) occurred at a few locations.

**Filter 2: Cropping disturbance**. Predictable, historical, seasonal times of local cropping system disturbances (CSD) provided the second filter of seedling recruitment opportunity. Generally, the CSD period increased with decreasing latitude, with some exceptions. The total duration of the CSD period was over twice as long in the south as that in the north. After layby there occurred a mid-season period of CSD inactivity at all locations, the post-layby (PL) period, when cropping activity ceases. This post-layby period was from JW 21-34, with durations of 2-13 weeks, depending on location. No CSD's occurred at any location during JW 26-29, excepting the UK

**Seedling emergence structure: time, number**. Recruitment at each locality possessed seasonal structure (time, number) consisting of 2-4 discrete seasonal cohorts, a consequence of habitat and disturbance limitations. The spring cohort (JW 12-24) was the largest recruitment investment, with consistent local times of seasonal onset. A short gap in recruitment occurred between the spring and summer cohorts at all locations, for both populations, and coincided with layby and summer solstice (LSS), the PL period, crop canopy closure, and a period of increasing heat and potentially decreasing moisture. This LSS gap was characterized by the consistent absence of seedling emergence during JW 25-26 in undisturbed, and in a small number of experimentally disturbed, instances. The gap was unexpected at this favorable recruitment time of the season. The specific timing of the recruitment hiatus varied with location (latitude), population and experimental disturbance. The short cessation of seedling recruitment may be a consequence of a unique seasonal change in light, the most important environmental signal regulating *C. album* seed germination. Photoperiod length and intensity increases before, and then decreases after, summer solstice, a threshold event, a solar switch. The coincidence of all these anthrogenic, astronomical and environmental factors at the LSS may be an important basis by which *C. album* seedling emergence is 'fine-tuned' consistent with local opportunity. Late recruitment consisted of the second largest cohort in the summer (JW 25-37). This cohort was smaller than, longer and more variable than, and consistently separated from, the spring cohort. Recruitment during the autumn cohort (burial year, following year) occurred at only a few locations, in low numbers. Winter recruitment only occurred in warm, irrigated, multi-cropped Portugal.

**Emergence cohorts**. Local recruitment opportunity was seized by *C. album* structured in seasonal cohorts whose timing was consistent with habitat and cropping disturbance periodicity. Distinct recruitment cohorts arose around the temporal interfaces separating CSD operations. This may be an adaptive means by which *C. album* searches for, and exploits, recruitment opportunity just prior to, and after, predictable disturbances. These patterns coincide with intervening periods of no disturbance, opportunity: spring and summer cohorts are separated by the LSS gap; the early summer cohort at the beginning of the post-layby (PL) period of mid-season opportunity; and at harvest. Recruitment periodicity differed by population at each local common nursery. Experimental disturbance affected the recruitment cohort structure in several contradictory ways. In some locations it was extended, others shortened, in others stimulated the appearance of new late season cohorts

**Seed control of emergence**. The control of *C. album* seedling emergence is contained in the heteroblastic traits of its locally adapted seeds, and is stimulated by a complex interaction of light (photoperiod duration, intensity and quality), heat (growth; oxygen-water and nitrate-water solubility), water, nitrate and oxygen signals inherent in the local environment. Our observations of complex recruitment patterns occurring at critical cropping times is strong evidence that *C. album* possesses a flexible and sensitive germination regulation system adaptable to opportunity in many different Eurasian and North American agricultural habitats.

**Summer solstice light signal**. Herein we show for the first time a potential role for summer solstice as a rapid astronomical signal, a threshold switch, regulating photoperiod length, as well as the direction of change in diurnal length (longer, shorter). This solstice photoperiod switch may provide a rapid light signal stimulating or inhibiting recruitment, a dependable seasonal signal ensuring the separation of seedling cohorts before and after layby and canopy closure.

**CONTENTS**

# 1 INTRODUCTION

Common lambsquarters, fat hen (*Chenopodium album*) (CHEAL) is a widespread and troublesome weed in agricultural and disturbed habitats throughout the north temperate regions of the world. Seed germination and seedling emergence from the soil seed pool are critical threshold life history events immediately preceding assembly in agricultural communities and subsequent interference with crop productivity. How do successful *Chenopodium album* seedlings assemble with crop plants to form local agricultural communities over such a diverse range of habitats? What forces of nature are responsible for the seedling emergence patterns and quantities observed across the wide geographic and environmental range of adaption this highly successful weed displays?

One of the most important events in a plant's life history is the time of seed germination and seedling emergence, the resumption of embryo growth and plant development. Emergence timing is crucial, it is when the individual plant assembles in the local community and begins its struggle for existence with neighbors. Resumption of growth at the right time in the community allows the plant to seize and exploit local opportunity at the expense of neighbors, allowing development to reproduction and replenishment of the local soil seed pool at abscission. Soil seed pools are the source of all future local annual weed infestations, and the source of enduring occupation of a locality.

Community assembly of crops and weeds in agroecosystems, and its consequences, is a complex set of phenomena (Dekker et al., 2003; 2014). Accurate predictions of the time of weed interference, weed control tactic timing, crop yield losses due to weeds, and replenishment of weed seed to the soil seed pool, require information about how agricultural communities assemble and interact. Despite attempts at description (e.g. Booth and Swanton, 2002), little is known about the rules of community assembly. Their elucidation may remain an empirically intractable problem.

Despite this, there exist two opportunities to understand agroecosystem community assembly during the recruitment phase (e.g. seedling emergence) that predicate future interactions with other plants. The first advantage derives from the annual disturbance regime in agricultural fields that eliminates above ground vegetation (e.g. winter kill, tillage including seedbed preparation, early season herbicide use). Understanding community assembly is most tractable when starting each growing year with a field barren of above-ground vegetation and possessing only dormant underground propagules (e.g soil seed and bud pools), a typical situation in much of world agriculture.

The second advantage derives from the observation that the time of emergence of a particular plant from the soil relative to its neighbors (i.e. crops, other weeds) is the single most important determinate of subsequent weed control tactic use, competition, crop yield losses and weed seed fecundity. Seedling recruitment is the first assembly step in these disturbed agricultural communities, and is therefore the foundation upon which all that follows is based. Information predicting recruitment therefore may be the single most important life history behavior in weed management.

## 1.1 Seed Germinability and the Soil Environment

Successful *C. album* seedling recruitment in a wide diversity of habitats appears to be a complex dynamic interaction between the after-ripening effects of the soil environment (light, temperature, nitrate, moisture level) and the germinability states of individual seeds in the local soil pool (Altenhofen, 2009; Altenhofen and Dekker, 2013). *C. album* recruitment is an integrated response to all five of these parameters appropriate to the locality. Successful timing of seedling emergence therefore involves an adaptation in sensitivity to each of these environmental signals appropriate to the specific field being exploited. Observation of only some of the parameters controlling emergence has limited our understanding of *C. album* seed behaviors. Of the 24 studies on *C. album* germination reviewed by Altenhofen (2009), only 12% included five or more parameters, 21% included four, 29% included three, and 38% included only one or two. Additionally, only 4 of the studies compared multiple populations of *C. album*. [*add Alistair Murdoch study*] Controlled environment studies of all five parameters influencing *C. album* germination are difficult to conduct. Observing the influence of all five on seedling recruitment in field conditions presents intractable experimental obstacles. It is precisely for this reason that past attempts to model environmental control of seedling emergence have failed: light-heat-$H_2O$-$NO^2$-seed germinability is complex and dynamic. Hydro-thermal models of germination regulation provide insight for simpler, non-dormant, crop seeds but are insufficient to understand weedy adaptation to the wide range of agricultural habitats exploited by *C. album*. [*ref, maybe Murdoch*] There exists an alternative method to understand regulation of seedling emergence. Recruitment can also be studied effectively by observing the integrated effects of environment, seed dormancy and the particular cropping system of a locality: exploiting local opportunity.

## 1.2 Recruitment Seeks Local Opportunity

*C. album* possesses the ability for fine-scale germinability adaptation to local resources and conditions by means of adaptive change in the sensitivity of seeds to light-heat-$H_2O$-$NO^2$ signals in the soil. Local adaptation also results from adaptive behaviors to predictable disturbances (e.g. tillage, herbicides, frozen winter soil) and neighboring organisms (e.g crops, other weed species) with which it interacts. *C. album* plant life history events, as for any weed species, conform to the times of mortality risk from any source. Seizing and exploiting local opportunity is an integrated recruitment response by *C. album* to these local conditions, resources, disturbances and neighbor interactions. Together these local forces of nature act as recruitment filters, the survivors of which assemble with crops to form the local plant community.

**1.2.1 Opportunity filter 1: habitat**. Natural selection acts as a recruitment filter in the first instance during times of limitations in locally available resources (light, water, nitrate) and conditions (heat): limits to the length of the growing season. The duration of the local season of seedling emergence provides an estimate of the times of favorable environmental resources and conditions for growth: between the winter-preplant and post-harvest-winter interfaces.

**1.2.2 Opportunity filter 2: local disturbance**. Natural selection acts as a filter in the second instance during times of favorable growth: particular seasonal times when local recruitment is limited by high mortality from predictable cropping system disturbances (tillage, herbicides). Two types of opportunity exist for successful seedling emergence during the growing season. First, successful recruitment phenotypes are those that emerge and survive at favorable times (e.g. post-layby, post-harvest), avoiding predictable cropping disturbances. The second type of successful recruitment phenotypes are those that emerge during times when both the risk of mortality and the potential payback in fecundity are high (e.g. early season emergence). Emergence during these periods of higher mortality risk is a hedge-bet strategy for fitness (*refs*) favoring those individuals able to survive, escape, tolerate or avoid disturbances. The potential payback for such risk is much greater seed fecundity at season's end.

Filters of local opportunity act together, channeling *C. album* seedlings to the observed location, time and magnitude of recruitment. Soil seed pools therefore are a historical memory of past successful phenotypes. Current and future seedlings emerging from the soil are the survivors of historical disturbances.

## 1.3 Continental Observations of *C. album* Seedling Emergence

Herein are reported the observations made of *C. album* seedling emergence in Europe and North America. These observations are utilized to allow deductive evaluation (falsification) of the validity of opportunity space-time as the primary restraint on local seedling recruitment. Therefore two opposing sets of hypotheses are posed:

### 1.3.1 Opportunity hypotheses:

**1**: Seedling emergence of a local *C. album* population will occur near the onset of seasonal periods of low mortality allowing weed growth and seed production (seizing opportunity)

**2**: Seedling emergence of a local *C. album* population will occur early in the growing season when historical disturbance mortality has been high but potential seed fecundity of current survivors is also high (hedge-bet opportunity)

**3**: Seedling emergence of a displaced *C. album* population will occur at times consistent with the environmental signals appropriate for success at its past location, but potentially at times and numbers inappropriate to its new location (displacement of opportunity)

**4**: Seedling emergence of a local *C. album* population disturbed early in the current season will be displaced to either periods of low mortality (seizing opportunity) or disturbed periods with high fecundity returns (hedge-bet opportunity) (current disturbance opportunity)

### 1.3.2 Null hypothesis:

1: Seedling emergence of a local *C. album* population will occur at any period of the calendar year (random recruitment)

2: Seedling emergence of a local *C. album* population will occur at any time during the growing season (random seasonal recruitment)

3: Seedling emergence of a local *C. album* population will occur at the same times regardless of geographic location or time of seasonal disturbances (fixed germination stimuli recruitment)

## 1.4 Representation of Seedling Recruitment

**1.4.1 Weeds as a complex adaptive, soil-seed communication system**. The nature of weeds is a complex adaptive, soil-seed communication system. The nature of weedy life history is an adaptable, changeable system in which complex behaviors emerge when self-similar plant components self-organize into functional traits possessing biological information about spatial structure and temporal behavior (Dekker, 2014a, b). Weedy plant spatial structure is the foundation for emergent life history behavior: self-similar timing of life history processes regulated by functional traits expressed via environment-plant communication. The nature of weeds is an environment-biology communication system. Biology is information. Information comes via evolution, an ongoing exchange between organism and environment. Information is physical. Biology is physical information with quantifiable complexity.

### 1.4.2 Analysis of variance, inference and weed complexity.

> "On the other hand, when we find departure from normality, this may indicate certain forces, such as selection, affecting the variable under study." "… skewness …" "… bimodality …" (Sokal and Rohlf, 1981)

Several crucial assumptions are made when analysis of variance statistics are utilized with normal distributions of population observations: observations are drawn from normally distributed populations and errors are normally distributed within each treatment population and are independent of each other; observations are random samples from the populations and independence of error effects; and, the variances of the populations are homogeneous (Kirk, 1982). For the empirical scientists these assumptions of normally distributed events in nature present some fundamental problems. There is often good reason to suspect that the additivity of treatment and environmental effects and the independent distribution of experimental errors with a common variance are false (Cochran and Cox, 1957).

There exists a need to know the bell curve intimately and identify where it can and cannot hold. Empirical natural scientists, users of the bell curve, need to justify its use, not the opposite.

The Gaussian, and such as the Poisson law, are the only class of distributions that are sufficiently described by the standard deviation and the mean. Standard deviations are a measure of the degree of risk and randomness. Standard deviation is just a number that you scale things to, a matter of mere correspondence if phenomenon are Gaussian. Standard deviations do not exist outside the Gaussian; if they do exist they do not matter or explain much. Other statistical notions that do not have significance outside the Gaussian include *correlation* and *regression*. A single measure for randomness, standard deviation, cannot be used to describe risk. A simple characterization of uncertainty does not exist in complex natural systems. There is a need to understand randomness fully.

Benoît Mandelbrot has argued that normal distributions do not properly capture empirical and "real world" distributions. There are other forms of randomness that can be used to model extreme changes in risk and randomness. Traditional "bell curves" are inadequate for measuring risk, such curves disregard the possibility of sharp jumps or discontinuities. Traditional approaches are based on random walks, contrasting with a world primarily driven by random jumps. Tools designed for random walks address the wrong problem (Mandelbrot and Taleb, 2006). Mandelbrotian randomness is fractal randomness, uncertainty. It links randomness to the geometry of nature: fractal geometry. Fractality is the repetition of geometric patterns at different scales, revealing smaller and smaller versions of themselves. Small parts resemble, to some degree, the whole. Significant differences in the definition of an objects dimension connote a high degree of structural complexity. There are many processes in nature forming structures in which, at decreasing scales, similar principles of formation persist (e.g. branching processes in biological growth occur stepwise in increasing detail). Parts of an object has similar properties as the whole, the property of scale invariance and self-similarity (Stoyan and Stoyan, 1994) (e.g seed germination heteroblasty). Fractality is the repetition of geometric patterns at different scales, revealing smaller and smaller versions of themselves. Small parts resemble, to some degree, the whole (Taleb, 2009). There is no qualitative change when an object changes size. Fractals generate pictures of ever increasing complexity by using a deceptively minuscule recursive rule; a rule that can be reapplied to itself infinitely. The fractal has numerical and statistical measures that are (somewhat) preserved across scales; the ratio is the same, unlike the Gaussian. The shapes are never the same, yet they bear an affinity to one another, a strong family resemblance. Self-affinity might be a more appropriate term for natural complexity than strictly mathematically self-similar. This character of self-affinity implies that one deceptively short and simple rule of iteration can be used, either by a computer or, more randomly, by Mother Nature, to build shapes of seemingly great complexity. It is how nature works.

**1.4.3 Representing seedling emergence**. Weed seed recruitment biology is the consequence of natural selection for seizing and exploiting local opportunity spacetime by a population. Seedling emergence behavior has been shaped by selection forces dominating the individual cropping fields studied. *Chenopodium album* behavior presented in this paper demonstrates that observations or emergence behavior are not drawn from normally distributed populations. Crucial assumptions are violated. Errors are not normally distributed within each treatment population, and are not independent of each other, quite the contrary. Seedling recruitment observations are not random samples from the two populations and error effects are not independent; the variances of the populations are not homogeneous. The additivity of emergence treatment and environmental effects, and the independent distribution of experimental errors with a common variance, are false. Analysis of variance statistics are not the appropriate basis of seedling emergence models (Schutte et al., 2013).

How then can seedling emergence be represented, modeled? ANOVA statistics assumes behaviors not apparent in this data. At the beginning we are left, as with all complex adaptive systems, with a historical approach: presenting behavior on its own terms. Weed seedling recruitment behavior is the annual life history behavior of a local population seizing and exploiting opportunity created by the cropping practices of the agro-habitat and the environmental resources and conditions it contains. The approach taken is unusual in the weed recruitment literature. It is existential rather than empirical. Understanding seedling recruitment requires observing emergence in a locality, for a population, as a unique entity. This is the way evolution works. Herein emergence is presented as historical calendar, without ANOVA statistics. Emergence calendars allowed magnitude, duration and pattern to be revealed. They also allowed the complex *C. album* story to be revealed.

## 2 METHODS AND MATERIALS

Experiments on *Chenopodium album* were established in 2005 at 11 different locations in Western Europe and North America, ranging from N 60$^{o}$ to N 37$^{o}$ latitude and W 89$^o$ to E 23$^o$ longitude (Table 1), an indication of the broad ecological and geographic distribution of this cosmopolitan weed species. Seed populations were collected at all sites and buried to form local soil seed pool nurseries from which seedling emergence data was collected.

**Table 1**. Local *Chenopodium album* 2005 seed population collection site information (country, latitude, longitude, elevation above sea level) and time (Julian week; calendar date) of field seed harvest and collection; [1]Canada population mixture (%) of two local populations.

| **Seed Collection Site** | | | | **Time of Seed Collection** | |
|---|---|---|---|---|---|
| **Country** | **Latitude** | **Longitude** | **Elevation** | **JW** | **Date** |
| Finland; Jokioinen | N 60° 54’ 19.41“ | E 23° 29’ 56.68“ | 105 m | JW 36 | 7SEP05 |
| Norway; As, Akershus | N 59° 39’ 43“ | E 10° 43’ 35“ | 90 m | JW 36 | 9SEP05 |
| Sweden; ?? | N 58° 29’ ?“ | E 15° 29’ ?“ | ?? m | JW 36 | 8SEP05 |
| Denmark; Flakkeberg:<br>1: DEN-COM<br>2: DEN-LOCAL 2 | 1: N 55° 19’ 21“<br>2: N 55° 19’ 21“ | 1: E 11° 23’ 22“<br>2: E 11° 23’ 22“ | 31 m | 1: JW 34-35<br>2: JW 41 | 1: 22-31AUG05<br>2: 10OCT05 |
| United Kingdom; ?? | N 52° 12’ 18“ | W 1° 36’ 00“ | 49 m | JW 42 | 17OCT05 |
| Czech Republic; Ceske Budejovice | N 48° 58’ 30“ | E 14° 27’ 14“ | 387m | JW 40-44 | [?October?] |
| Canada; Quebec[1]:<br>1: Ste-Clotilde (73%)<br>2: Ste-Bruno (27%) | 1: N 45° 10’ 03“<br>2: N 45° 33’ 01“ | 1: W 73° 40’ 50“<br>2: W 73° 21’ 07“ | 1: 53 m<br>2: 61m | 1: JW 38; 40<br>2: JW 41 | 1: 21SEP-4OCT05<br>2: 11OCT05 |
| Italy; Umbria, Perugia, Papiano | N 42° 57’ 25“ | E 12° 22’ 36“ | 164 m | JW 41 | 12OCT05 |
| United States; Pana, Illinois | N 39° 20’ 19.03“ | W 89° 05’ 28.22“ | 202 m | JW 43 | 26OCT05 |
| Portugal; ?? | N 38° 35’ 17.82“ | W 8° 57’ 28.92“ | 43 m | JW 43 | 25OCT05 |
| Spain; Cordoba | N 37° 49’ 49.59“ | W 4° 54’ 52.53“ | 90 m | JW 32 | 7AUG05 |

## 2.1 The Soil Seed Pool

**2.1.1 Local seed collection**. Seeds from locally adapted *C. album* were collected at all experimental sites from well-established populations in an agricultural field near the experimental burial nursery (Table 1). They were harvested by shaking ripe plants into a paper or cellophane bag, thus collecting only the mature seeds that dehisced naturally at abscission. The collected seeds were hand cleaned and unwanted material was removed by a blower or similar method. Seeds were dried in a room with daylight (but not direct sunlight) at ambient temperatures for one week (20-25° C) and subsequently stored in airtight containers at 4$^o$ C until burial.

**2.1.2 Common Denmark seed population**. A common population was also collected at Flakkeberg, Denmark for burial at all 11 seed burial nursery sites for comparison with the locally adapted seeds. The common seedlot (DENCOM) was collected from August 22-31, 2005 in obtain enough ripe seeds for all eleven burial nurseries. The seeds of the common seedlot were subsequently sent to HerbiSeed (New Farm, Mire Lane, West End, Twyford, England, UK, RG10 0NJ; herbiseed.com). HerbiSeed dried (22$^o$ C, cool air extraction for 8 hours) cleaned the seed, then split the seeds into eleven equal portions, packed them in sealed containers which were distributed to the burial nurseries.

## 2.2 Seed Burial Nurseries

**2.2.1 Local seed burial nursery sites**. Two *C. album* seed populations (DENCOM; locally adapted) were buried at each of the 11 burial nurseries in the autumn of 2005 (Table 2). The burial nursery at each location was placed in full sunlight, and at least 2 m from any wall, hedge or other obstruction that might alter the local microclimate. A small fence with a maximum mesh size of 9 mm was installed around the burial samples, either individually or around the whole area, in order to exclude macrofauna.

**Table 2**. Local *Chenopodium album* seed burial nursery site location information (country, latitude, longitude, elevation above sea level) and time of local seed burial in 2005 (Julian week; calendar date).

| Seed Burial Site | | | | Seed Soil Burial Time | |
|---|---|---|---|---|---|
| **Country** | **Latitude** | **Longitude** | **Elevation** | **JW** | **Date** |
| Finland; Jokioinen | N 60° 48’ 41.75“ | E 23° 28’ 43.87“ | 104m | JW 42 | 19OCT05 |
| Norway; As, Akersus | N 59° 40’ 16“ | E 10° 46’ 17“ | 100m | JW 43 | 24OCT05 |
| Sweden; ?? | N 58° 26’ ?“ | E 15° 28’ ?“ | ?? m | JW 50 | 10DEC05 |
| Denmark; Flakkeberg | N 55° 19’ 28“ | E 11° 23’ 21“ | 31m | JW 44 | 3NOV05 |
| United Kingdom; ?? | N 52° 12’ 18“ | W 1° 36’ 00“ | 49m | JW 43 | 27OCT05 |
| Czech Republic; Ceske Budejovice | N 48° 58’ 28“ | E 14° 27’ 28“ | 386m | JW 43 | 28OCT05 |
| Canada; L’Acadie, Quebec | N 45° 17’ 45“ | W 73° 21’ 12“ | 44m | JW 43 | 28OCT05 |
| Italy; Umbria, Perugia, Papiano | N 42° 57’ 25“ | E 12° 22’ 36“ | 164m | JW 44 | 4NOV05 |
| United States; Urbana, Illinois | N 40.12° ’ ?“ | W 80.26° ’ ?“ | 222m | JW 45 | 7NOV05 |
| Portugal; ?? | N 38° 42’ 34.32“ | W 9° 10’ 57.12“ | 49m | JW 45 | 12NOV05 |
| Spain; Sevilla | N 37° 21’ 6.69“ | W 5° 56’ 19.85 “ | 30m | JW 45 | 11NOV05 |

**2.2.2 Burial soil substrate**. The two populations (DENCOM, locally adapted) were buried in a soil substrate made locally at each location according to the recipe for seed compost (gardeningdata.co.uk/soil/john_innes), comprised of 2:1:1 parts by volume of loam, unfertilized peat and sand, respectively. The loam was sourced locally, and the sand and peat were purchased. At some locations the loam was sterilized. Loam and peat was sieved through a 9 mm sieve before mixing. For each cubic metre of mix 0.6 kg ground limestone and 1.2 kg superphosphate was added. After mixing the substrate was spread out to dry at room temperature for easier handling. All the finished substrate was sieved through a 2.00 mm sieve, and only what passed through the sieve was used.

**2.2.3 Seed burial**. Seed was buried in a 5 cm deep, 25 cm diameter hole dug in the soil for each sample. To contain the seed-soil-mixture within the hole, a fine mesh at least 40 cm diameter was placed into the holes and extended at least 2 cm above the surface. Most nurseries used a double layer of Lutrasil Thermoselect 18 (nonwovens-group.com/bereiche/Lutrasil_18). For each sample 1000 seeds were mixed thoroughly with 2.5 L of dry soil substrate (Sweden, 4000 seeds per sample; Spain, 2000 seeds per sample). The mixing was done by enclosing soil and seeds in a closed container or plastic bag and shaking this around in all directions for at least five minutes. This mixture was then placed on top of the cloth/mesh in the hole. The soil was compacted with a disc and a weight on top, until it was level with the soil surface. The soil was watered with deionised water. Each sample was watered with 100 ml on each of the first three days after initiation, corresponding to approximately 6 mm precipitation. At the Danish location, 12.6 mm precipitation occurred after the second irrigation of 100 ml, and the third was not carried out.

**2.2.4 Burial nursery establishment**. Each local seed burial nursery consisted of 24 samples for counting seedling emergence: twelve samples were established for each of two populations (DENCOM, locally adapted) randomized in two blocks. Six additional soil mixture samples, without seeds, were included for determination of soil water content and seedling emergence from unsterilized loam. Samples were included to anticipate the onset of seedling emergence of DENCOM by loosely covering the sample with a glass plate in the early spring after snow and frost had ceased, allowing room for ventilation. Soil temperature was obtained at least hourly using a logger buried at a depth of 2 cm. In addition, air temperature and precipitation were recorded at nearby meteorological stations, except in Sweden, where precipitation was recorded manually and air temperature was not recorded, and in the Czech Republic, where only air temperature was recorded. In Portugal samples were irrigated with 14.5 mm every day from 18 April 2006 to 31 August 2006. In Spain the samples were irrigated with 10.2 mm on 18, 22 and 25 May 2006. In Italy, samples were irrigated with 20 mm on 11 July 2006.

## 2.3 Soil Disturbance

Each local seed burial nursery included six treatments (one untreated; five disturbances, D1-5) for each of two populations (DENCOM, locally adapted) replicated twice, for a total of 24 samples utilized for counting seedling emergence.

**2.3.1 Soil disturbance methods**. Each early season soil disturbance treatment was carried out by carefully lifting each sample out of the hole in the surrounding local soil and removing the cloth/mesh containing the soil/seed mixture. Then, outside in full daylight, the soil-seed mixture was sieved (4 mm grate; Spain, 2 mm) and returned to its original the mesh-lined hole in the local surrounding soil. Numbers of all emerged and unemerged seedlings were recorded and removed. Prior to soil disturbance , emergence numbers for all sampleswere pooled to represent undisturbed samples up to that date, as many as 12 replicates for these pooled observations.

**2.3.2 Timing of soil disturbances**. Seedling emergence was observed from undisturbed and early season disturbance (5 times) samples (Table 3).

**Table 3**. Timing of experimental early season soil disturbances (D1-5) by location, date, Julian week (JW), or accumulated degree-days (DD, $^{o}$ C) after the first disturbance (D-1); see methods and materials for explanations.

| Location | Latitude | Early Season Soil Disturbances | | | | | | | | | |
|---|---|---|---|---|---|---|---|---|---|---|---|
| | | D-1 | | D-2 | | D-3 | | D-4 | | D-5 | |
| | | Date | JW | DD | JW | DD | JW | DD | JW | DD | JW |
| Finland | N 60°48' | 28MAY06 | 22 | 32 | 22 | 82 | 23 | 143 | 24 | 225 | 25 |
| Norway | N 59°40' | 27APR06 | 17 | 66 | 19 | 104 | 19 | 199 | 22 | 290 | 23 |
| Sweden | N 58°26' | 13APR06 | 15 | 99 | 18 | 150 | 19 | 198 | 19 | 311 | 21 |
| Denmark | N 55°19' | 18APR06 | 16 | 53 | 17 | 109 | 19 | 210 | 20 | 382 | 23 |
| UK | N 52°12' | 31MAR06 | 13 | 36 | 14 | 71 | 16 | 110 | 16 | 154 | 17 |
| Czech Rep | N 48°58' | 18APR06 | 16 | 80 | 17 | 153 | 18 | 192 | 19 | 273 | 19 |
| Canada | N 45°17' | 3APR06 | 14 | 51 | 16 | 122 | 18 | 183 | 19 | 235 | 20 |
| Italy | N 42°57' | 14MAR06 | 11 | 142 | 14 | 191 | 15 | 332 | 17 | 432 | 18 |
| USA | N 40°12' | 15MAR06 | 11 | 23 | 13 | 89 | 15 | 124 | 15 | 229 | 16 |
| Portugal | N 38°42' | 22APR06 | 17 | 119 | 18 | 163 | 18 | 227 | 19 | 329 | 20 |
| Spain | N 37°21' | 1DEC05 | 48 | 855 | 10 | 1034 | 19 | 1119 | 20 | 1170 | 21 |

*Undisturbed*. The samples were undisturbed for the duration of the experiment.
*Disturbance 1 (D-1).* The first soil disturbance was timed for the onset of the first seedling to emerge in any experimental sample. The criteria for this event varied between locations, an indication of the broad geographic conditions in which *C. album* seedling emergence is adapted. Soil disturbance was initiated when the first seedling emerged, or under the glass plate cover (not observed in Italy, Portugal or Spain), or when farmers in surrounding fields began to till their fields (Italy). To predict the onset of emergence, a glass plate was placed on top of the extra seed sample in the early spring of 2006, allowing room for ventilation. The date when the first *C. album* seedlings were observed emerging under this glass plate was considered the onset of recruitment.
*Disturbance 2 (D-2).* The second soil disturbance was timed for the beginning of the spring seedling recruitment period, the first indications of significant numbers beginning to emerge denoting the start of the spring 'flush'.
*Disturbance 3 (D-3).* The third soil disturbance was timed for when 50 degree-days had accumulated after the second disturbance (D-2). The degree-days were calculated as:

$$\sum_i^t(((T_{min}+T_{max})/2)-3)$$

where $T_{min}$ is the daily minimum, and $T_{max}$ the daily maximum, air temperature; 3 is the base air temperature for *C. album*; *I* is the day of D-2; and *t* is the number of days after *i*. In Sweden, soil temperature was used for these calculations.
*Disturbance 4 (D-4).* The fourth soil disturbance was timed for when 100 degree-days had accumulated after the third disturbance (D-3).
*Disturbance 5 (D-5).* The fifth soil disturbance was timed for when 200 degree-days had accumulated after the fourth disturbance (D-4).

Local experimental conditions mitigated the timing and implementation of these disturbance timings. This was most apparent in Spain where significant numbers of seedlings emerged in the autumn of burial, 2005. As such, the first disturbance (D-1) in Spain occurred on 1 December 2005.

Despite these mitigations, and the experimental necessity to adapt these general rules for local timing of disturbance, several overarching generalizations can be made. Very generally, the onset of seedling emergence (D-1) occurred earlier with decreasing latitude. Subsequent disturbances (D-2 to D-5) occurred generally 0-3 weeks apart (excepting Spain, D-2, D-3), providing a stable basis for comparison of disturbance timings within and between geographic locations (Table 3).

**2.3.3 Cropping system calendar**. The typical, historical, times of predictable agricultural disturbances (e.g. seedbed preparation, planting, primary weed control tactic, crop harvest) or winter period (frozen or cold soil, seasonal conditions not conducive to growth) for each of the seed collection sites were determined to establish the conditions in which locally adapted weed populations emerged, and to which they were historically adapted (Table 4).

**Table 4**. Historical (1995-2004) times (JW, Julian week) and qualities of cropping systems, cropping practices and disturbances for each local *Chenopodium album* seed population collection site; [1]weed control (tillage, herbicide application) time.

| Seed Population Collection Site | Crop Rotation and System | Seedbed Preparation | Planting Time | Primary Weed Control Time[1] | Crop Harvest Time |
|---|---|---|---|---|---|
| Finland; Jokioinen | Spring cereals; Grassland | JW 18-22 | JW 18-22 | JW 24-25 | 1-JW 34-36 |
| Norway, As, Akershus | 1-Spring cereals<br>2-Swedes<br>3-Winter wheat | 1-JW 16-19<br>2-JW 16-19<br>3-JW 36-37 | 1-JW 16-19<br>2-JW 16-19<br>3-JW 36-37 | 1-JW 21-24<br>2-JW 21-24<br>3-JW 17-19 | 1-JW 32-37<br>2-JW 36-39<br>3-JW 30-34 |
| Sweden; ?? | 1-Spring cereals<br>2-Oilseed rape | JW 16-18 | JW 16-18 | JW 22-24 | 1-JW 33-35<br>2-JW 31-33 |
| Denmark; Flakkeberg:<br>1-DENCOM<br>2-LOCAL | 1-Spring cereals<br>2-Winter wheat | 1-JW 14-16<br>2-JW 36-40 | 1-JW 14-16<br>2-JW 36-40 | 1-JW 17-20<br>2-JW 39-44 | 1-JW 32-34<br>2-JW 30-32 |
| United Kingdom; ?? | Vegetable | JW 10-19 | JW 10-19 | JW 12-20: Herbicide<br>JW 24-32: Hand hoe | JW 33-37 |
| Czech Republic; Ceske Budejovice | 1-Spring cereals<br>2-Swede rape | 1-JW 14-18<br>2-JW 40-41 | 1-JW 21-22<br>2-38-39 | 1-JW 14-19<br>2-JW 40-44 | JW 31-37 |
| Canada , Quebec:<br>1-Sainte-Cotilde<br>2-Saint-Bruno | 1-Onions<br>2-Sweet maize | 1-JW 17-19<br>2-JW 19-21 | 1-JW 17-19<br>2-JW 19-21 | 1-JW 17-23<br>2-JW 21-25 | 1-JW 35-37<br>2-JW 31-36 |
| Italy; Umbria, Perugia, Papiano | Maize; Soybean; Sunflower; Vegetables | JW 6-17 | JW 7-18 | JW 6-23 | JW 33-43 |
| United States; Pana, Illinois | 1-Maize<br>2-Forage | 1-JW 16-18<br>2-JW 35-37 | 1-JW 16-18<br>2-JW 35-37 | JW 17-18 | JW 40-41 |
| Portugal; ?? | 1-Potato; irrigated<br>2-Maize; irrigated | JW 7-15 | JW 11-17 | JW 11-22 | JW 35-44 |
| Spain; Cordoba | 1-Maize; irrigated<br>2-Cotton; irrigated<br>3-Wheat | 1-JW 5-13<br>2-JW 10-17<br>3-JW 43-48 | 1-JW 6-13<br>2-JW 16-20<br>3-JW 48-52 | 1-JW 16-20<br>2-JW 16-20<br>3-JW 5-10 | 1-JW 40-44<br>2-JW 40-44<br>3-JW 31-35 |

### 2.4 Seedling Emergence

**2.4.1 Recruitment magnitude**. *Chenopodium album* seedling emergence (number per sample) was recorded on at least a weekly basis for the duration of these experiments, with the exception of times when the ground was frozen or covered with snow. At peak recruitment times, emergence was monitored 2-3 times a week. Emerged *C. album* were counted, recorded and cut off at soil surface. At counting, a ring with a diameter 4 cm less than the sample (that is, 21 cm) was used as the sample area. Only *C. album* emerging within this inner ring were counted (of 700 possible) to avoid edge effects (e.g. drying out, light penetration) influencing the number of emerged seedlings. To allow comparisons between all locations, emergence numbers for Sweden were divided by 4 (2800 seeds in the central 0.7 $m^2$ sampling area;) and by 2 in Spain (1400 seeds in the central 0.7 $m^2$ sampling area). The *C. album* seedlings emerging outside the ring were also counted, cut off and recorded separately (data not reported).

**2.4.2 Recruitment pattern**. Data presented assumes that any gap in emergence of one Julian week (JW) is evidence of discontinuous seedling emergence of that population at that location during that seasonal time: separate seedling cohorts. Similarly, only emergence numbers of one mean seedling per JW per treatment are reported.

# 3 RESULTS

## 3.1 *Chenopodium album* Seedling Emergence

*Chenopodium album* seedling emergence is a function of the opportunity available to a local population at a particular time, cumulatively an annual life history cycle, and place. The setting within which *Chenopodium album* evolution occurs is the population at a specific location. An understanding of seedling recruitment is provided by observations of where and when local opportunity space-time is seized and exploited. Parsing aggregate behavior allows additional insights into the complexity of *C. album* seedling emergence. Emergence can be parsed into population, location and soil disturbance parameters. Each emergence parameter is revealed by its seedling emergence cohort structure (magnitude, number, duration).

## 3.2 Aggregate Seedling Emergence

Aggregated seedling emergence observation were parsed into population, location and soil disturbance parameters.

**3.2.1 Aggregate emergence**. Cumulative annual *C. album* seedling emergence is provided in calendar 18. Almost continuous recruitment occurred in some nursery in each Julian week, when accumulated over population (LOCAL, DEN-COM), location (10), and soil disturbance (D1-5). Exceptional, brief periods of no emergence occurred in the winter (JW52-1, 7-8) and autumn (JW43) of 2005-2006.

**3.2.2 Population emergence**. Parsing aggregate behavior by population provided more temporal detail into the structure of winter and autumn no emergence periods (calendar 19, bottom). The LOCAL population had more exceptional no emergence periods in both the winter and autumn, compared to fewer in DEN-COM in both seasons.

**3.2.3 Location emergence**. Parsing emergence by location provided more detailed information about recruitment structure timing (calendar 19, top). A more complex and detailed pattern of seedlings seizing opportunity in patterns unique to each location was provided. The winter and autumn emergence cohorts disappeared in many northern locations. Continuous annual emergence did not occur at any individual location. Periods of no emergence were considerably longer in all locations relative to the overall aggregate behavior. The duration of individual seasonal emergence cohorts was influenced by several factors other than latitude, including heat (e.g. shorter in the north), moisture (e.g. less in arid Spain) and seed pool size (e.g. Sweden).

**3.2.4 Soil disturbance emergence**. Seedling emergence in undisturbed and disturbed soil did not occur within identical seasonal times. Therefore direct cohort structure comparisons between these regimes are confounded (calendar 20, 21, 22).

What lies underneath this aggregated view of seedling recruitment? Comparisons of local and the introduced population (DEN-COM) seedling recruitment within and between locations are influenced by soil disturbance and seasonal time. Seedling recruitment cohort structure consists of seedling magnitude (number of emerged seedlings), annual cohort number, cohort duration (time, JW; period, number of weeks), and seasonal cohort pattern.

## 3.3 Seedling Emergence Cohort Structure

### 3.3.1 Seedling Recruitment Magnitude

Seedling recruitment magnitude varied by population, location, disturbance and season.

Soil disturbance. DEN-COM population recruitment in undisturbed soils was usually greater than LOCAL. In disturbed soils either population could be greater at a location. DEN-COM emergence was usually greater in the earliest (D1), while LOCAL was usually greater in later (D2-5) disturbances. Peak emergence of both populations was often correlated with latitude: earlier disturbance times north, later times south.

Season. Seedling emergence magnitude was greatest in the spring, and less so in the summer and autumn. In the spring each disturbed population had greater magnitude at some locations (DEN-COM, 5; LOCAL, 4). When DEN-COM magnitude was greater than LOCAL in the spring, it was often less so in the subsequent summer. In the summer LOCAL was usually greater (6 sites) than DEN-COM (1 site).

**3.3.1.1 Undisturbed soil**. DEN-COM population seedling emergence was greater than local at all locations (8) except Sweden and Spain in undisturbed soil (table 5), and in the spring (calendars 1-11).

**3.3.1.2 Disturbed soil**. Soil disturbance did not consistently affect differences in emergence between the two populations (table 5). DEN-COM magnitude was usually greater than LOCAL in Finland, Denmark, Canada and UK. LOCAL emergence was usually greater in Norway, Sweden, Italy, Czech Republic and Portugal. When populations differed at several times at a location, DEN-COM was greater earlier, while LOCAL was greater in the later disturbance times (Norway, Czech, Canada, Italy; except UK). Maximum emergence magnitude occurred earlier in northern-most locations (D1-2) than in southern-most nurseries (D3-4) in both populations. DEN-COM peak emergence at mid-latitude Canada and Italy was earlier (D2) than that of LOCAL (D3-4).

**3.3.1.3 Season**. In undisturbed soil, the greatest magnitude of seedling emergence occurred in the spring, at which time DEN-COM was greatest at most locations (except Sweden, Portugal) (table 6). In disturbed soil, the most emergence occurred in the spring, less in summer, the least in autumn (table 7). DEN-COM spring emergence magnitude was greater than LOCAL in all or most disturbance regimes in Finland, Denmark, Czech Rep., Canada and UK; but only in Finland in the summer. Local spring emergence magnitude was greater than DEN-COM in all or most disturbance regimes in Norway, Sweden, Italy and Portugual; in the summer magnitudes were also greater in these same countries as well as in Czech Rep. and UK. In the spring each disturbed population had greater magnitude at some locations (DEN-COM, 5; LOCAL, 4). In the summer LOCAL was usually greater (6 sites) than DEN-COM (1 site).

### 3.3.2 Seedling Recruitment Cohort Number

Seedling recruitment can be characterized by the number of discrete cohorts that emerge in a growing season from one population at one location. A recruitment cohort is that group of seedlings (at least one per JW) that emerge in continuous weeks. Cohorts are separated by a gap of at least one JW in which no emergence occurs. Cohort number was the least sensitive structural parameter studied, easily confounded by seed burial number (e.g. Sweden) or growing conditions (e.g. Spain).

Undisturbed soil. Population cohort number in undisturbed soil was similar, or favored either DEN-COM or LOCAL, depending on location. LOCAL tended to have similar or smaller cohort numbers compared to DEN-COM. Some tendency to increased cohort number with decreased latitude was observed with both populations, possibly due to longer favorable season duration.

Disturbed soil. Most often observed was LOCAL cohort number greater than DEN-COM. DEN-COM peak cohort number occurred at later or similar disturbance times to the north, earlier to the south, compared to LOCAL.

**3.3.2.1 Undisturbed soil**. With several exceptions, cohort number increased with decreasing latitude for both populations in undisturbed soil (table 8). Low undisturbed soil cohort number in Sweden may be due to more continuous emergence arising from four times as many seed buried. Low cohort number in Spain was probably due to very dry conditions. Cohort number in undisturbed soil were similar (4 sites: Finland, Sweden, Denmark, Canada and UK), DEN-COM was greater (4 sites, mostly southern: Norway, Italy, Portugal and Spain), or LOCAL was greater in Czech Republic, relative to the other population.

**3.3.2.2 Disturbed soil**. LOCAL cohort number in disturbed soil was often greater than DEN-COM (5 sites: Denmark, Czech Rep., Italy, UK, Spain); sometimes less (2 sites: Finland, Portugal) or equivocal (2 sites: Norway, Canada) (table 8). Comparison of the time of disturbance with greatest number of cohorts was often correlated with latitude. DEN-COM had peak cohort number at later or similar disturbance times than LOCAL at Italy and northern locations, while LOCAL had peak numbers at later disturbance times in UK and Portugal.

**3.3.3 Seedling Recruitment Duration**

The seasonal duration of seedling emergence is characterized by two parameters, period and time. Period duration is the Julian weeks (JW) within which emergence occurred, including intervening weeks of no emergence between cohorts; the recruitment season length. Time duration is the number of weeks in which some emergence occurred, not including intervening weeks with no emergence between cohorts; the total time of active seedling recruitment.

Seedling emergence season duration is the consequence of both locally available resources and conditions (heat-winter, moisture, light, nitrate) and cropping disturbances: local opportunity spacetime. Season length is not a simple latitude-heat gradient.

Undisturbed soil. Recruitment season length, and the number of weeks in which emergence occurred, were both inversely related to latitude for both populations with several local heat-moisture exceptions. Spain was an exception due to the lack of seasonal moisture. UK was an exception due to the moderating gulf stream influences on that northerly island. Emergence time duration had no clear latitude pattern of dominance of one population longer than the other at all locations. Emergence period duration had an unusual latitude pattern consisting of three groups: DEN-COM longer at the extreme north and south; LOCAL longer at mid-latitudes; population similarity at mid-north and mid-south localities.

Disturbed soil. In disturbed soil, no consistent pattern of seedling emergence duration period and latitude was observed. Some tendency to longer duration period with decreasing latitude for the LOCAL population in disturbed soil was observed. The range of disturbed soil emergence duration periods for both populations was similar (JW11-46); with the exception of short Spain. The maximum duration period occurred in the first disturbance regime (D1) in both populations in all locations; the least in later disturbances.

In disturbed soil, the range of seedling emergence duration times in the LOCAL was greater than that in the DEN-COM population. Generally LOCAL seedling emergence duration time was inversely related to location latitude in disturbed soil. Generally DEN-COM duration times were similar (5-9 JW) with latitude in disturbed soil. Apparently, LOCAL is able to exploit locally available soil disturbance opportunity more completely with changes in location latitude than is DEN-COM. All locations had some disturbance regimes with population duration time differences. At most locations LOCAL emergence duration time was longer in some disturbance regimes, while DEN-COM duration time was longer only in the extreme latitude locations (Finland, Spain). Similar LOCAL and DEN-COM emergence duration times was frequently observed in disturbed soil. The maximum duration time occurred in the first disturbance regime (D1) in both populations in all locations; the least in later disturbances.

**3.3.3.1 Duration period**.

**3.3.3.1.1 Undisturbed soil**. The duration within which seedling emergence occurred in undisturbed soil was inversely related to latitude; except in Denmark and Spain (LOCAL) and Czech Republic (DEN-COM) (table 9). If intervening cold winter period is not considered, the LOCAL population in Denmark follows this generalization. Spain was an exception due to the lack of seasonal moisture for both populations.

Within a location, duration period was similar (5 sites), DEN-COM longer (2 sites), or LOCAL longer (3 sites), relative to the other population (table 9). A latitudinal pattern of population recruitment duration period differences in undisturbed soil was observed. There exist three approximate clusters of population differences: DEN-COM period was longer at the extreme northern (Finland) and southern (Spain) locations; LOCAL longer at the mid-latitudes (Denmark, Czech Rep., Canada); and similar population duration periods in mid-north (Norway, Sweden) and mid-south (Italy, Portugal) locations.

**3.3.3.1.2 Disturbed soil**. In disturbed soil, no consistent pattern of seedling emergence duration period and latitude was observed, as with duration time (table 10). The range of emergence duration periods for disturbance regimes (D1-5) in both populations was similar: LOCAL (JW11-46) and DEN-COM (JW 11-45); with the exception of Spain which was short due to lack of seasonal moisture. Italy had the longest recruitment duration periods in all disturbances for both populations. The disturbed DEN-COM population had no clear pattern of period with latitude. Exceptions included Norway, Denmark, Czech Republic, Canada, UK all with short duration periods. Duration period in disturbed soil for the LOCAL population exhibited some tendency to increase with decreasing latitude. A notable exception was Canada, with the shortest periods in most disturbance regimes. No clear dominance of one population in disturbed soil emergence duration period was observed. Of fifty (50) possible population comparisons (within 10 locations, each with 5 disturbance regimes) of emergence duration period: the most frequent was LOCAL longer than DEN-COM (22: Norway, Denmark, Czech Rep., UK). Duration period was similar in 16 (Sweden, Italy, Portugal). Disturbed period in DEN-COM was longer than LOCAL in Finland. The disturbance regime with maximum emergence duration period was the first (D1) in both populations in all locations. The longest duration period was never in the latest (D5) with DEN-COM, and never in late regimes (D3-4) with LOCAL, populations over all locations.

**3.3.3.2 Duration time**.

**3.3.3.2.1 Undisturbed soil**. The number of weeks in which some seedling emergence occurred in undisturbed soil was usually inversely related to latitude: increasing duration time with decreasing latitude (table 9). Spain was an exception due to the lack of seasonal moisture. UK was an exception due to the moderating gulf stream influences on that northerly island. Duration time for DEN-COM in Canada was also somewhat shorter to those locations immediately north and south. Within a location, duration time was similar (4 sites), DEN-COM longer (4 sites), or LOCAL longer (2 sites), relative to the other population.

**3.3.3.2.2 Disturbed soil**. The range of seedling emergence duration times in disturbed soil was greater in the LOCAL (0-17 JW) than in DEN-COM (0-11 JW) population (table 10). Generally LOCAL seedling emergence duration time was inversely related to location latitude for all five disturbance regimes. Exceptions included some disturbance regimes with shorter time in Denmark, Canada, Portugal, and Spain; and longer in Italy. Generally DEN-COM duration times were similar (5-9 JW) with location latitude for all five disturbance regimes. Exceptions included some disturbance regimes with longer time in Italy and UK; shorter in Norway, Denmark, Canada, and Spain (lack of seasonal moisture). Of fifty (50) possible population comparisons (within 10 locations, each with 5 disturbance regimes) of emergence duration time: the most frequent was LOCAL similar to DEN-COM (25 instances). All locations had some disturbance regimes with population duration time differences. At most locations (8 of 10) LOCAL emergence duration time was longer than DEN-COM within some of the five disturbance regimes. DEN-COM duration time was longer than LOCAL in 6 disturbance regimes, primarily in the extreme latitude locations (Finland, 4; Spain, 1). The disturbance regime with maximum emergence duration time was the first (D1) in both populations in all locations. The longest duration period was never in the latest (D5) with DEN-COM, or late (D-4) with LOCAL, populations over all locations.

### 3.3.4 Seedling Recruitment Pattern

Seasonal cohort pattern of seedling emergence was revealed by comparison of the two populations within a location and disturbance. Four comparable seasonal cohorts were observed: autumn of seed burial (2005), and three seasons of the following year (2006; spring, summer, autumn). The winter cohort occurred in only three nurseries. Seedling emergence in undisturbed and disturbed soil did not occur within identical seasonal times, therefore direct cohort structure comparisons between these regimes are not appropriate. Indirect comparisons of soil disturbance are possible within the same seasonal cohort by the change in emergence timing (early, late) among all locations (summary 1-3).

<u>Burial autumn cohort</u>. The burial autumn cohort arose in five locations in undisturbed, only one (Spain) in disturbed soil, generally from the more southerly of ten locations. DEN-COM pattern shifted either earlier or later than LOCAL in all burial autumn cohorts.

<u>Spring cohort</u>. The spring seedling emergence cohort was the most important, occurring in all locations in both undisturbed and disturbed soil. In undisturbed soil LOCAL was most likely to occur at later or similar times, than DEN-COM among locations. In disturbed soil DEN-COM was most likely to be similar to, or occur earlier, than LOCAL. The effect of disturbance made it more likely DEN-COM would occur at similar and earlier times, LOCAL earlier and later, compared to undisturbed soil.

<u>Summer cohort</u>. The summer seedling emergence cohort was the second most important seasonal cohort, occurring in all locations except Spain in both undisturbed and disturbed soil. In undisturbed soil, any population difference was possible, except LOCAL did not occur later than DEN-COM. In disturbed soil LOCAL was most likely to be earlier or later than DEN-COM. The effect of disturbance made it more likely LOCAL would occur later than DEN-COM compared to undisturbed soil. The disturbance effect also made it less likely populations would be similar, or that DEN-COM occurred earlier or later than LOCAL, compared to undisturbed soil. Disturbance had opposite effects on spring and summer cohorts. The spring DEN-COM cohort was more likely to be similar, more or less than LOCAL; in the summer less likely to change in these directions. In the summer, LOCAL was more likely to be later than DEN-COM; in the spring less likely.

<u>Autumn cohort</u>. The autumn seedling emergence cohort was of lesser importance, and occurred in only four locations in both undisturbed and disturbed soil. In undisturbed soil any population difference was possible, except LOCAL did not occur later than DEN-COM; the same likelihoods found in the undisturbed summer cohort. In disturbed soil it was equally likely for DEN-COM to occur earlier or later, or LOCAL later. The effect of disturbance was to make LOCAL more likely to occur later, less likely to occur earlier or be similar, compared to DEN-COM in undisturbed soil.

The combined effect of disturbance on 2006 cohorts was no change in similar populations; DEN-COM contracting (less early and late); LOCAL shifting to later times (less early, more late); compared to cohort patterns in undisturbed soil.

**3.3.4.1 Burial autumn cohort**. The burial autumn cohort arose in five of ten locations in undisturbed soil, generally from more southerly nurseries (table 11). DEN-COM occurred either earlier or later than LOCAL in all burial autumn cohorts. The undisturbed DEN-COM cohort pattern differed from LOCAL in all five locations: two earlier, in three later. The burial autumn cohort occurred only in Spain in disturbed soil due to very early accumulation of heat units dictating experimental soil disturbance (see methods and materials) (table 11). In that instance DEN-COM emerged earlier than LOCAL.

**3.3.4.2 Spring cohort**. Seedling emergence in the spring was the most important seasonal cohort, occurring in all locations in both undisturbed and disturbed soil (table 11, summary 1). In undisturbed soil LOCAL was most likely to occur later, or occur at a similar time, compared to DEN-COM among locations in the spring cohort (summary 1). In disturbed soil DEN-COM was most likely to be similar to, or occur earlier, relative to LOCAL in the spring cohort. With disturbance, DEN-COM was more likely to occur at similar and earlier times, and LOCAL less likely to occur earlier and later, compared to undisturbed soil.

**Summary 1**. Spring seedling recruitment pattern changes in cohort timing (similar, earlier, later; number of locations) of populations (DEN-COM, LOCAL).

| **SPRING Cohort: Changes in Seedling Recruitment Pattern** | | | |
|---|---|---|---|
| Population Comparisons | Soil Disturbance: Change in Number of Locations | | |
| | Un-disturbed | Change | Disturbed |
| Similar | 3 | more → | 5 |
| DEN-COM earlier | 1 | more → | 3 |
| DEN-COM later | 1 | = | 1 |
| LOCAL earlier | 2 | less → | 1 |
| LOCAL later | 4 | less → | 2 |
| Total | 11 | | 12 |
| Absent Locations | 0 | | 0 |

**3.3.4.3 Summer cohort**. Seedling emergence in the summer was the second most important seasonal cohort, occurring in all locations except Spain in both undisturbed and disturbed soil (table 11, summary 2). In undisturbed soil any population difference was possible (similar, earlier, later), except LOCAL later than DEN-COM, which did not occur (summary 2). At most locations, LOCAL was most likely to be earlier or later than DEN-COM in disturbed soil. With disturbance LOCAL was most likely to occur later than DEN-COM compared to undisturbed soil. Disturbance was also associated with less likelihood of populations being similar, or DEN-COM occurring earlier of later than LOCAL, than undisturbed. The opposite effect of disturbance on populations was observed in the summer compared to the spring cohort. In the spring DEN-COM was more likely to be similar, more or less than LOCAL; in the summer less likely to change in these directions. In the summer, LOCAL was more likely to be later than DEN-COM; in the spring less likely.

**Summary 2**. Summer seedling recruitment pattern changes in cohort timing (similar, earlier, later; number of locations) of populations (DEN-COM, LOCAL).

| **SUMMER Cohort: Changes in Seedling Recruitment Pattern** | | | |
|---|---|---|---|
| Population Comparisons | Soil Disturbance: Change in Number of Locations | | |
| | Un-disturbed | Change | Disturbed |
| Similar | 3 | less → | 2 |
| DEN-COM earlier | 3 | less → | 0 |
| DEN-COM later | 2 | less → | 1 |
| LOCAL earlier | 3 | = | 3 |
| LOCAL later | 0 | more → | 5 |
| Total | 11 | | 11 |
| Absent Locations | 1 | | 1 |

**3.3.4.4 Autumn cohort**. Seedling emergence in the autumn cohort was of lesser importance, and occurred in only four of ten locations in both undisturbed and disturbed soil (table 11, summary 3). In undisturbed soil all population differences were equally possible (similar, earlier, later), except LOCAL later than DEN-COM, which did not occur (summary 3); the same likelihoods as in the undisturbed summer cohort. In disturbed soil it was equally likely that DEN-COM would occur earlier or later, or LOCAL later, at one of the four locations. The effect of disturbance was to make LOCAL more likely to occur later, less likely to occur earlier, and less likely to be similar to DEN-COM, compared to undisturbed soil.

**Summary 3**. Autumn seedling recruitment pattern changes in cohort timing (similar, earlier, later; number of locations) of populations (DEN-COM, LOCAL).

| **AUTUMN Cohort: Changes in Seedling Recruitment Pattern** | |
|---|---|
| Population | Soil Disturbance: Change in Number of |

| Comparisons | Locations | | |
|---|---|---|---|
| | Un-disturbed | Change | Disturbed |
| Similar | 1 | less → | 0 |
| DEN-COM earlier | 1 | = | 1 |
| DEN-COM later | 1 | = | 1 |
| LOCAL earlier | 1 | less → | 0 |
| LOCAL later | 0 | more → | 2 |
| Total | 4 | | 4 |
| Absent Locations | 6 | | 6 |

**3.3.4.5 All cohorts**. The net effect of disturbance in the combined spring, summer and autumn cohorts compared to that in undisturbed soil is not change in the occurrence of populations being similar (summary 4). DEN-COM cohort pattern contracts: less likely to be earlier and later than LOCAL. LOCAL pattern shifts: less earlier and more later than DEN-COM.

**Summary 4**. Spring, summer and autumn total seedling recruitment pattern changes in cohort timing (similar, earlier, later; number of locations) of populations (DEN-COM, LOCAL).

| **SPRING-SUMMER-AUTUMN Cohorts: Changes in Seedling Recruitment Pattern** | | | |
|---|---|---|---|
| Population Comparisons | Soil Disturbance: Change in Number of Locations | | |
| | Un-disturbed | Change | Disturbed |
| Similar | 7 | = | 7 |
| DEN-COM earlier | 5 | less → | 4 |
| DEN-COM later | 4 | less → | 3 |
| LOCAL earlier | 6 | less → | 4 |
| LOCAL later | 4 | more → | 9 |
| Total differences | 26 | | 27 |
| Absent Locations | 7 | | 7 |

**3.3.4.6 Undisturbed soil populations**.

**3.3.4.6.1 DEN-COM in Denmark and other locations**. Comparisons of changes in seasonal seedling emergence cohort patterns of DEN-COM in Denmark (D-C/Den) were made with both populations (DEN-COM, LOCAL) at all locations.

Burial autumn, winter cohorts. Seedling emergence cohorts absent in most locations compared to D-C/Den (calendars 12-13). LOCAL and DEN-COM present only in southern-most locations, and then later, compared to D-C/Den.

Spring cohort. Seedling emergence cohorts absent in all locations in some weeks compared to D-C/Den (calendars 12-13). The absence of emergence in LOCAL and DEN-COM emergence patterns was associated with shifts to earlier emergence in early spring at all locations south of Denmark; as well as shifts to later emergence in late spring at most locations north and south of Denmark. Emergence was absent, without shifts, in extreme north (Finland) and south (Spain) locations in both populations compared to D-C/Den.

Summer, autumn cohorts. LOCAL and DEN-COM seedling emergence cohorts occurred later than those of D-C/Den at most locations in the summer, and at some locations in the autumn.

**3.3.4.6.2 Populations at a location**. Few differences in seedling emergence cohort patterns between LOCAL and DEN-COM populations were observed: local and the population introduced from Denmark exploited opportunity in a similar manner (calendar 14, summary 5). When they occurred, the most likely difference between LOCAL compared to DEN-COM at a location was to be absent, or a change in emergence pattern in the summer cohort.

**Summary 5**. Difference in LOCAL seedling emergence pattern (absent, earlier, later) compared to the DEN-COM population within 10 locations and 5 seasonal cohorts.

| | LOCAL Difference versus DEN-COM | | | |
|---|---|---|---|---|
| Cohort | Absent | Early | Late | Total of 30 |
| Burial Autumn | 4 | 0 | 0 | 4 |
| Winter | 1 | 2 | 1 | 4 |
| Spring | 2 | 1 | 3 | 6 |
| Summer | 4 | 1 | 6 | 11 |
| Autumn | 2 | 0 | 1 | 3 |
| Total of 50 | 13 | 4 | 10 | |

**3.3.4.7 Disturbed soil populations**. [calendars 15-16]

**3.3.4.7.1 Effect of disturbance on seasonal cohort pattern**. No emergence in either population predominated in the autumn cohort, and to a much lesser extent in the summer, and infrequently in the spring cohort (summary 6-7).

<u>Local population</u>. Among all locations, soil disturbance affected the LOCAL population timing pattern of seedling emergence in several ways in the several seasonal cohorts (table 12, summary 6). Delayed LOCAL seedling emergence was the most likely outcome with soil disturbance, and the primary outcome in the spring and summer cohorts. Emergence absent in disturbed soil when present in undisturbed soil was the second most likely outcome of LOCAL, and the primary outcome in the autumn cohort. Early LOCAL emergence in disturbed compared to undisturbed soil occurred in some fewer instances, and was the least likely outcome in the spring cohort. A pattern of similar LOCAL emergence timing in undisturbed and disturbed soil was the least likely outcome in the summer and autumn cohorts. No emergence in either population predominated in the autumn cohort, and to a much lesser extent in the summer.

**Summary 6**. Effect of disturbance on LOCAL population seasonal cohort timing pattern (similar, earlier, later than undisturbed) and occurrence (disturbed absent, no cohort of either population) (table 12).

| **LOCAL Seasonal Cohort Seedling Recruitment Pattern** | | | | |
|---|---|---|---|---|
| Disturbed compared to Undisturbed | Disturbance differences: Number of Locations | | | |
| | Spring | Summer | Autumn | Total |
| Disturbed SIMILAR | 5 | 9 | 0 | 14 |
| Disturbed EARLIER | 4 | 16 | 4 | 24 |
| Disturbed LATER | 36 | 25 | 4 | 65 |
| Disturbed ABSENT | 20 | 20 | 9 | 49 |
| No cohort | 0 | 9 | 44 | 53 |

DEN-COM population. Among all locations, soil disturbance affected the DEN-COM population timing pattern of seedling emergence in several ways in the several seasonal cohorts (table 12, summary 7). Emergence absent in disturbed soil when present in undisturbed soil was the primary outcome of DEN-COM, and the primary outcome in the summer and autumn cohort. Delayed DEN-COM seedling emergence was the second most likely outcome with soil disturbance, and the primary outcome in the spring cohort. Early DEN-COM emergence in disturbed compared to undisturbed soil was the least likely outcome in the spring and summer cohorts. A pattern of similar DEN-COM emergence timing in undisturbed and disturbed soil occurred in some fewer instances, and was the least likely outcome in the autumn cohort.

**Summary 7**. Effect of disturbance on DEN-COM population seasonal cohort timing pattern (similar, earlier, later than undisturbed) and occurrence (disturbed absent, no cohort of either population) (table 12).

| **DEN-COM Seasonal Cohort Seedling Recruitment Pattern** | | | | |
|---|---|---|---|---|
| Disturbed compared to Undisturbed | Disturbance differences: Number of Locations | | | |
| | Spring | Summer | Autumn | Total |
| Disturbed SIMILAR | 12 | 9 | 2 | 23 |
| Disturbed EARLIER | 1 | 8 | 3 | 12 |
| Disturbed LATER | 29 | 8 | 3 | 40 |
| Disturbed ABSENT | 12 | 20 | 13 | 45 |
| No cohort | 4 | 14 | 32 | 50 |

**3.3.4.7.2 Effect of disturbance timing on emergence pattern**. No emergence in either population occurred relatively consistently among the disturbance regimes, D1-5 (summary 8-9).

LOCAL population. Among all locations, soil disturbance affected the LOCAL population timing pattern of seedling emergence in several ways in the several soil disturbance regimes (D1-5; table 12, summary 8). Delayed LOCAL seedling emergence was the primary outcome with soil disturbance. Emergence absent in disturbed soil when present in undisturbed soil was the second most likely outcome with soil disturbance of LOCAL. Early LOCAL emergence in disturbed compared to undisturbed soil occurred in some fewer instances. A pattern of similar LOCAL emergence timing in undisturbed and disturbed soil was the least likely outcome.

**Summary 8**. Effect of disturbance on LOCAL population seasonal cohort timing pattern (similar, earlier, later than undisturbed) and occurrence (disturbed absent, no cohort of either population) (table 12).

| **LOCAL Disturbed Seedling Recruitment Pattern** | | | | | | |
|---|---|---|---|---|---|---|
| Disturbed compared to Undisturbed | Disturbance differences: Number of Locations | | | | | |
| | D1 | D2 | D3 | D4 | D5 | Total |
| Disturbed SIMILAR | 7 | 2 | 1 | 2 | 1 | 13 |
| Disturbed EARLIER | 4 | 4 | 4 | 4 | 7 | 23 |
| Disturbed LATER | 10 | 14 | 12 | 13 | 16 | 65 |
| Disturbed ABSENT | 8 | 11 | 12 | 8 | 8 | 47 |
| No cohort | 11 | 11 | 11 | 11 | 9 | 53 |

DEN-COM population. Among all locations, soil disturbance affected the DEN-COM population timing pattern of seedling emergence in several ways in the several soil disturbance regimes (table 12, summary 9). Emergence absent in disturbed soil when present in undisturbed soil was the most likely outcome with soil disturbance of DEN-COM. Delayed DEN-COM seedling emergence was the second most likely outcome with soil disturbance. A pattern of similar DEN-COM emergence timing in undisturbed and disturbed soil occurred in some fewer instances. Early DEN-COM emergence in disturbed compared to undisturbed soil was the least likely outcome.

**Summary 9**. Effect of disturbance on DEN-COM population seasonal cohort timing pattern (similar, earlier, later than undisturbed) and occurrence (disturbed absent, no cohort of either population) (table 12).

| **DEN-COM Disturbed Seedling Recruitment Pattern** | | | | | | |
|---|---|---|---|---|---|---|
| Disturbed compared to Undisturbed | Disturbance differences: Number of Locations | | | | | |
| | D1 | D2 | D3 | D4 | D5 | Total |
| Disturbed SIMILAR | 8 | 3 | 3 | 4 | 4 | 22 |
| Disturbed EARLIER | 1 | 2 | 2 | 2 | 5 | 12 |
| Disturbed LATER | 6 | 10 | 7 | 9 | 8 | 40 |
| Disturbed ABSENT | 8 | 9 | 11 | 9 | 8 | 45 |
| No cohort | 10 | 11 | 10 | 10 | 9 | 50 |

**3.3.4.8 Local recruitment cohort patterns**.

Cohort emergence pattern structure by location: recruitment cohort pattern; population patterns; experimental disturbance (table 14).

**3.3.4.8.1 Finland**. *C. album* seedling emergence in Finland commenced at the end of the planting period in this spring cereal-grassland cropping system. Four distinct recruitment cohorts arose around the temporal interfaces separating CSD operations: first at planting-weed control; second after summer solstice in the early post-layby period; third at harvest; and the fourth in the autumn. The structure of seedling emergence cohorts at Finland may be an adaptive means by which *C. album* searches for, and exploits, recruitment opportunity just prior to, and after, predictable cropping system disturbances. These times coincide with intervening periods of no disturbance, opportunity spacetime. DEN-COM recruitment at the Finnish common nursery was greater than that of the local population in all cohorts. Late emergence was greater, and lasted for a longer time, in DEN-COM relative to the local population. Experimental disturbances extended the duration of each cohort, and resulted in the appearance of two new late season cohorts.

*Recruitment cohort pattern*. Recruitment of *C. album* seedlings began at the end of planting (JW 22) (table 14). From this time until after layby (JW 26), seedlings emerged in largest two cohorts of the season. The late spring 1st cohort ended just after summer solstice when weed control ceased (JW25), with the exception of two experimental disturbances (D2-3) in the local population which bridged the solstice-layby (S-L) gap (JW 26). The early summer 2nd cohort occurred at the beginning of the post-layby (P-L) period (JW 27-29), except in undisturbed DEN-COM. The late summer 3rd cohort (JW 34-38) emerged during harvest, except in the undisturbed and D1 of the local population.

*Population patterns*. DEN-COM recruitment was numerically greater than that of the local population in all cohorts. The local and DEN-COM populations exhibited similar recruitment behavior early in the season, but not later. Late emergence was greater, and lasted for a longer time in DEN-COM relative to the local population, especially in the later timed experimental disturbances. Late summer 3rd cohort DEN-COM emergence was 2-3 weeks longer than that of the local population. The appearance of this 3rd cohort in the undisturbed DEN-COM may have been the absent 2nd undisturbed cohort, a delayed response to the late season Finish environment. An autumnal 4th DEN-COM cohort (JW 41) emerged very late in the season (or, delayed 3rd cohort), not present in the local population.

*Experimental disturbance*. In general, experimental disturbance extended the numbers and duration of each *C. album* cohort, the early cohort started earlier while the later cohorts ended later. They also stimulated the appearance of two new late season cohorts (local, at harvest; DEN-COM, post-harvest). Experimental disturbance caused emergence to commence one week earlier (JW 22; D1-2) than that observed in the undisturbed in both populations. Experimental disturbance extended the duration of all DEN-COM cohorts, including the appearance of an autumnal 4th cohort. In the local population, those disturbances did not extend the duration of the early summer 1st cohort. However, those disturbances did result in the appearance of a late summer 3rd cohort (JW 34-35) in the local population. This local 3rd cohort was not present in the undisturbed.

**3.3.4.8.2 Sweden**. [Compare greater seed burial number per plot in Sweden (4X) and Spain (2X); saturation hypothesis: burying greater numbers of seeds in the soil:

1. DOES increase recruitment numbers, filling the same cohort spacetime
2. DOES NOT cause seedlings to emerge at new seasonal times]

Four times the amount of seed (800) was buried in individual plots at the Swedish location compared to the other locations (200 seeds per plot). These added seed numbers per area allow comparisons to determine if this resulted in a more complex recruitment structure in this spring cereal cropping system. Despite the added seeds, the recruitment cohort structure was simpler than that observed at other locations. The added seed occupied more potential opportune emergence times during crop establishment phase (1st cohort), while the late summer 2nd cohort was restricted to harvest time, anticipating post-harvest opportunity. The interval between these two cohorts, the P-L period, commenced with the coincidence of summer solstice and layby. The late spring DEN-COM cohort began recruitment one week earlier than the local when disturbed. Experimental disturbance also extended the local 1st cohort by two weeks, the DEN-COM population by four weeks.

*Recruitment cohort pattern*. Two C. album cohorts were observed in Sweden, late spring and late summer. The 1st cohort began before seedbed preparation and ended at layby, just after weed control operations ceased, at summer solstice (JW15-25). The 2nd cohort occurred at harvest (JW 32-33). Both populations, disturbed and undisturbed, were similar and all confined to this period. No recruitment occurred during the post-layby period (JW 24-31) that separated the two cohorts.

*Population patterns*. Recruitment in the 1st cohort in both populations was similar in terms of numbers and pattern, with the exception that DEN-COM began one week before the local population.

*Experimental disturbance*. The experimentally disturbed 1st cohort of the local population continued to emerge for two weeks after, and DEN-COM four weeks after, the undisturbed.

**3.3.4.8.3 Denmark**. Denmark was the only location where two locally-adapted populations (DEN-COM, DEN-LOC2) could be compared to determine how local recruitment varies in a common nursery. The most apparent differences between the two locally-adapted Danish populations was the numerical size of DEN-COM was greater than that of DEN-LOC2 in most instances, and that the late emergence in DEN-LOC2 was of longer duration (to JW 41) than DEN-COM (to JW 29). This lack of late season DEN-COM emergence in may indicate it was historically exposed to spring-winter cropping system practices in which extended cropping operations late in the season deter that late recruitment (e.g. Canada, Norway). The recruitment cohort structure of Danish populations was more complex (6 cohorts) than that observed at other locations, and occurred over the entire year (JW46 to JW 40). Distinct recruitment cohorts arose around the temporal interfaces separating CSD operations, coinciding with intervening periods of no disturbance, opportunity spacetime. Periods of no recruitment provided opportunity to the late spring cohort which was separated from early summer 4th cohort by the summer solstice-layby (S-L) gap (JW 24-25). The late spring 3rd, and early summer 4th, cohorts coincided with the post-layby (P-L) period (JW 21-29) of mid-season opportunity. Local recruitment behavior of DEN-COM in Denmark provided the basis for comparisons with the other nine common nursery locations.

*Recruitment cohort pattern*. Recruitment of local Danish *C. album* populations began in the autumn of burial (JW46) and continued to the following autumn (JW 40) in six separate cohorts in this spring-winter cereal cropping system. The autumnal burial year 1st cohort (JW 46-49) recruited low numbers of seedlings in both local populations. Two spring cohorts were observed in the undisturbed plots of both local populations. The early spring 2nd cohort (JW 16-19) occurred between the end of planting and the weed control period in the spring. The late spring 3rd cohort (JW 22-24) occurred from the weed control period to layby. This cohort was the largest despite its relatively short duration and coincided with the P-L period of opportunity. The 2nd and 3rd cohorts were separated by a recruitment hiatus in JW22 in the DEN-LOC2, and in many instances in the DEN-COM, population. The spring cohorts were separated from the early summer cohort by a gap in recruitment, the summer solstice-layby (S-L) gap. This absence of seedling emergence occurred on JW24 (except DEN-COM, undisturbed), and on JW25 (except D5), in both populations. The early summer 4th cohort (JW 25-29) occurred in the last half of the P-L period (JW 24-29) in the D5 plots of both populations. No emergence occurred in DEN-COM after JW 29. Two very small cohorts occurred late in the season in the DEN-LOC2 population. The late summer 5th cohort (JW 32-35) coincided with the spring harvest and pre-autumnal planting periods. The early autumn 6th cohort (JW 40) occurred during the winter crop planting and weed control periods.

*Population patterns*. DEN-LOC2 exhibited two closely timed spring cohorts (2nd and 3rd). This intervening spring hiatus (JW 20) was also observed in DENCOM (undisturbed, D4-5), which was otherwise continuous (D1-3; JW 16-23). The most apparent difference in the two locally-adapted Danish populations was in the extended late emergence in DEN-LOC2 (to JW 41), which in DEN-COM ended JW 29. This lack of late season emergence in DEN-COM may indicate it was historically exposed to spring-winter cropping system practices in which extended operations late in the season deter that late recruitment (e.g. Canada, Norway). The numerical size of DEN-LOC2 was less than that in the DEN-COM population, except in the latest 5th and 6th cohorts.
*Experimental disturbance*. All experimental disturbances of the DEN-COM population resulted in the 3rd cohort ending one week earlier than in the undisturbed plots. In both populations experimental disturbance D5 resulted in the appearance of an early summer 4th cohort. Experimental disturbance of DEN-LOC2 extended the duration of the 5th cohort (D3-4 only), and resulted in the appearance of a new early autumn 6th cohort (D5 only).

**3.3.4.8.4 Canada**.

*Recruitment cohort pattern*. No late emergence: from Diane, 6.10: "My comment as to why there is no emergence in the fall relate to the nature of the heavy clay we find at this location and the large amount of rain both in the spring and again from august onward. This create a solid crust barrier at the soil surface which prevents any further emergence eventhough the soil humidity and temperature may be optimal. The ability to form this barrier is confirmed by the high soil crusting index of 0.76 of this soil (maximum =1). Once the barrier has been created and it will not be broken until soil heaving from the frost action occurs in winter or else it is physically broken by human intervention (fall plowing). By august the soil surface looks like ciment."

**3.3.4.8.5 Italy**. Long and complex but fits interface theory; excellent long SL gap example.

*Recruitment cohort pattern*.

-Spring cohort 1 start on spring equinox, JW 12, both populations; similar numbers in both populations except local D5 spanning LSS gap. Largest cohort investment. Some instances experimental disturbance delayed, others extended, duration of cohort compared to undisturbed.

-Late season consisted of 5 discrete cohorts beginning at the end of the PL period until post-harvest (JW 28-46) in this highly favorable southern location with a long growing season. Cohort structure more complex than others, especially the DEN-COM recruitment response to Umbrian conditions. In the undisturbed: two very small late season cohorts: 2nd mid-summer cohort (JW 30) during the late PL period-pre-harvest; last, 3rd mid-autumn cohort (JW 44-45) during post-harvest period

**3.3.4.8.6 Spain**.

[low recruitment behavior example of first and only effort under dry conditions, tip of heteroblastic distribution; dry, decrease; extra seeds, increase; what do you do when its very dry, what is your limited recruitment number hedgebet going to be?]

[Compare greater seed burial number per plot in Sweden (4X) and Spain (2X); saturation hypothesis: burying greater numbers of seeds in the soil:
1. DOES increase recruitment numbers, filling the same cohort spacetime
2. DOES NOT cause seedlings to emerge at new seasonal times]

## 3.4 Seedling Recruitment: Between Localities

### 3.4.1 Undisturbed Soil

**3.4.1 Local population**. Burial year emergence occurred in both Denmark populations, but was absent at all other nurseries (except Portugal) (Calendar 13). Spring seedling emergence of LOCAL populations occurred earlier than DEN-COM at all nurseries south of Denmark (except UK). To the north, emergence was often absent in the spring cohort. Summer and autumn emergence occurred at all nurseries (except DEN-COM and Spain).

**3.4.2 DEN-COM population**. Burial year emergence of DEN-COM occurred in Denmark, UK, Portugal and Spain, but was absent at all other nurseries (Calendar 13). Relative to DEN-COM, earlier in the UK and later in UK, Portugal and Spain. Spring seedling emergence of DEN-COM populations occurred earlier than Denmark at all nurseries south of Denmark (except Sweden). To the north, emergence was often absent in the spring cohort. Summer and autumn DEN-COM emergence occurred at all nurseries (except DEN-COM and Spain).

**3.4.3 Local and DEN-COM**. Burial year emergence, when it occurred, was often absent in the LOCAL population (Calendar 14). Seedling emergence in undisturbed soil at a particular nursery was usually similar between the LOCAL and DEN-COM populations, but some instances of absent, earlier or later than DEN-COM emergence occurred in the LOCAL populations regardless of latitude.

### 3.5 Seedling Recruitment: Within Locality

#### 3.5.1 Denmark

Local variation between two local populations was apparent at the Denmark site (Calendar 1, Tables 5, 6). Seedling emergence season length was JW 16-48.

**3.5.1.1 Undisturbed soil**. Both local populations had burial autumn and spring cohorts. DEN-COM had a summer cohort, unlike LOCAL.

**3.5.1.2 Disturbed soil**. The LOCAL spring cohort had earlier and later emergence, with more numbers. The LOCAL summer cohort had later emergence. The DEN-COM spring cohort had later emergence, with more numbers.

#### 3.5.2 Finland

Seedling emergence season length was JW 22-41 (Calendar 2, Tables 5, 6).

**3.5.2.1 Undisturbed soil**. Both populations had a spring cohort, but with greater numbers with DEN-COM. The LOCAL population had an early summer cohort, unlike DEN-COM. The DEN-COM population had a late summer cohort, unlike LOCAL.

**3.5.2.2 Disturbed soil**. Both spring and early summer LOCAL cohorts had earlier and later emergence. The DEN-COM spring cohort had earlier and later emergence, with more numbers. The DEN-COM late summer cohort had later emergence, with more numbers.

#### 3.5.3 Norway

Seedling emergence season length was JW 17-46 (Calendar 3, Tables 5, 6).

**3.5.3.1 Undisturbed soil**. Both LOCAL and DEN-COM populations had early spring, late spring and early summer cohorts. The early spring emergence numbers were greater in DEN-COM than LOCAL. The DEN-COM population had an autumn cohort, unlike LOCAL.

**3.5.3.2 Disturbed soil**. The late spring cohort had earlier emergence and greater numbers in both LOCAL and DEN-COM populations. The LOCAL population early summer cohort had earlier and later emergence, unlike DEN-COM.

#### 3.5.4 Sweden

Seedling emergence season length was JW 15-33 (Calendar 4, Tables 5, 6).

[Results were confounded by artefact: 4x numbers buried]

4000 seeds per sample (4x); 2800 possible seeds in the 0.7 $m^2$ central sample area; data presented divided by 4 to make all nursery observations comparable.

The additional buried seed may have simplified the cohort structure by the emergence of seeds for the entire spring period.

**3.5.4.1 Undisturbed soil**. Both LOCAL and DEN-COM populations had spring and summer cohorts.

**3.5.4.2 Disturbed soil**. The spring cohort for both LOCAL and DEN-COM populations had later emergence, and often less numbers.

### 3.5.5 Czech Republic

Seedling emergence season length was JW 15-35 (Calendar 5, Tables 5, 6).

**3.5.5.1 Undisturbed soil**. Both LOCAL and DEN-COM populations had spring and summer cohorts. The spring emergence numbers were greater in DEN-COM than LOCAL.

**3.5.5.2 Disturbed soil**. The spring cohort for both LOCAL and DEN-COM populations had later emergence, but the LOCAL population had greater numbers while DEN-COM had similar or less numbers. The LOCAL summer population had earlier and later emergence with somewhat greater numbers. The DEN-COM summer population had some later emergence.

### 3.5.6 Canada

Seedling emergence season length was JW 14-31 (Calendar 6, Tables 5, 6).

[Results were confounded by artefact: two different LOCAL populations (Ste-/Clotilde, 73%; Ste-Bruno, 27%) collected 1-3 weeks apart, and buried in the L'Acadie common nursery]

**3.5.6.1 Undisturbed soil**. The LOCAL population had spring and summer cohorts. The DEN-COM had burial autumn and spring cohorts. The spring emergence numbers were greater in DEN-COM than LOCAL.

**3.5.6.2 Disturbed soil**. The spring cohort for both LOCAL and DEN-COM populations had later emergence. The LOCAL population had greater numbers while DEN-COM had greater or lesser numbers.

### 3.5.7 Italy

Seedling emergence season length was JW 12-46 (Calendar 7, Tables 5, 6).

**3.5.7.1 Undisturbed soil**. Both LOCAL and DEN-COM populations had spring, summer and autumn cohorts. DEN-COM summer and autumn cohorts occurred at more times than those of LOCAL.

**3.5.7.2 Disturbed soil**. The spring cohort for both LOCAL and DEN-COM populations had later emergence and greater numbers. Both summer and autumn cohorts for LOCAL and DEN-COM populations had later emergence, but those of LOCAL occurred more frequently. The DEN-COM early summer cohort emergence occurred earlier.

### 3.5.8 USA, Illinois

Seedling emergence season began on JW 11 (Calendar 8, Tables 5, 6).

[Results were confounded by two artefacts: unemerged ('white threads') and emerged seedlings counted on disturbance day were combined. No emergence data taken after June 20, 2006, JW26.]

**3.5.8.1 Undisturbed soil**. Both LOCAL and DEN-COM populations had spring cohorts; but with greater numbers with DEN-COM.

**3.5.8.2 Disturbed soil**. The spring cohort for both LOCAL and DEN-COM populations had later emergence; similar numbers with LOCAL; similar or lesser numbers with DEN-COM.

### 3.5.9 UK

Seedling emergence season length was JW 10-49 (Calendar 9, Tables 5, 6).

[Seedling recruitment responses anomalous: similar to more southerly location; possible consequence of gulf stream warming of British Isles.]

**3.5.9.1 Undisturbed soil**. Both populations had spring, and numerous summer, cohorts. The LOCAL population had a small winter cohort, unlike DEN-COM. The DEN-COM population had burial autumn cohorts, unlike LOCAL.

**3.5.9.2 Disturbed soil**. The LOCAL population early spring and summer cohort emergence occurred late, while the late spring cohort occurred early. DEN-COM early summer emergence sometimes occurred early.

**3.5.10 Portugal**

Seedling emergence season length occurred for the entire year (Calendar 10, Tables 5, 6).

[Results were confounded by artefact: irrigation; consequence was almost ideal conditions (heat, moisture) for continuous seedling emergence.]

**3.5.10.1 Undisturbed soil**.

Seedling emergence in both populations occurred at several times in all four seasons.

**3.5.10.2 Disturbed soil**. Greater numbers emerged in the spring cohort with both populations. Emergence was greater in one instance in the summer cohort of LOCAL. Emergence was greater sometimes in the spring cohort of DEN-COM.

**3.5.11 Spain**

Seedling emergence season length occurred for the entire year (Calendar 11, Tables 5, 6).

[Results were confounded by artefact: very dry, un-irrigated site, despite heat conditions favourable to year-long recruitment.]

2000 seeds per sample (2x); 1400 possible seeds in the 0.7 $m^2$ central sample area; data presented divided by 2 to make all nursery observations comparable.

# 4 DISCUSSION

## 4.1 The Risk of Mortality and Seedling Recruitment

Weedy plant life history, as for any organism, conforms to the time of mortality risk from either the environment (e.g. frozen soil in the winter) or cropping systems practices (e.g. herbicide use). *C. album* life histories are constrained by the environmental and cropping system practices of a locality. It is hypothesized that *C.* album seedling recruitment timing and magnitude have adapted to these local conditions and are expressed in emergence behavior.

The life history trajectories of *C. album* plants from embryogenic induction of variable dormancy capacity among seeds of a population and their dispersal to the soil, until germination, emergence and the resumption of autotrophic growth in agricultural communities is determined by the risk of mortality. The primary risks of mortality during this life history arise from environmental and cropping systems disturbances (density-independent mortality) and the effects of neighbors as the season progresses (density-dependent mortality). If this is true, then the patterns and magnitude of *C. album* recruitment should be a reflection of the historical survivors selected by these risks in each locality where they have adapted and evolved.

## 4.2 Seasonal Filters of Seedling Recruitment

There exist seasonal filters of local seedling recruitment opportunity. Seedling emergence opportunity arises from the duration of the local *C. album* season and cropping system disturbance limitations. When does *C. album* emerge? And when does it not emerge?

**4.2.1 Seasonal filter 1: Habitat**. Seedling recruitment of *C. album* was prevented at most locations by environmental conditions unfavorable to plant growth. The dominant inhibition was the winter period when temperatures were unfavorable to plant growth. In Spain and Portugal these conditions were not as apparent. In Spain the lack of moisture prevented growth for much of the year, while irrigation in Portugal locations allowed seedling emergence for most of the year. At all other more northerly locations emergence was prevented during these cold periods.

LQ seedling recruitment opportunity arises first from the duration of the local season. The duration of local seasonality (JW; range) of seedling emergence by latitude provides an estimate of available environmental signal space: heat-growth/gas solubility, light-photoperiod/quality/flux; a direct estimate of winter-preplant and post-harvest-winter interfaces. In particular habitats it also may provide estimates of water/water-nitrate/water-oxygen signals.

**4.2.1.1 Latitude seasonality**. Generally, the duration of recruitment (undisturbed, experimentally disturbed) of both local and Denmark common (DEN-COM) populations increased with decreasing latitude, north-to-south (table 19, local; table 20, DEN-COM). This latitudinal season duration phenomena was most apparent at the onset of recruitment in the winter-spring. Recruitment of both populations commenced at similar times (JW) at a particular location. This latitudinal phenomena was apparent, with fewer numbers and greater variability over longer seasonal times, at the end of the season in summer-autumn. Autumnal recruitment in the burial year (2005) occurred in a minority of locations: Denmark (both populations), UK and irrigated Portugal.

The environmental soil signals present in localities change with north-south latitude and affect the recruitment of seedlings to the extent external conditions modulate inherent qualities of the seed. One of the striking features of *C. album* seedling recruitment over this wide range of latitudes (61° N to 37° N) was the common expression of pattern and magnitude when local populations, and the common Danish population, are compared at the different burial sites. *C.* album seedling recruitment pattern and magnitude among populations (11) at different locations (10) display a common hedge-bet theme and structure for exploiting available agricultural opportunity space. Despite this commonality in response of this weedy species, latitude environment did affect seedling recruitment.

The duration of annual and seasonal seedling emergence increased with decreasing latitude in most instances. Exceptions occurred in the UK (Atlantic Ocean gulf stream warming effects), USA (data not collected after JW26) and Spain (drought conditions). For this reason, the duration of late season (summer, autumn) recruitment increases with decreasing latitude in most instances. Similarly, the number of discrete late season emerging cohorts usually increases with decreasing latitude. The peak spring emergence period occurs later, both beginning and end, occurs later in the season with increasing latitude in most locations, and for similar reasons as season recruitment duration.

**4.2.1.2 Local population seasonality**. The period of recruitment of the local population (undisturbed, experimentally disturbed) occurred from JW 2-37, with season durations from 12-37 weeks at the 11 locations (table 19).

**4.2.1.2.1 Recruitment onset interface**. This latitudinal season duration phenomena was most apparent at the onset of recruitment in the winter-spring. There were two exceptions. UK recruitment onset was earlier relative to that of latitude neighbors, and may be a reflection of the off-shore warming-moderating influence of the gulf stream current. Spain due to dry conditions.

**4.2.1.2.2 Recruitment end interface**. This latitudinal phenomena was apparent, with fewer numbers and greater variability over longer seasonal times, at the end of the season in summer-autumn. There were two exceptions. Canada recruitment ended very early relative to that of latitude neighbors in Europe. Spain due to dry conditions.

**4.2.1.2.3 Recruitment in autumn of burial year**. Autumnal recruitment in the burial year (2005) occurred in a minority of locations: Denmark (both populations) and UK populations, possibly a reflection of the off-shore warming-moderating influence of the gulf stream current at those sites. And at irrigated Portugal in the south. The presence of a highly germinable fraction in those local Denmark, UK and Portugal populations indicates a greater investment in early season recruitment relative to that of their latitudinal neighbors.

**4.2.1.3 Denmark common population seasonality**. In general, north of Denmark the DEN-COM recruitment season was longer than that of the local population. South of Denmark the DEN-COM season was shorter, and ended sooner, than the local population. Recruitment of both populations commenced at similar times (JW) at a particular location.

The period of recruitment of the Denmark common population (DEN-COM; undisturbed, experimentally disturbed) occurred from JW 3-41, with durations at individual locations from 1-35 weeks (table 20). The DEN-COM recruitment season was longer than the local population at all locations north of, and shorter at all those south of, Denmark. The onset of seasonal recruitment was consistently similar between local and DEN-COM populations at a particular location, with the exception of dry Spain when it commenced later. Seasonal periodicity differences between populations were most apparent with the ending time (JW) of recruitment. Local population recruitment ended later than DEN-COM at all locations south of Denmark. North of Denmark, the end of seasonal recruitment varied at each location between populations.

#### 4.2.2 Seasonal Filter 2: Cropping Disturbance

Predictable, historical, seasonal times of local cropping system disturbances (table 3, 14, 18) provide the second filter of seedling recruitment opportunity. Table 14 represents CSD as occurring over an extended period. At any individual locality in a field these cropping operations only occur on one day, for a moment in time. Neighbors emerging one day or after this moment can experience quite different subsequent fates: before, mortality; after, opportunity.

Cropping system disturbances occur at a locality over a period of time, while any individual site at those localities experiences these disturbances at a single, instantaneous time. Opportunity occurs after, mortality before, these disturbances. Cohort emergence is a collection of individual plant hedge-bets, all searching spacetime to occur after disturbance.

**4.2.2.1 Cropping system disturbance seasonality**. Generally, the CSD period increased with decreasing latitude with some exceptions. The total duration of the CSD period over twice as long in the south (e.g. Spain, 38 weeks) as that in the north (e.g. Finland, 18 weeks). Notable exceptions to this latitudinal response included locations with late CSD periods due to winter cereal crop establishment (Denmark, to JW 44; Norway, to JW 39). The UK CSD period commencing earlier that expected (JW 10) at that latitude, possibly due to gulf stream warming stimulating early season growth.

**4.2.2.2 Cropping system disturbance periodicity**. The CSD activity at all locations included a mid-season period of inactivity in which no cropping disturbances occurred (table 3). Layby is a common term used in the USA to indicate that time in the cropping system when cultivation tillage and herbicide applications cease, when the crop canopy closure shades the soil from direct solar radiation (often a strong stimulus to LQ germination) (Hawkins, 1962). Depending on location, this post-layby period (PLP) varied from JW 21-34, with durations of 2-13 weeks (ave. 7 weeks). Notably, no CSD’s occurred at any location during JW 26-29, excepting the UK and Canada. Relatively continuous cropping practices in intensive vegetable systems in the UK and Canada occurred during this PLP at the other locations. The mid-season hiatus began within a relatively narrow period around summer solstice (JW 21-26), and ended over an extended period (JW 23-34) with late season disturbances (e.g. harvest, winter planting), both depending on location. The early CSD (e.g. planting, weed control) period (JW 5-25) preceeding this midseason hiatus began from JW 5-18: earliest at the southmost (Spain, N 37°, JW 5), last at the most northern (Finland, N 61°, JW 19), location. At all locations the early period ended over a much shorter period (JW 20-25) leading up to the summer solstice, and appeared to be independent of latitude. The late CSD (e.g. harvest, winter planting) period (JW 30-44) following the PLP began (JW 30-35) and ended (JW 35-44) over an extended period, depending on location.

**4.2.2.3 Local recruitment cohort patterns and cropping disturbance**

**4.2.2.3.1 Recruitment cohort pattern**. Distinct recruitment cohorts arose around the temporal interfaces separating CSD operations. This behavior may be an adaptive means by which *C. album* searches for, and exploits, recruitment opportunity just prior to, and after, predictable cropping system disturbances. These patterns coincide with intervening periods of no disturbance: recruitment gap periods of opportunity: late spring and early summer cohorts separated by the LSS gap; the early summer cohort followed by the post-layby (P-L) period of mid-season opportunity.

**4.2.2.3.2 Population patterns**. DEN-COM recruitment was greater than that of the local population in all cohorts (FIN; DEN). DEN-COM recruitment was similar to that of the local population in all cohorts (SWE). Local recruitment was greater than that of DEN-COM (latest cohorts only): DEN (DEN-LOC2).

**4.2.2.3.3 Experimental disturbance**. Recruitment cohort durations extended the duration of each cohort (FIN), extended the duration of the late spring cohort (SWE; DEN), shortened the duration of the late spring cohort (DEN), or resulted in the appearance of new late season cohorts (FIN; DEN).

## 4.3 Seasonal Recruitment Periodicity

Recruitment opportunity at each locality possessed structure (time, number), a consequence of the two previous, sequential filters of seedling emergence. *C. album* recruitment structure reveals 2-4 discrete seasonal cohorts at each location, for both local and DEN-COM populations. Most commonly 2 cohorts were observed, less frequently 1 or 3, and 4 only at Portugal.

Those seedling that emerge early in the season experience two opposing forces which result in the majority of seedling emergence occurring in the spring. Seedlings that emerge early and survive will accumulate the maximum amount of biomass, thus produce more seed, by taking advantage of the longer growth period. They also face the greatest chance of mortality, primarily due to farming system practices associated with crop establishment in the spring. The hedge-bet strategy revealed by *C. album* was that early emergence was worth the high risk of mortality: most of the recruitment was in the spring. This period of maximal emergence was continuous at most locations for both populations studied (common Danish and local). When peak spring emergence was discontinuous, two periods were observed. Often the earlier spring period was greater than that later in the spring.

Two periods of relatively low disturbance occur in the annual cropping systems of Europe and North America when favorable germination and growth conditions exist. These include the period after weed control practices cease in late spring or early summer until harvest (called the ‘layby’ period in North America), and the post-harvest period ending with the first late-season killing frosts. Presumably seedlings emerging during the harvest period will be small and many may escape death during and after harvesting operations. Both of these seasonal periods are times when normally no production practices occur, when weed emergence and growth are undisturbed except by neighbors (density-dependent mortality risk), and when the risk of density-independent mortality is relatively low. It is hypothesized this is an opportunity space that *C. album* recruitment exploits.

**4.3.1 Early recruitment**. The spring cohort (JW 12-24) was the largest recruitment investment, with consistent times of seasonal onset (see above).

**4.3.2 Layby-summer solstice (LSS) recruitment gap**. A recruitment gap of relatively short duration was observed between the spring and summer cohorts at all locations, and coincided with the summer solstice and PLP. The LSS recruitment gap was observed at all locations, for both populations. It was characterized by the consistent absence of seedling emergence in the JW 22-28 period. The LSS gap occurred during JW 25-26, depending on location. The specific timing of the recruitment hiatus varied with location (latitude), population and experimental disturbance.

The LSS gap in recruitment never occurred in any location in the experimentally undisturbed plots. *C. album* emergence during the LSS period was restricted to experimentally disturbed plots, and only in populations at or near their local home location. No emergence occurred during JW 25 or 26 at the UK, Czech Republic or Spain. Some infrequent recruitment occurred in this solstice period in the experimentally disturbed local populations (FIN, NOR, DEN (2), CAN, ITALY, PORT). DEN-COM at non-Danish locations emerged during the LSS gap only in two locations immediately to the north of Denmark (Norway, Sweden).

The occurrence of the LSS gap was unexpected at this favorable time of the growing season. The recommencement of recruitment shortly after was even more difficult to explain. In addition to the coincidence of summer solstice, PLP and canopy closure this is a period of increasing heat and potentially decreasing moisture. The short duration cessation of seedling recruitment may be a consequence of a unique change in the most important environmental signal regulating seed germination, light (Altenhofen, 2009), that occurs at this time of the growing season. Prior to summer solstice photoperiod duration and intensity increase, afterwards they steadily decrease. Photoperiod changes at summer solstice is therefore a threshold event, a solar switch. Changing photoperiod is a signal that regulates flowering in other weed species (e.g. decreasing photoperiod in weedy *Setaria* (Dekker, 2003). The coincidence of all these anthrogenic, astronomical and environmental factors at the LSS may be an important basis for fine-tuning LQ seedling emergence consistent with local opportunity.

Late emergence in the summer and autumn occurred at most locations. Unlike the continuous emergence pattern observed in the spring, these late emergence cohorts were several and discrete and occurred over a longer period of time. Two periods in which no or little late recruitment occurred were identified at many locations. These include the period from when weed control practices cease ('layby') until harvest, and the post-harvest period ending with the first late-season killing frosts. Both of these seasonal periods are times when little or no farming operations occur, and the risk of density-independent mortality is low. These opportunity spaces were exploited by both populations, primarily in the post-layby period, and to a lesser extent in the post-harvest period. Seedlings usually emerged at the beginning of these periods when it did occur, taking advantage of the lowered risks of mortality (density-independent during both periods; density-dependent in the post-harvest period).

**4.3.3 Late recruitment**. Followed by this, the second largest cohort was in the summer-autumn (JW 25-37). This cohort was numerically smaller than, temporally longer and more variable than, consistently separated from, the spring cohort. Apparently LQ recruitment opportunity spacetime is more fragmented in this period.

**4.3.4 Burial autumn recruitment**. Infrequent locations; short duration; consistent; third largest cohort, low numbers.

**4.3.5 Winter recruitment**. Only occurred in warm, irrigated, multi-cropping Portugal. Smallest cohort.

## 4.4 Seed Germination Regulation

Is seedling recruitment inherent or environmental? Seed dormancy heterogeneity (heteroblasty) is the 'blueprint' that guides the magnitude and timing of seedling recruitment in other weed species (e.g. weedy *Setaria*; Jovaag et al., 2007a, b, c). It is hypothesized that *C. album* seedling recruitment behavior (pattern and magnitude) will be a consequence of inherent seed dormancy capacity among seeds of a population (nature) interacting with the local environmental conditions of the soil in which they are buried (nuture).

Seedling recruitment of *C.* album is the consequence of environmental modification of inherent dormancy capacity differences between and with populations. In some instances environment appears to dominate control of seed behavior, in others endogenous control. Evidence for fine-scale adaptation to local conditions by local populations is also revealed in these studies. Clues as to the relative contributions provided by inherent dormancy capacity qualities of the *C. album* seed and that of the soil environment on recruitment are revealed by the responses of the common Danish population at non-Danish burial nurseries. Presumably, environmental cues in the soil dominate when the common Danish population responds in a similar manner to that of the local population. Differences between the common Danish population seedling emergence pattern and magnitude from that of the local may indicate the role played by inherent qualities of the seed. Patterns instilled in the common Danish population and expressed at non-Danish locations, unlike those of the local population, are a clear indication of inherent dormancy heterogeneity controlling recruitment behavior.

The first clear indication of inherent qualities regulating recruitment pattern and magnitude comes from the differential recruitment of the two Danish populations taken from the same field in Flakkeberg. In terms of burial year, spring and late season behaviors differed. The local population was more dormant overall as indicated by its consistently lower magnitude. The local Danish population emerged somewhat later in the burial year, spring recruitment ended earlier, and late emergence occurred, all relative to the common Danish population. These differences in behavior indicate a fine scale adaptation in *C. album* to local conditions.

Recruitment in the burial year, at all four non-Danish locations where it did occur, appeared to be due to endogenous factors instilled in the common Danish seed during it development on the parent plant. The common Danish, and not the local, population emerged in the burial year in Canada, Spain and the UK. It also emerged with the local population in Portugal, but in much larger numbers.

The time pattern responses of the common Danish population was often similar to the local population at many sites. Spring emergence of both populations usually began at similar times at all localities indicating environmental control. The role of both environmental and inherent factors was observed in the time those spring peak periods of recruitment ended: similar times (environment) at 5 locations; earlier (4 locations) or later (2 locations) indicating endogenous control. The role of both inherent and external factors was also revealed in late season emergence patterns and magnitudes.

Inherent dormancy qualities were also apparent in the larger magnitude of total emergence of the common Danish population at all locations except Sweden and Portugal. Seen at a finer temporal scale, both environment and seed qualities interacted to produce differences between local and common Danish populations within seasonal periods. Only in Norway (common Danish greater) and Portugal (local population greater) were consistent magnitude differences observed in every seasonal period.

In most burial nurseries both local and common Danish populations took advantage of the late season no-low disturbance periods by initiating recruitment to seize these opportunity spaces. The occurance of this common pattern was an indication of environmental opportunity. But, the fact that patterns of both populations at almost all the localities indicate a general, inherent, recruitment pattern to the species.

A more specific, population-scale, indication of inherent emergence regulation and of fine-scale local population adaptation was revealed when Danish (local only) and Spanish (common Danish only) populations took advantage of the late season hiatus. Similarly, a USA (common Danish only) population seized the post-layby hiatus opportunity (note: this an artifact of not taking data after ca. JW30).

## 5 SEEDLING EMERGENCE CALENDARS

### 5.1 Calendars 1-11, A-B

**A**: Calendar of historical times (Julian week, JW) of cropping system practices and disturbances (see Table 4), winter and opportunity spacetime (see text); *Chenopodium album* seedling emergence magnitude (mean numbers per JW of 700 possible buried seeds) with time (2005, 2006) in undisturbed and disturbed soil (5 events, D-1-5; see Table 3) for local and common Denmark (DEN-COM) populations.

| KEY: | | |
|---|---|---|
| | | emergence: prior to disturbance |
| | | emergence: during, after disturbance |
| | | JW of disturbance; no emergence occurred |
| | | JW of disturbance; emergence occurred |
| | 66 | number of seedlings emerging |

**B**: Calendar of historical cropping disturbances, winter and opportunity; *Chenopodium album* seedling emergence cohort magnitude (% of 700 buried seeds; emergence less than 1% = 1%) with time in undisturbed and disturbed soil; seedling recruitment cohorts and potential seed fecundity; seedling investment, % of buried seed emerged.

| KEY: | | |
|---|---|---|
| | | undisturbed soil seedling emergence |
| | | disturbed; SAME emergence time as undisturbed |
| | | disturbed; EARLIER than undisturbed |
| | | disturbed; LATER than undisturbed |
| | | JW of disturbance; no emergence occurred |

Calendar 1: Denmark
Calendar 2: Finland
Calendar 3: Norway
Calendar 4: Sweden
Calendar 5: Czech Republic
Calendar 6: Canada
Calendar 7: Italy
Calendar 8: USA
Calendar 9: UK
Calendar 10: Portugal
Calendar 11: Spain

**Calendar 1A**: Denmark

| DENMARK | N 55° E11° | 2005 C.E. | | | | | 2006 C.E. | | | | | | | | | YEAR-SEASON-MONTH-JULIAN WEEK | | | | | | | | | | | | | | | | | | | | 2006 C.E. | |
|---|---|---|---|---|---|---|---|---|---|---|---|---|---|---|---|---|---|---|---|---|---|---|---|---|---|---|---|---|---|---|---|---|---|---|---|---|---|
| Flakkeberg | | AUTUMN | | | | | SPRING | | | | | | | | | | | | SUMMER | | | | | | | | | | | | | | AUTUMN | | | | |
| | | NOV | | DEC | | | | APRIL | | | | MAY | | | | JUNE | | | | | JULY | | | | AUGUST | | | | SEPT | | | | OCT | | | NOV | |
| POPULATION | DISTURBANCE | 46 | 47 | 48 | 49 | | 13 | 14 | 15 | 16 | 17 | 18 | 19 | 20 | 21 | 22 | 23 | 24 | 25 | 26 | 27 | 28 | 29 | 30 | 31 | 32 | 33 | 34 | 35 | 36 | | 39 | 40 | 41 | | 44 | 45 |
| | WINTER | | | | | ■ | ■ | | | | | | | | | | | | | | | | | | | | | | | | | | | | | | |
| | Seedbed Prep | | | | | | | ■ | ■ | ■ | | | | | | | | | | | | | | | | | | | | ■ | ■ | ■ | ■ | | | | |
| Spring cereals | Planting | | | | | | | ■ | ■ | ■ | | | | | | | | | | | | | | | | | | | | ■ | ■ | ■ | ■ | | | | |
| Winter wheat | Weed Control | | | | | | | | | | ■ | ■ | ■ | ■ | | | | | | | | | | | | | | | | | | ■ | ■ | ■ | ■ | ■ | |
| | Harvest | | | | | | | | | | | | | | | | | | | | | | | ■ | ■ | ■ | ■ | ■ | | | | | | | | | |
| | OPPORTUNITY | ■ | ■ | ■ | ■ | | | | | | | | | | ■ | ■ | ■ | ■ | ■ | ■ | ■ | ■ | ■ | | | | | | ■ | | | | | | | | ■ |
| | UNDISTURBED | 1 | | 1 | | | | | | 1 | 4 | 3 | 4 | | | 9 | 4 | 1 | | | | | | | | | | | | | | | | | | | |
| DENMARK | Disturbance-1 | 1 | | 1 | | | | | | | 2 | 27 | 63 | 1 | 2 | 12 | 2 | | | | | | | | | | | | | | | | | | | | |
| COMMON: | Disturbance-2 | | | | 2 | | | | | 2 | 4 | | 14 | 6 | 19 | 24 | 5 | | | | | | | | | | | | | | | | | | | | |
| DEN-COM | Disturbance-3 | 2 | | | 1 | | | | | 1 | 5 | 2 | 2 | 26 | 19 | 100 | 4 | | | | | | | | | | | | | | | | | | | | |
| | Disturbance-4 | 1 | | | 4 | | | | | 1 | 3 | 4 | 7 | | 2 | 105 | 8 | | | | | | | | | | | | | | | | | | | | |
| | Disturbance-5 | 1 | | | | | | | | | 4 | 3 | 3 | | | 14 | 8 | | 14 | 36 | 5 | | 1 | | | | | | | | | | | | | | |
| | UNDISTURBED | | | 2 | | | | | | 1 | 1 | 1 | 1 | | | 2 | | | | | | | | | | 1 | | | | | | | | | | | |
| | Disturbance-1 | | | | | | | | | | 1 | 1 | 9 | | 1 | 1 | 2 | | | | | | | | | 1 | | | | | | | | | | | |
| DENMARK: | Disturbance-2 | 1 | | | | | | | | 1 | 1 | | | | 16 | 6 | | | | | | | | | | 1 | | | | | | | | | | | |
| LOCAL | Disturbance-3 | | | | | | | | | | | 1 | | | 3 | 33 | 7 | | | | | | | | | | | | 1 | | | | | | | | |
| | Disturbance-4 | | | 1 | | | | | | 1 | | | | | | 34 | 7 | | | | | | | | | | 1 | | | | | | | | | | |
| | Disturbance-5 | | | 1 | | | | | | | | 1 | | | | | 1 | | 5 | 20 | 3 | 1 | 3 | | | | | | | | | | 1 | | | | |

**Calendar 1B**:  Denmark

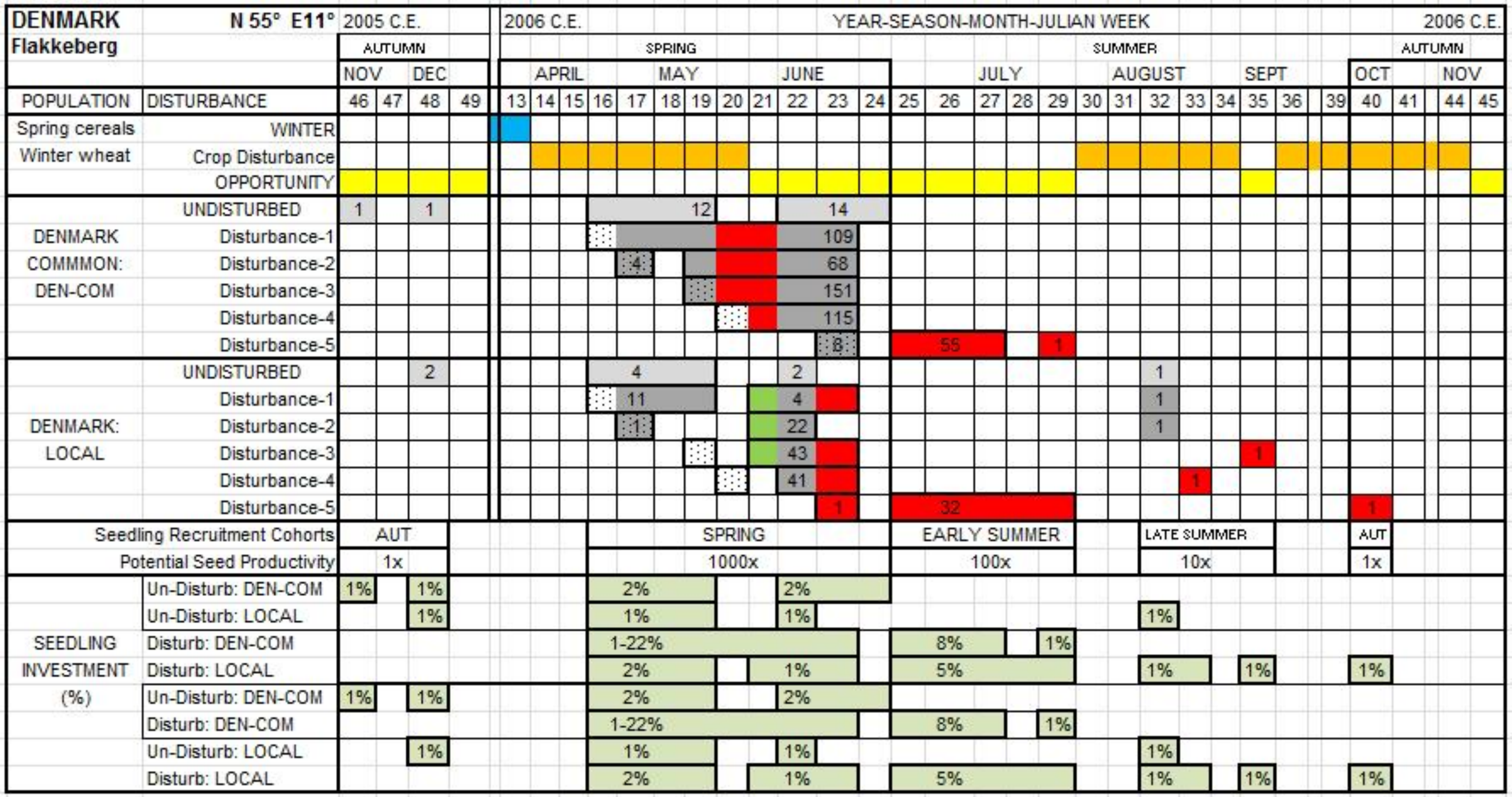

| DENMARK | N 55° E11° | 2005 C.E. | | | | 2006 C.E. | | | | | | | | | | | | | | YEAR-SEASON-MONTH-JULIAN WEEK | | | | | | | | | | | | | 2006 C.E. | |
|---|---|---|---|---|---|---|---|---|---|---|---|---|---|---|---|---|---|---|---|---|---|---|---|---|---|---|---|---|---|---|---|---|---|---|
| Flakkeberg | | | AUTUMN | | | | | | | | SPRING | | | | | | | | | | | | | SUMMER | | | | | | | | | AUTUMN | |
| | | NOV | | DEC | | | | APRIL | | | MAY | | | | JUNE | | | | | JULY | | | | | AUGUST | | | SEPT | | | OCT | | NOV | |
| POPULATION | DISTURBANCE | 46 | 47 | 48 | 49 | 13 | 14 | 15 | 16 | 17 | 18 | 19 | 20 | 21 | 22 | 23 | 24 | 25 | 26 | 27 | 28 | 29 | 30 | 31 | 32 | 33 | 34 | 35 | 36 | 39 | 40 | 41 | 44 | 45 |
| Spring cereals | WINTER | | | | | | | | | | | | | | | | | | | | | | | | | | | | | | | | | |
| Winter wheat | Crop Disturbance | | | | | | | | | | | | | | | | | | | | | | | | | | | | | | | | | |
| | OPPORTUNITY | | | | | | | | | | | | | | | | | | | | | | | | | | | | | | | | | |
| | UNDISTURBED | 1 | | 1 | | | | | | | | 12 | | | | 14 | | | | | | | | | | | | | | | | | | |
| DENMARK | Disturbance-1 | | | | | | | | | | | | | | | 109 | | | | | | | | | | | | | | | | | | |
| COMMMON: | Disturbance-2 | | | | | | | | | 4 | | | | | | 68 | | | | | | | | | | | | | | | | | | |
| DEN-COM | Disturbance-3 | | | | | | | | | | | | | | | 151 | | | | | | | | | | | | | | | | | | |
| | Disturbance-4 | | | | | | | | | | | | | | | 115 | | | | | | | | | | | | | | | | | | |
| | Disturbance-5 | | | | | | | | | | | | | | | 8 | | | 55 | | | 1 | | | | | | | | | | | | |
| | UNDISTURBED | | | 2 | | | | | | 4 | | | | | 2 | | | | | | | | | | 1 | | | | | | | | | |
| | Disturbance-1 | | | | | | | | | 11 | | | | | 4 | | | | | | | | | | 1 | | | | | | | | | |
| DENMARK: | Disturbance-2 | | | | | | | | | 1 | | | | | 22 | | | | | | | | | | 1 | | | | | | | | | |
| LOCAL | Disturbance-3 | | | | | | | | | | | | | | 43 | | | | | | | | | | | | | 1 | | | | | | |
| | Disturbance-4 | | | | | | | | | | | | | | 41 | | | | | | | | | | | 1 | | | | | | | | |
| | Disturbance-5 | | | | | | | | | | | | | | | 1 | | | 32 | | | | | | | | | | | | 1 | | | |
| Seedling Recruitment Cohorts | | | AUT | | | | | | | | | | SPRING | | | | | | | EARLY SUMMER | | | | | | LATE SUMMER | | | | | AUT | | | |
| Potential Seed Productivity | | | 1x | | | | | | | | | | 1000x | | | | | | | 100x | | | | | | 10x | | | | | 1x | | | |
| | Un-Disturb: DEN-COM | 1% | | 1% | | | | | | 2% | | | | | 2% | | | | | | | | | | | | | | | | | | | |
| | Un-Disturb: LOCAL | | | 1% | | | | | | 1% | | | | | 1% | | | | | | | | | | 1% | | | | | | | | | |
| SEEDLING | Disturb: DEN-COM | | | | | | | | | 1-22% | | | | | | | | | 8% | | | 1% | | | | | | | | | | | | |
| INVESTMENT | Disturb: LOCAL | | | | | | | | | 2% | | | | | 1% | | | | 5% | | | | | | 1% | | | 1% | | | 1% | | | |
| (%) | Un-Disturb: DEN-COM | 1% | | 1% | | | | | | 2% | | | | | 2% | | | | | | | | | | | | | | | | | | | |
| | Disturb: DEN-COM | | | | | | | | | 1-22% | | | | | | | | | 8% | | | 1% | | | | | | | | | | | | |
| | Un-Disturb: LOCAL | | | 1% | | | | | | 1% | | | | | 1% | | | | | | | | | | 1% | | | | | | | | | |
| | Disturb: LOCAL | | | | | | | | | 2% | | | | | 1% | | | | 5% | | | | | | 1% | | | 1% | | | 1% | | | |

**Calendar 2A**: Finland

| FINLAND | N 61° E23° | 2006 C.E. | | | | | | | | | YEAR-SEASON-MONTH-JULIAN WEEK | | | | | | | | | | | | | | | | | | 2006 C.E. |
|---|---|---|---|---|---|---|---|---|---|---|---|---|---|---|---|---|---|---|---|---|---|---|---|---|---|---|---|---|---|
| Jokioinen | | | | | | SPRING | | | | | | | | | | | SUMMER | | | | | | | | | AUTUMN | | | |
| | | | | MAY | | | | JUNE | | | | | JULY | | | | AUGUST | | | | SEPT | | | | | | OCTOBER | | |
| POPULATION | DISTURBANCE | 16 | 17 | 18 | 19 | 20 | 21 | 22 | 23 | 24 | 25 | 26 | 27 | 28 | 29 | 30 | 31 | 32 | 33 | 34 | 35 | 36 | 37 | 38 | 39 | 40 | 41 | 42 | 43 |
| | WINTER | | | | | | | | | | | | | | | | | | | | | | | | | | | | |
| Spring Cereals | Seedbed Prep | | | | | | | | | | | | | | | | | | | | | | | | | | | | |
| Grassland | Planting | | | | | | | | | | | | | | | | | | | | | | | | | | | | |
| | Weed Control | | | | | | | | | | | | | | | | | | | | | | | | | | | | |
| | Harvest | | | | | | | | | | | | | | | | | | | | | | | | | | | | |
| | OPPORTUNITY | | | | | | | | | | | | | | | | | | | | | | | | | | | | |
| FINLAND: LOCAL | UNDISTURBED | | | | | | | | 27 | 2 | | | 1 | | | | | | | | | | | | | | | | |
| | Disturbance-1 | | | | | | | 5 | 9 | 9 | | | 1 | | | | | | | | | | | | | | | | |
| | Disturbance-2 | | | | | | | 22 | 1 | 21 | 1 | 1 | 3 | | | | | | | | 1 | | | | | | | | |
| | Disturbance-3 | | | | | | | | 21 | | 4 | 1 | 12 | | | | | | | 1 | | | | | | | | | |
| | Disturbance-4 | | | | | | | | 30 | 4 | | | 7 | | | | | | | 1 | | | | | | | | | |
| | Disturbance-5 | | | | | | | | 26 | 1 | | | 4 | | | | | | | 1 | | | | | | | | | |
| FINLAND: DEN-COM | UNDISTURBED | | | | | | | | 96 | 8 | | | | | | | | | | 2 | 3 | | | | | | | | |
| | Disturbance-1 | | | | | | | 13 | 14 | 37 | | | 1 | | | | | | | 2 | | 1 | | | | | | | |
| | Disturbance-2 | | | | | | | 70 | 7 | 111 | 22 | | 14 | | | | | | | | 1 | 1 | | | | | | | |
| | Disturbance-3 | | | | | | | | 96 | | 13 | | 79 | | 3 | | | | | 5 | 8 | 2 | 4 | | | | 1 | | |
| | Disturbance-4 | | | | | | | | 95 | 14 | | | 43 | | 8 | | | | | 11 | 16 | 1 | 3 | 1 | | | 3 | | |
| | Disturbance-5 | | | | | | | | 108 | 4 | | | 22 | 2 | 7 | | | | | 14 | 11 | 1 | 1 | | | | 1 | | |

**Calendar 2B**: Finland

| FINLAND Jokioinen | N 61° E23° | 2006 C.E. | | | | | | | | | | | YEAR-SEASON-MONTH-JULIAN WEEK | | | | | | | | | | | | | | | | 2006 C.E. |
|---|---|---|---|---|---|---|---|---|---|---|---|---|---|---|---|---|---|---|---|---|---|---|---|---|---|---|---|---|---|
| | | | | | SPRING | | | | | | | | | | | | SUMMER | | | | | | | | AUTUMN | | | | |
| | | | | MAY | | | | JUNE | | | | | JULY | | | | AUGUST | | | | SEPTEMBER | | | | | OCTOBER | | | |
| POPULATION | DISTURBANCE | 16 | 17 | 18 | 19 | 20 | 21 | 22 | 23 | 24 | 25 | 26 | 27 | 28 | 29 | 30 | 31 | 32 | 33 | 34 | 35 | 36 | 37 | 38 | 39 | 40 | 41 | 42 | 43 |
| Spring Cereals & Grassland | WINTER | | | | | | | | | | | | | | | | | | | | | | | | | | | | |
| | Crop Disturbance | | | | | | | | | | | | | | | | | | | | | | | | | | | | |
| | OPPORTUNITY | | | | | | | | | | | | | | | | | | | | | | | | | | | | |
| FINLAND: LOCAL | UNDISTURBED | | | | | | | | 27 | | | | 1 | | | | | | | | | | | | | | | | |
| | Disturbance-1 | | | | | | | | 23 | | | | 1 | | | | | | | | | | | | | | | | |
| | Disturbance-2 | | | | | | | | 49 | | | | | | | | | | | | 1 | | | | | | | | |
| | Disturbance-3 | | | | | | | | 21 | | | | 17 | | | | | | | 1 | | | | | | | | | |
| | Disturbance-4 | | | | | | | | | 4 | | | 7 | | | | | | | 1 | | | | | | | | | |
| | Disturbance-5 | | | | | | | | | | | | 4 | | | | | | | 1 | | | | | | | | | |
| FINLAND: DEN-COM | UNDISTURBED | | | | | | | | 104 | | | | | | | | | | | 5 | | | | | | | | | |
| | Disturbance-1 | | | | | | | | 64 | | | | 1 | | | | | | | 2 | | 1 | | | | | | | |
| | Disturbance-2 | | | | | | | | 210 | | | | 14 | | | | | | | | 2 | | | | | | | | |
| | Disturbance-3 | | | | | | | | 96 | | 13 | | 79 | | 3 | | | | | | 19 | | | | | | 1 | | |
| | Disturbance-4 | | | | | | | | | 14 | | | 43 | | 8 | | | | | | 32 | | | | | | 3 | | |
| | Disturbance-5 | | | | | | | | | | | | 31 | | | | | | | | 27 | | | | | | 1 | | |
| | Seedling Recruitment Cohorts | | | | | | | | LATE SPRING | | | | EARLY SUMMER | | | | | | | | | LATE SUMMER | | | | | AUT | | |
| | Potential Seed Productivity | | | | | | | | 1000x | | | | 100x | | | | | | | | | 10x | | | | | 1x | | |
| SEEDLING INVESTMENT (%) | Un-Disturb: LOCAL | | | | | | | | 4% | | | | 1% | | | | | | | | | | | | | | | | |
| | Un-Disturb: DEN-COM | | | | | | | | 15% | | | | | | | | | | | 1% | | | | | | | | | |
| | Disturb: LOCAL | | | | | | | 1-7% | | | | | | | | | | | | 1% | | | | | | | | | |
| | Disturb: DEN-COM | | | | | | | 2-30% | | | | | 1-12% | | | | | | | 1-5% | | | | | | | 1% | | |
| | Un-Disturb: LOCAL | | | | | | | | 4% | | | | 1% | | | | | | | | | | | | | | | | |
| | Disturb: LOCAL | | | | | | | 1-7% | | | | | | | | | | | | 1% | | | | | | | | | |
| | Un-Disturb: DEN-COM | | | | | | | | 15% | | | | | | | | | | | 1% | | | | | | | | | |
| | Disturb: DEN-COM | | | | | | | 2-30% | | | | | 1-12% | | | | | | | 1-5% | | | | | | | 1% | | |

**Calendar 3A**: Norway

| NORWAY | N 60° E 11° | 2006 C.E. | | | | | | | | | | | YEAR-SEASON-MONTH-JULIAN WEEK | | | | | | | | | | | | | | | 2006 C.E. |
|---|---|---|---|---|---|---|---|---|---|---|---|---|---|---|---|---|---|---|---|---|---|---|---|---|---|---|---|---|
| As, Akersus | | | | | | | SPRING | | | | | | | | | | | | SUMMER | | | | | | | | AUTUMN | | | |
| | | APRIL | | | MAY | | | | JUNE | | | | | JULY | | | | AUGUST | | | | SEPT | | | | | OCT | NOV |
| POPULATION | DISTURBANCE | 15 | 16 | 17 | 18 | 19 | 20 | 21 | 22 | 23 | 24 | 25 | 26 | 27 | 28 | 29 | 30 | 31 | 32 | 33 | 34 | 35 | 36 | 37 | 38 | 39 | 40 | 46 |
| | WINTER | | | | | | | | | | | | | | | | | | | | | | | | | | | |
| Spring cereals | Seedbed Prep | | | | | | | | | | | | | | | | | | | | | | | | | | | |
| Swedes | Planting | | | | | | | | | | | | | | | | | | | | | | | | | | | |
| Winter wheat | Weed Control | | | | | | | | | | | | | | | | | | | | | | | | | | | |
| | Harvest | | | | | | | | | | | | | | | | | | | | | | | | | | | |
| | OPPORTUNITY | | | | | | | | | | | | | | | | | | | | | | | | | | | |
| | UNDISTURBED | | | 2 | 7 | 8 | | 1 | 8 | 7 | | | 6 | | | | | | | | | | | | | | | |
| | Disturbance-1 | | | | 3 | 4 | | 1 | 5 | 1 | | | 1 | | | | | | | | | | | | | | | |
| NORWAY: | Disturbance-2 | | | 2 | 6 | 8 | | 20 | 183 | 12 | 2 | | 21 | 1 | | 1 | | | | | | | | | | | | |
| LOCAL | Disturbance-3 | | | | 5 | 6 | | 1 | 83 | 63 | 1 | 2 | 23 | | | | | | | | | | | | | | | |
| | Disturbance-4 | | | | 10 | 10 | | 1 | 3 | 12 | 6 | 9 | 40 | | 1 | | | | | | | | | | | | | |
| | Disturbance-5 | | | | 6 | 6 | | | 8 | | 2 | 2 | 62 | 13 | | | | | | | | | | | | | | |
| | UNDISTURBED | | | 3 | 28 | 58 | | 2 | 10 | | | 2 | 14 | 2 | | | | | | | | | | | | | | 3 |
| | Disturbance-1 | | | 2 | 6 | 26 | | 3 | 9 | 1 | | | | | | | | | | | | | | | | | | |
| NORWAY: | Disturbance-2 | | | 2 | 28 | 59 | | 22 | 112 | 18 | 1 | 2 | 12 | 1 | | | | | | | | | | | | | | |
| DEN-COM | Disturbance-3 | | | 3 | 19 | 56 | | | 66 | 34 | 1 | | 7 | 2 | | | | | | | | | | | | | | |
| | Disturbance-4 | | | 5 | 27 | 58 | | | 14 | 2 | | 2 | 19 | | | | | | | | | | | | | | | |
| | Disturbance-5 | | | 2 | 29 | 56 | | 1 | 9 | | | 1 | 29 | 5 | | | | | | | | | | | | | | |

**Calendar 3B**: Norway

| NORWAY As, Akersus | N 60° E 11° | 2006 C.E. | | | | | | | | | | | | YEAR-SEASON-MONTH-JULIAN WEEK | | | | | | | | | | | | | | | 2006 C.E. |
|---|---|---|---|---|---|---|---|---|---|---|---|---|---|---|---|---|---|---|---|---|---|---|---|---|---|---|---|---|---|
| | | | | | | SPRING | | | | | | | | | | | SUMMER | | | | | | | | AUTUMN | | | | |
| | | APRIL | | | MAY | | | | JUNE | | | | | JULY | | | AUGUST | | | | | SEPTEMBER | | | | OCT | | | NOV |
| POPULATION | DISTURBANCE | 15 | 16 | 17 | 18 | 19 | 20 | 21 | 22 | 23 | 24 | 25 | 26 | 27 | 28 | 29 | 30 | 31 | 32 | 33 | 34 | 35 | 36 | 37 | 38 | 39 | 40 | 46 | 47 |
| Spring Cereals | WINTER | | | | | | | | | | | | | | | | | | | | | | | | | | | | |
| Swedes | Crop Disturbance | | | | | | | | | | | | | | | | | | | | | | | | | | | | |
| Winter Wheat | OPPORTUNITY | | | | | | | | | | | | | | | | | | | | | | | | | | | | |
| | UNDISTURBED | | | | | 17 | | | 16 | | | | 6 | | | | | | | | | | | | | | | | |
| | Disturbance-1 | | | | | 7 | | | 7 | | | | 1 | | | | | | | | | | | | | | | | |
| NORWAY: | Disturbance-2 | | | | | 8 | | | 217 | | | | 21 | | | 1 | | | | | | | | | | | | | |
| LOCAL | Disturbance-3 | | | | | 6 | | | 173 | | | | | | | | | | | | | | | | | | | | |
| | Disturbance-4 | | | | | | | | 70 | | | | | | 1 | | | | | | | | | | | | | | |
| | Disturbance-5 | | | | | | | | | | | | 79 | | | | | | | | | | | | | | | | |
| | UNDISTURBED | | | | | 89 | | | 12 | | | | 18 | | | | | | | | | | | | | | | 3 | |
| | Disturbance-1 | | | | | 34 | | | 13 | | | | | | | | | | | | | | | | | | | | |
| DEN-COM | Disturbance-2 | | | | | 59 | | | 168 | | | | | | | | | | | | | | | | | | | | |
| | Disturbance-3 | | | | | 56 | | | 101 | | | | 9 | | | | | | | | | | | | | | | | |
| | Disturbance-4 | | | | | | | | 18 | | | | 21 | | | | | | | | | | | | | | | | |
| | Disturbance-5 | | | | | | | | | | | | 35 | | | | | | | | | | | | | | | | |
| Seedling Recruitment Cohorts | | | | | SPRING | | | | LATE SPRING | | | | | EARLY SUMMER | | | | | | | | | | | | | | AUT | |
| Potential Seed Productivity | | | | | 1000X | | | | 1000x | | | | | 100X | | | | | | | | | | | | | | 1x | |
| | Un-Disturb: LOCAL | | | | 1% | | | | 1% | | | | 1% | | | | | | | | | | | | | | | | |
| | Un-Disturb: DEN-COM | | | | 6% | | | | 1% | | | | 1% | | | | | | | | | | | | | | | 1% | |
| SEEDLING | Disturb: LOCAL | | | | 1% | | | | 1-17% | | | | | | | | | | | | | | | | | | | | |
| INVESTMENT | Disturb: DEN-COM | | | | 2-4% | | | | 1-12% | | | | | | | | | | | | | | | | | | | | |
| (%) | Un-Disturb: LOCAL | | | | 1% | | | | 1% | | | | 1% | | | | | | | | | | | | | | | | |
| | Disturb: LOCAL | | | | 1% | | | | 1-17% | | | | | | | | | | | | | | | | | | | | |
| | Un-Disturb: DEN-COM | | | | 6% | | | | 1% | | | | 1% | | | | | | | | | | | | | | | 1% | |
| | Disturb: DEN-COM | | | | 2-4% | | | | 1-12% | | | | | | | | | | | | | | | | | | | | |

**Calendar 4A**: Sweden

| SWEDEN | N 58° E 15° | 2006 C.E. | | | | | | | | | | | | YEAR-SEASON-MONTH-JULIAN WEEK | | | | | | | | | | | 2006 C.E. | |
|---|---|---|---|---|---|---|---|---|---|---|---|---|---|---|---|---|---|---|---|---|---|---|---|---|---|---|
| | | | | | | | SPRING | | | | | | | | | | | | | SUMMER | | | | | | |
| | | | APRIL | | | | MAY | | | | JUNE | | | | | JULY | | | | | AUGUST | | | SEPT | | |
| POPULATION | DISTURBANCE | 13 | 14 | 15 | 16 | 17 | 18 | 19 | 20 | 21 | 22 | 23 | 24 | 25 | 26 | 27 | 28 | 29 | 30 | 31 | 32 | 33 | 34 | 35 | 36 | 37 |
| | WINTER | ■ | ■ | | | | | | | | | | | | | | | | | | | | | | ■ | ■ |
| Spring cereals Oilseed rape | Seedbed Prep | | | | ■ | ■ | ■ | | | | | | | | | | | | | | | | | | | |
| | Planting | | | | ■ | ■ | ■ | | | | | | | | | | | | | | | | | | | |
| | Weed Control | | | | | | | | | | ■ | ■ | ■ | | | | | | | | | | | | | |
| | Harvest | | | | | | | | | | | | | | | | | | | ■ | ■ | ■ | ■ | ■ | | |
| | OPPORTUNITY | | | ■ | | | | ■ | ■ | ■ | | | | ■ | ■ | ■ | ■ | ■ | ■ | | | | | | | |
| SWEDEN: LOCAL | UNDISTURBED | | | | 11 | 36 | 79 | 102 | 5 | 7 | | | | | | | | | | | 12 | 8 | | | | |
| | Disturbance-1 | | | | 37 | 60 | 76 | 103 | 6 | 10 | 2 | | | | | | | | | | 10 | 9 | | | | |
| | Disturbance-2 | | | | 12 | 45 | 76 | 42 | 6 | 8 | 1 | 2 | | | | | | | | | 10 | 6 | | | | |
| | Disturbance-3 | | | | 11 | 43 | 84 | 90 | 4 | 14 | 1 | | | | | | | | | | 11 | 9 | | | | |
| | Disturbance-4 | | | | 10 | 35 | 73 | 96 | 14 | 14 | 6 | | | | | | | | | | 8 | 8 | | | | |
| | Disturbance-5 | | | | 12 | 14 | 76 | 105 | 4 | 38 | 13 | 5 | | | | | | | | | 10 | 8 | | | | |
| SWEDEN: DEN-COM | UNDISTURBED | | | 2 | 13 | 26 | 69 | 84 | 5 | 7 | | | | | | | | | | | 5 | 10 | | | | |
| | Disturbance-1 | | | 1 | 41 | 61 | 80 | 82 | 5 | 11 | | | | | | | | | | | 5 | 8 | | | | |
| | Disturbance-2 | | | 2 | 12 | 26 | 64 | 32 | 3 | 9 | | | | | | | | | | | 9 | 16 | | | | |
| | Disturbance-3 | | | 1 | 13 | 25 | 67 | 72 | 3 | 11 | | | | | | | | | | | 6 | 7 | | | | |
| | Disturbance-4 | | | 2 | 12 | 27 | 75 | 86 | 15 | 13 | 6 | | | | | | | | | | 6 | 9 | | | | |
| | Disturbance-5 | | | 1 | 14 | 25 | 69 | 83 | 6 | 31 | 14 | 3 | 2 | 1 | | | | | | | 7 | 7 | | | | |

**Calendar 4B**: Sweden

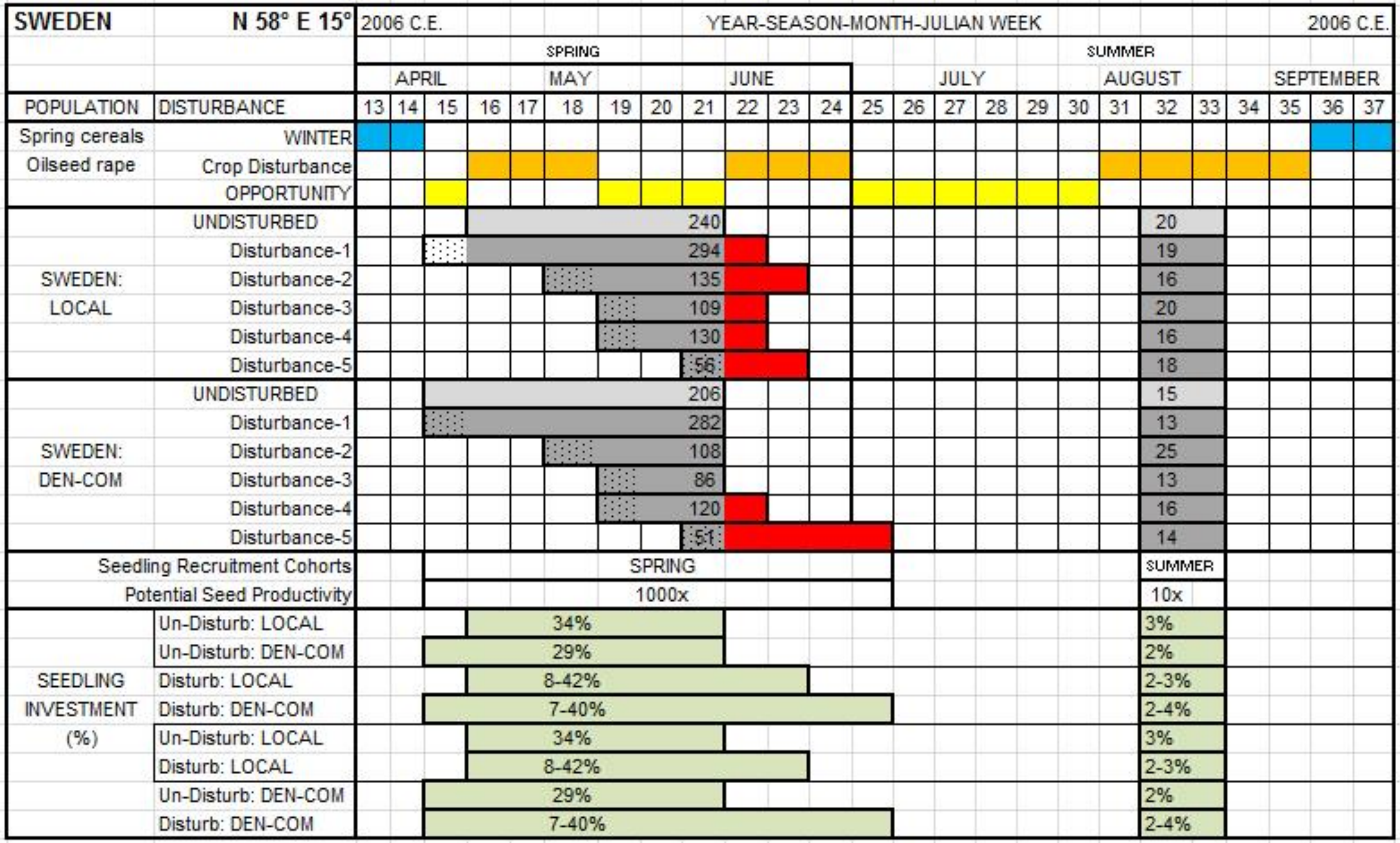

| SWEDEN | N 58° E 15° | 2006 C.E. | | | | | | | | | | | | YEAR-SEASON-MONTH-JULIAN WEEK | | | | | | | | | | | 2006 C.E. | |
|---|---|---|---|---|---|---|---|---|---|---|---|---|---|---|---|---|---|---|---|---|---|---|---|---|---|---|
| | | | | | | | SPRING | | | | | | | | | | | | | SUMMER | | | | | | |
| | | | APRIL | | | | MAY | | | | JUNE | | | | | JULY | | | | | AUGUST | | | | SEPTEMBER | |
| POPULATION | DISTURBANCE | 13 | 14 | 15 | 16 | 17 | 18 | 19 | 20 | 21 | 22 | 23 | 24 | 25 | 26 | 27 | 28 | 29 | 30 | 31 | 32 | 33 | 34 | 35 | 36 | 37 |
| Spring cereals | WINTER | | | | | | | | | | | | | | | | | | | | | | | | | |
| Oilseed rape | Crop Disturbance | | | | | | | | | | | | | | | | | | | | | | | | | |
| | OPPORTUNITY | | | | | | | | | | | | | | | | | | | | | | | | | |
| | UNDISTURBED | | | | | | | | | 240 | | | | | | | | | | | 20 | | | | | |
| | Disturbance-1 | | | | | | | | | 294 | | | | | | | | | | | 19 | | | | | |
| SWEDEN: | Disturbance-2 | | | | | | | | | 135 | | | | | | | | | | | 16 | | | | | |
| LOCAL | Disturbance-3 | | | | | | | | | 109 | | | | | | | | | | | 20 | | | | | |
| | Disturbance-4 | | | | | | | | | 130 | | | | | | | | | | | 16 | | | | | |
| | Disturbance-5 | | | | | | | | | 56 | | | | | | | | | | | 18 | | | | | |
| | UNDISTURBED | | | | | | | | | 206 | | | | | | | | | | | 15 | | | | | |
| | Disturbance-1 | | | | | | | | | 282 | | | | | | | | | | | 13 | | | | | |
| SWEDEN: | Disturbance-2 | | | | | | | | | 108 | | | | | | | | | | | 25 | | | | | |
| DEN-COM | Disturbance-3 | | | | | | | | | 86 | | | | | | | | | | | 13 | | | | | |
| | Disturbance-4 | | | | | | | | | 120 | | | | | | | | | | | 16 | | | | | |
| | Disturbance-5 | | | | | | | | | 51 | | | | | | | | | | | 14 | | | | | |
| | Seedling Recruitment Cohorts | | | | | | | | SPRING | | | | | | | | | | | | SUMMER | | | | | |
| | Potential Seed Productivity | | | | | | | | 1000x | | | | | | | | | | | | 10x | | | | | |
| | Un-Disturb: LOCAL | | | | | | 34% | | | | | | | | | | | | | | 3% | | | | | |
| | Un-Disturb: DEN-COM | | | | | | 29% | | | | | | | | | | | | | | 2% | | | | | |
| SEEDLING | Disturb: LOCAL | | | | | | 8-42% | | | | | | | | | | | | | | 2-3% | | | | | |
| INVESTMENT | Disturb: DEN-COM | | | | | | 7-40% | | | | | | | | | | | | | | 2-4% | | | | | |
| (%) | Un-Disturb: LOCAL | | | | | | 34% | | | | | | | | | | | | | | 3% | | | | | |
| | Disturb: LOCAL | | | | | | 8-42% | | | | | | | | | | | | | | 2-3% | | | | | |
| | Un-Disturb: DEN-COM | | | | | | 29% | | | | | | | | | | | | | | 2% | | | | | |
| | Disturb: DEN-COM | | | | | | 7-40% | | | | | | | | | | | | | | 2-4% | | | | | |

**Calendar 5A**:  Czech Republic

| CZECH | N 48° E58° | 2006 C.E. | | | | | | | | | | | | | | | YEAR-SEASON-MONTH-JULIAN WEEK | | | | | | | | | | | | | | | | | 2006 C.E. |
|---|---|---|---|---|---|---|---|---|---|---|---|---|---|---|---|---|---|---|---|---|---|---|---|---|---|---|---|---|---|---|---|---|---|---|
| Ceske Budejovice | | | | | | | | SPRING | | | | | | | | | | | | SUMMER | | | | | | | | | | | AUTUMN | | | |
| | | | | APRIL | | | MAY | | | | JUNE | | | | | | JULY | | | | AUGUST | | | SEPT | | | | | | OCTOBER | | | NOV | |
| POPULATION | DISTURBANCE | 13 | 14 | 15 | 16 | 17 | 18 | 19 | 20 | 21 | 22 | 23 | 24 | 25 | 26 | 27 | 28 | 29 | 30 | 31 | 32 | 33 | 34 | 35 | 36 | 37 | 38 | 39 | 40 | 41 | 42 | 43 | 44 | 45 |
| | WINTER | ■ | | | | | | | | | | | | | | | | | | | | | | | | | | | | | | | | ■ |
| | Seedbed Prep | | ■ | ■ | ■ | ■ | ■ | | | | | | | | | | | | | | | | | | | | | | ■ | ■ | | | | |
| Spring cereals | Planting | | | | | | | | | ■ | ■ | | | | | | | | | | | | | | | | ■ | ■ | ■ | | | | | |
| Swede rape | Weed Control | | ■ | ■ | ■ | ■ | ■ | ■ | | | | | | | | | | | | | | | | | | | | | ■ | ■ | ■ | ■ | ■ | |
| | Harvest | | | | | | | | | | | | | | | | | | | ■ | ■ | ■ | ■ | ■ | ■ | ■ | | | | | | | | |
| | OPPORTUNITY | | | | | | | | ■ | | | ■ | ■ | ■ | ■ | ■ | ■ | ■ | ■ | | | | | | | | | | | | | | | |
| | UNDISTURBED | | | 1 | 11 | 7 | 7 | 2 | | 1 | 1 | | | | | | 3 | | | | 2 | | | | | | | | | | | | | |
| | Disturbance-1 | | | | 3 | | 49 | 9 | 5 | 2 | 4 | 4 | | | | 1 | | 1 | | | | | | | | | | | | | | | | |
| CZECH: | Disturbance-2 | | | | 17 | 3 | 30 | 83 | 4 | | 5 | 2 | | | | 1 | | 1 | 1 | | | | 1 | 1 | | | | | | | | | | |
| LOCAL | Disturbance-3 | | | 1 | 8 | 6 | 5 | 10 | 12 | 8 | 31 | 34 | 3 | | | 11 | 3 | | | | | 1 | | 1 | | | | | | | | | | |
| | Disturbance-4 | | | | 9 | 10 | 6 | | 6 | 11 | 11 | 12 | 3 | | | 8 | 3 | | 1 | | 1 | | 1 | | | | | | | | | | | |
| | Disturbance-5 | | | | 11 | 5 | 8 | | | 5 | 4 | 21 | | | | 4 | 2 | 1 | | | | 1 | | | | | | | | | | | | |
| | UNDISTURBED | | | 6 | 77 | 17 | 22 | 4 | 2 | 2 | 1 | | | 1 | | 1 | 1 | | | | | | | | | | | | | | | | | |
| | Disturbance-1 | | | 1 | 25 | 19 | 84 | 10 | 3 | 2 | 1 | | | | | | | | | | | | | | | | | | | | | | | |
| CZECH: | Disturbance-2 | | | 4 | 57 | 16 | 110 | 60 | 4 | 1 | 3 | 3 | | | | | | | | | | | | | | | | | | | | | | |
| DEN-COM | Disturbance-3 | | | 7 | 75 | 13 | 14 | 29 | 22 | 23 | 12 | 7 | | | | 1 | | | | | | | | | | | | | | | | | | |
| | Disturbance-4 | | | 5 | 111 | 14 | 19 | 2 | 5 | 8 | 8 | 15 | 5 | | | 6 | 1 | 1 | | | | | | | | | | | | | | | | |
| | Disturbance-5 | | | 3 | 57 | 21 | 24 | 4 | | 1 | 8 | 4 | 2 | | | 4 | | | | | | | | | | | | | | | | | | |

**Calendar 5B**: Czech Republic

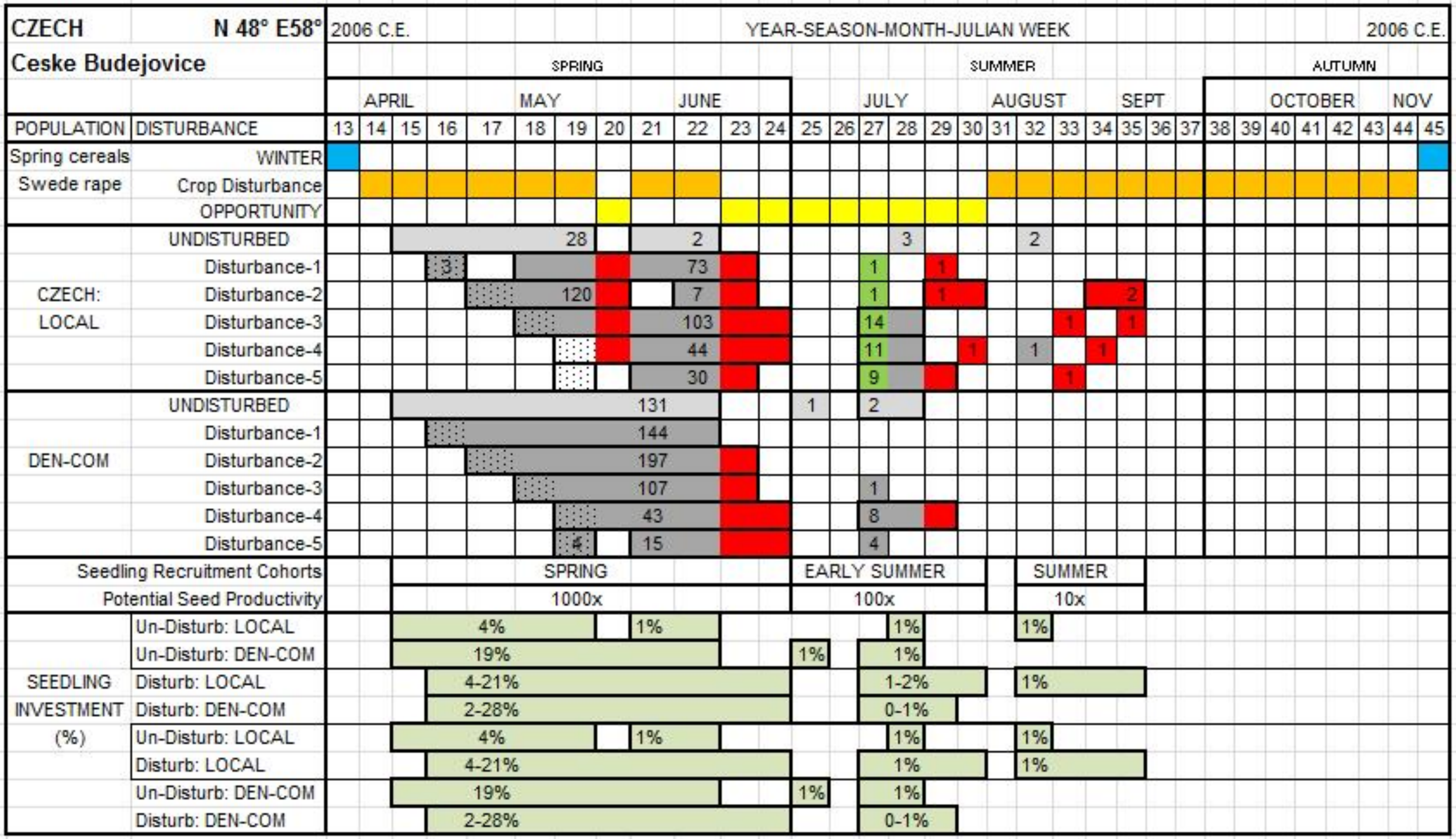

| CZECH | N 48° E58° | 2006 C.E. | | | | | | | | | | | | | | | YEAR-SEASON-MONTH-JULIAN WEEK | | | | | | | | | | | | | | | | 2006 C.E. | |
|---|---|---|---|---|---|---|---|---|---|---|---|---|---|---|---|---|---|---|---|---|---|---|---|---|---|---|---|---|---|---|---|---|---|---|
| Ceske Budejovice | | | | | | | | SPRING | | | | | | | | | | | | SUMMER | | | | | | | | | | | AUTUMN | | | |
| | | | APRIL | | | | MAY | | | | JUNE | | | | | | JULY | | | | AUGUST | | | SEPT | | | | | | OCTOBER | | | NOV | |
| POPULATION | DISTURBANCE | 13 | 14 | 15 | 16 | 17 | 18 | 19 | 20 | 21 | 22 | 23 | 24 | 25 | 26 | 27 | 28 | 29 | 30 | 31 | 32 | 33 | 34 | 35 | 36 | 37 | 38 | 39 | 40 | 41 | 42 | 43 | 44 | 45 |
| Spring cereals | WINTER | | | | | | | | | | | | | | | | | | | | | | | | | | | | | | | | | |
| Swede rape | Crop Disturbance | | | | | | | | | | | | | | | | | | | | | | | | | | | | | | | | | |
| | OPPORTUNITY | | | | | | | | | | | | | | | | | | | | | | | | | | | | | | | | | |
| | UNDISTURBED | | | | | | | 28 | | | 2 | | | | | | 3 | | | | 2 | | | | | | | | | | | | | |
| | Disturbance-1 | | | | 3 | | | | | | 73 | | | | | 1 | | 1 | | | | | | | | | | | | | | | | |
| CZECH: | Disturbance-2 | | | | | | | 120 | | | 7 | | | | | 1 | | 1 | | | | | | 2 | | | | | | | | | | |
| LOCAL | Disturbance-3 | | | | | | | | | | 103 | | | | | 14 | | | | | | 1 | | 1 | | | | | | | | | | |
| | Disturbance-4 | | | | | | | | | | 44 | | | | | 11 | | | 1 | | 1 | | 1 | | | | | | | | | | | |
| | Disturbance-5 | | | | | | | | | | 30 | | | | | 9 | | | | | | 1 | | | | | | | | | | | | |
| | UNDISTURBED | | | | | | | | | 131 | | | | 1 | | 2 | | | | | | | | | | | | | | | | | | |
| | Disturbance-1 | | | | | | | | | 144 | | | | | | | | | | | | | | | | | | | | | | | | |
| DEN-COM | Disturbance-2 | | | | | | | | | 197 | | | | | | | | | | | | | | | | | | | | | | | | |
| | Disturbance-3 | | | | | | | | | 107 | | | | | | 1 | | | | | | | | | | | | | | | | | | |
| | Disturbance-4 | | | | | | | | | 43 | | | | | | 8 | | | | | | | | | | | | | | | | | | |
| | Disturbance-5 | | | | | | | 4 | | 15 | | | | | | 4 | | | | | | | | | | | | | | | | | | |
| | Seedling Recruitment Cohorts | | | | | | | SPRING | | | | | | | | EARLY SUMMER | | | | | | SUMMER | | | | | | | | | | | | |
| | Potential Seed Productivity | | | | | | | 1000x | | | | | | | | 100x | | | | | | 10x | | | | | | | | | | | | |
| | Un-Disturb: LOCAL | | | | | 4% | | | | 1% | | | | | | | 1% | | | | 1% | | | | | | | | | | | | | |
| | Un-Disturb: DEN-COM | | | | | 19% | | | | | | | | 1% | | | 1% | | | | | | | | | | | | | | | | | |
| SEEDLING | Disturb: LOCAL | | | | | 4-21% | | | | | | | | | | | 1-2% | | | | 1% | | | | | | | | | | | | | |
| INVESTMENT | Disturb: DEN-COM | | | | | 2-28% | | | | | | | | | | | 0-1% | | | | | | | | | | | | | | | | | |
| (%) | Un-Disturb: LOCAL | | | | | 4% | | | | 1% | | | | | | | 1% | | | | 1% | | | | | | | | | | | | | |
| | Disturb: LOCAL | | | | | 4-21% | | | | | | | | | | | 1% | | | | 1% | | | | | | | | | | | | | |
| | Un-Disturb: DEN-COM | | | | | 19% | | | | | | | | 1% | | | 1% | | | | | | | | | | | | | | | | | |
| | Disturb: DEN-COM | | | | | 2-28% | | | | | | | | | | | 0-1% | | | | | | | | | | | | | | | | | |

**Calendar 6A**:  Canada

| CANADA N 45° W 73° L'Acadie, Quebec | | 2005 C.E. | | | 2006 C.E. | | | | | | | | | | | YEAR-SEASON-MONTH-JULIAN WEEK | | | | | | | | | | | | | | 2006 C.E. |
|---|---|---|---|---|---|---|---|---|---|---|---|---|---|---|---|---|---|---|---|---|---|---|---|---|---|---|---|---|---|---|
| | | AUTUMN | | | SPRING | | | | | | | | | | | | | | | SUMMER | | | | | | | | | | |
| | | NOV | | | APRIL | | | | MAY | | | | JUNE | | | | JULY | | | | | | AUGUST | | | | SEPT | | | |
| POPULATION | DISTURBANCE | 45 | 46 | 47 | 13 | 14 | 15 | 16 | 17 | 18 | 19 | 20 | 21 | 22 | 23 | 24 | 25 | 26 | 27 | 28 | 29 | 30 | 31 | 32 | 33 | 34 | 35 | 36 | 37 | 38 |
| | WINTER | ■ | ■ | ■ | ■ | | | | | | | | | | | | | | | | | | | | | | | | | ■ |
| Onions, Sweet maize | Seedbed Prep | | | | | | | | ■ | ■ | ■ | ■ | ■ | | | | | | | | | | | | | | | | | |
| | Planting | | | | | | | | ■ | ■ | ■ | ■ | ■ | | | | | | | | | | | | | | | | | |
| | Weed Control | | | | | | | | ■ | ■ | ■ | ■ | ■ | ■ | ■ | ■ | ■ | | | | | | | | | | | | | |
| | Harvest | | | | | | | | | | | | | | | | | | | | | | ■ | ■ | ■ | ■ | ■ | ■ | ■ | |
| | OPPORTUNITY | | | | | □ | □ | □ | | | | | | | | | | □ | □ | □ | □ | □ | | | | | | | | |
| CANADA: LOCAL | UNDISTURBED | | | | | 1 | 1 | 5 | 10 | 2 | 3 | 3 | 2 | 2 | | | | | | | | 2 | 1 | | | | | | | |
| | Disturbance-1 | | | | | 1 | | 1 | 10 | 5 | 3 | 8 | 2 | 1 | 1 | 2 | 1 | | | | | | | | | | | | | |
| | Disturbance-2 | | | | | 1 | | 5 | | 2 | 121 | 23 | | | 1 | 2 | | | | | | | | | | | | | | |
| | Disturbance-3 | | | | | | 2 | 3 | 8 | 2 | 6 | 143 | 4 | | | | | | | | | | | | | | | | | |
| | Disturbance-4 | | | | | 1 | | 7 | 15 | 3 | 8 | 309 | 58 | 1 | | | | | | | | | | | | | | | | |
| | Disturbance-5 | | | | | 1 | 1 | 7 | 7 | 1 | 3 | 5 | 71 | 23 | 4 | | | | 1 | | | | | | | | | | | |
| CANADA: DEN-COM | UNDISTURBED | | 1 | | | 3 | 4 | 42 | 37 | 11 | 17 | 5 | | | | | | | | | | | | | | | | | | |
| | Disturbance-1 | | | | | | | 14 | 38 | 8 | 13 | 5 | | | 1 | | | | | | | | | | | | | | | |
| | Disturbance-2 | | | | | 2 | 4 | 32 | 6 | 35 | 110 | 7 | 2 | | 1 | | | | | | | | | | | | | | | |
| | Disturbance-3 | | 1 | | | 3 | 3 | 45 | 40 | 10 | 36 | 131 | | 1 | | | | | | | | | | | | | | | | |
| | Disturbance-4 | | | | | | 3 | 45 | 40 | 15 | 9 | 147 | 8 | 9 | | 1 | | | | | | | | | | | | | | |
| | Disturbance-5 | | | | | | 1 | 37 | 44 | 8 | 11 | 15 | 57 | 12 | | | | | | | | | | | | | | | | |

**Calendar 6B**:  Canada

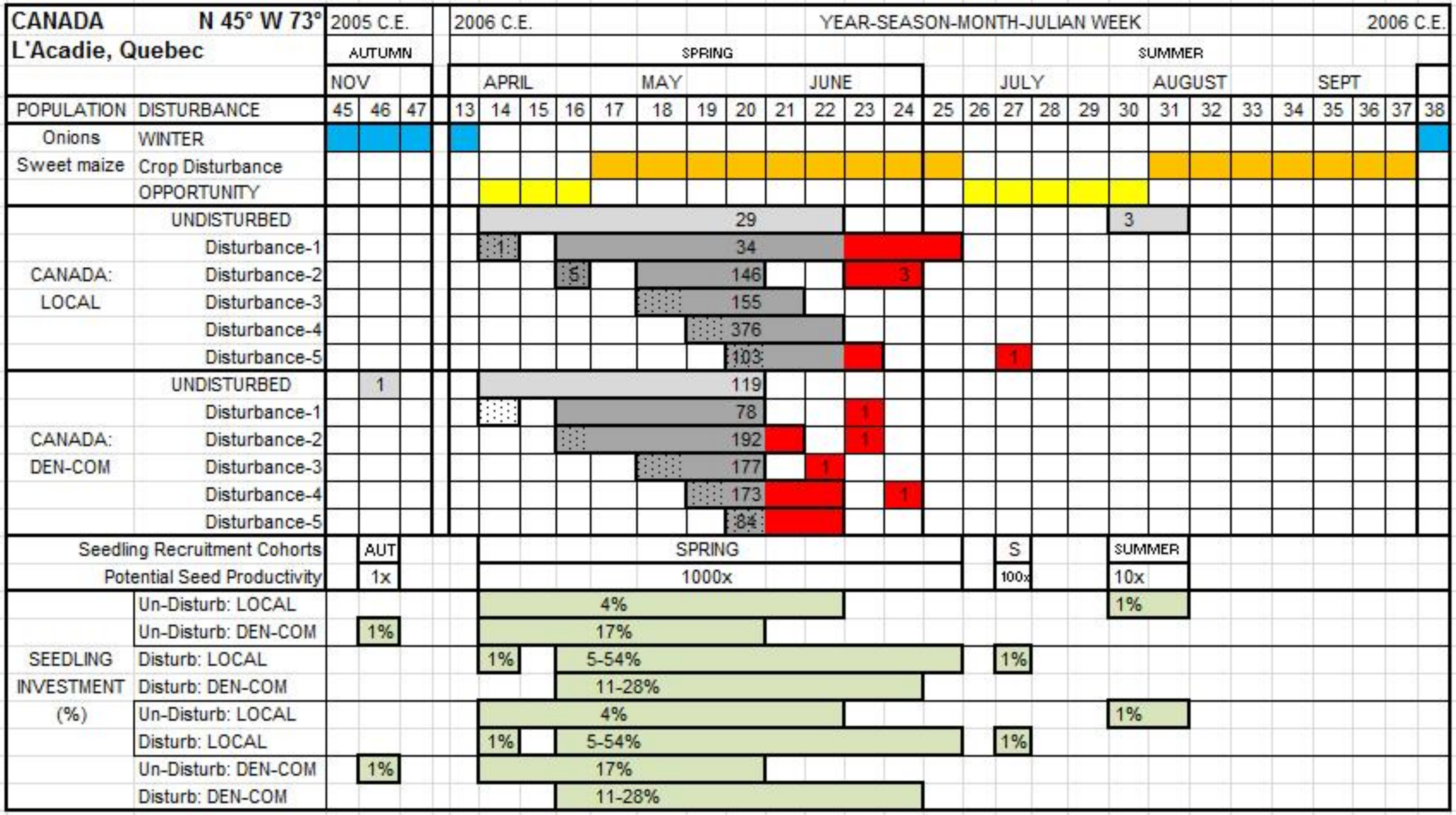

| | | | | | | | | | | | | | | | | | | | | | | | | | | | | | | |
|---|---|---|---|---|---|---|---|---|---|---|---|---|---|---|---|---|---|---|---|---|---|---|---|---|---|---|---|---|---|---|
| CANADA | N 45° W 73° | 2005 C.E. | | | 2006 C.E. | | | | | | | | | | | | | | YEAR-SEASON-MONTH-JULIAN WEEK | | | | | | | | | | 2006 C.E. | |
| L'Acadie, Quebec | | | AUTUMN | | | | | | | | SPRING | | | | | | | | | | | | SUMMER | | | | | | | |
| | | NOV | | | | APRIL | | | | MAY | | | | JUNE | | | | | JULY | | | | AUGUST | | | | SEPT | | | |
| POPULATION | DISTURBANCE | 45 | 46 | 47 | 13 | 14 | 15 | 16 | 17 | 18 | 19 | 20 | 21 | 22 | 23 | 24 | 25 | 26 | 27 | 28 | 29 | 30 | 31 | 32 | 33 | 34 | 35 | 36 | 37 | 38 |
| Onions | WINTER | | | | | | | | | | | | | | | | | | | | | | | | | | | | | |
| Sweet maize | Crop Disturbance | | | | | | | | | | | | | | | | | | | | | | | | | | | | | |
| | OPPORTUNITY | | | | | | | | | | | | | | | | | | | | | | | | | | | | | |
| | UNDISTURBED | | | | | | | | | | | 29 | | | | | | | | | | 3 | | | | | | | | |
| | Disturbance-1 | | | | | 1 | | | | | | 34 | | | | | | | | | | | | | | | | | | |
| CANADA: | Disturbance-2 | | | | | | | 5 | | | | 146 | | | | 3 | | | | | | | | | | | | | | |
| LOCAL | Disturbance-3 | | | | | | | | | | | 155 | | | | | | | | | | | | | | | | | | |
| | Disturbance-4 | | | | | | | | | | | 376 | | | | | | | | | | | | | | | | | | |
| | Disturbance-5 | | | | | | | | | | | 103 | | | | | | | 1 | | | | | | | | | | | |
| | UNDISTURBED | | 1 | | | | | | | | | 119 | | | | | | | | | | | | | | | | | | |
| | Disturbance-1 | | | | | | | | | | | 78 | | | 1 | | | | | | | | | | | | | | | |
| CANADA: | Disturbance-2 | | | | | | | | | | | 192 | | | 1 | | | | | | | | | | | | | | | |
| DEN-COM | Disturbance-3 | | | | | | | | | | | 177 | | 1 | | | | | | | | | | | | | | | | |
| | Disturbance-4 | | | | | | | | | | | 173 | | | | 1 | | | | | | | | | | | | | | |
| | Disturbance-5 | | | | | | | | | | | 84 | | | | | | | | | | | | | | | | | | |
| Seedling Recruitment Cohorts | | | AUT | | | | | | | | SPRING | | | | | | | | S | | | SUMMER | | | | | | | | |
| Potential Seed Productivity | | | 1x | | | | | | | | 1000x | | | | | | | | 100x | | | 10x | | | | | | | | |
| | Un-Disturb: LOCAL | | | | | | | | 4% | | | | | | | | | | | | | 1% | | | | | | | | |
| | Un-Disturb: DEN-COM | | 1% | | | | | | 17% | | | | | | | | | | | | | | | | | | | | | |
| SEEDLING | Disturb: LOCAL | | | | | 1% | | | 5-54% | | | | | | | | | | 1% | | | | | | | | | | | |
| INVESTMENT | Disturb: DEN-COM | | | | | | | | 11-28% | | | | | | | | | | | | | | | | | | | | | |
| (%) | Un-Disturb: LOCAL | | | | | | | | 4% | | | | | | | | | | | | | 1% | | | | | | | | |
| | Disturb: LOCAL | | | | | 1% | | | 5-54% | | | | | | | | | | 1% | | | | | | | | | | | |
| | Un-Disturb: DEN-COM | | 1% | | | | | | 17% | | | | | | | | | | | | | | | | | | | | | |
| | Disturb: DEN-COM | | | | | | | | 11-28% | | | | | | | | | | | | | | | | | | | | | |

**Calendar 7A**: Italy

| ITALY | N 43° E 12° | 2006 C.E. | | | | | | | | | | | | | | | | | | | | YEAR-SEASON-MONTH-JULIAN WEEK | | | | | | | | | | | | | | | | | | 2006 C.E. | |
| --- | --- | --- | --- | --- | --- | --- | --- | --- | --- | --- | --- | --- | --- | --- | --- | --- | --- | --- | --- | --- | --- | --- | --- | --- | --- | --- | --- | --- | --- | --- | --- | --- | --- | --- | --- | --- | --- | --- | --- | --- | --- | --- |
| Perugia, Umbria | | | WINTER | | | | | | | | | | SPRING | | | | | | | | | | | | | SUMMER | | | | | | | | | | | AUTUMN | | | | | |
| | | | | | | | | | APRIL | | | | MAY | | | | JUNE | | | | | JULY | | | | AUGUST | | | | SEPT | | | | | OCTOBER | | | | NOV | | | |
| POPULATION | DISTURBANCE | 5 | 6 | 7 | | 11 | 12 | 13 | 14 | 15 | 16 | 17 | 18 | 19 | 20 | 21 | 22 | 23 | 24 | 25 | 26 | 27 | 28 | 29 | 30 | 31 | 32 | 33 | 34 | 35 | 36 | 37 | 38 | 39 | 40 | 41 | 42 | 43 | 44 | 45 | 46 | 47 |
| | WINTER | | | | | | | | | | | | | | | | | | | | | | | | | | | | | | | | | | | | | | | | | |
| Maize | Seedbed Prep | | | | | | | | | | | | | | | | | | | | | | | | | | | | | | | | | | | | | | | | | |
| Soybeans | Planting | | | | | | | | | | | | | | | | | | | | | | | | | | | | | | | | | | | | | | | | | |
| Sunflower | Weed Control | | | | | | | | | | | | | | | | | | | | | | | | | | | | | | | | | | | | | | | | | |
| Vegetables | Harvest | | | | | | | | | | | | | | | | | | | | | | | | | | | | | | | | | | | | | | | | | |
| | OPPORTUNITY | | | | | | | | | | | | | | | | | | | | | | | | | | | | | | | | | | | | | | | | | |
| | UNDISTURBED | | | | | | 1 | 1 | 1 | 3 | 2 | 8 | 17 | 8 | 4 | | | | | | | | | | 2 | | | | | | | | | | | | | | 1 | 2 | | |
| | Disturbance-1 | | | | | | | 24 | 44 | 7 | 14 | 102 | 17 | 4 | 4 | | | | | | | | | | | 1 | | | | | | | | | | | | | 1 | | 1 | |
| ITALY: | Disturbance-2 | | | | | | | 1 | 2 | | 100 | 102 | 80 | 15 | 20 | 3 | | | | | | | | | | 1 | | | 2 | | 1 | | | | | | | | | | | |
| LOCAL | Disturbance-3 | | | | | | 1 | 1 | | | 3 | 101 | 172 | 39 | 19 | 2 | | | | | | | | | | 2 | | | 1 | | 2 | | | | 3 | | | | 1 | | | |
| | Disturbance-4 | | | | | | 2 | 1 | | | | 9 | 60 | 13 | 85 | 3 | | | | | | | 1 | 7 | | 1 | | | | | 3 | | | | 4 | | | | 1 | | 1 | |
| | Disturbance-5 | | | | | | | | 1 | | 1 | 7 | 11 | | 232 | 36 | 1 | 1 | 1 | 1 | 1 | 1 | 5 | | 1 | 4 | | | | | 6 | 1 | | | 12 | | | | | 1 | 1 | |
| | UNDISTURBED | | | | | | 6 | 10 | 10 | 14 | 6 | 5 | 4 | | | | | | | | | | | 1 | | | | | | | 1 | | | | 4 | | 2 | | 3 | | | |
| | Disturbance-1 | | | | | | 1 | 76 | 104 | 22 | 16 | 8 | 3 | 1 | | | | | | | | | | 1 | | | | | | | | | | | | | | | 1 | 1 | | |
| ITALY: | Disturbance-2 | | | | | | 5 | 20 | 17 | 9 | 172 | 80 | 22 | | 2 | | | | | | | | | | | 1 | | | | | | | | | | | 2 | | 1 | 2 | | |
| DEN-COM | Disturbance-3 | | | | | | 6 | 5 | 8 | 13 | 15 | 98 | 42 | 5 | 4 | | | | | | | | | | | | | | | | | | | | | | | | 1 | | | |
| | Disturbance-4 | | | | | | 8 | 19 | 13 | 10 | 8 | 4 | 10 | 2 | 4 | 2 | | | | | | | | | | | | | | | 2 | | | | | | 2 | | 2 | | | |
| | Disturbance-5 | | | | | | 5 | 3 | 9 | 13 | 2 | 3 | | | 62 | 4 | | | | | | | 8 | | 1 | | | | | | | | | | 2 | | 1 | | 1 | | | |

**Calendar 7B**: Italy

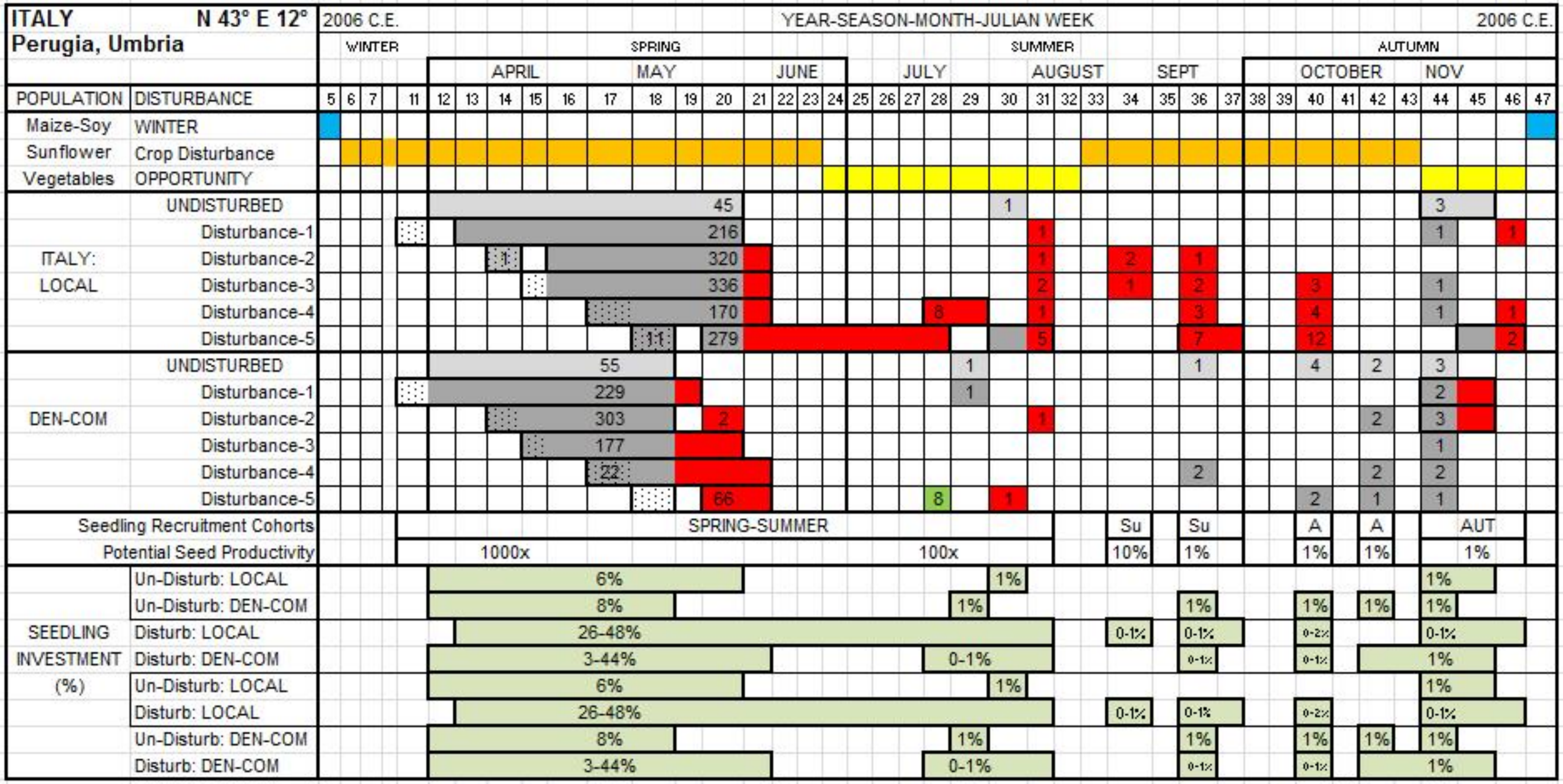

| POPULATION | DISTURBANCE | 5 | 6 | 7 | | 11 | 12 | 13 | 14 | 15 | 16 | 17 | 18 | 19 | 20 | 21 | 22 | 23 | 24 | 25 | 26 | 27 | 28 | 29 | 30 | 31 | 32 | 33 | 34 | 35 | 36 | 37 | 38 | 39 | 40 | 41 | 42 | 43 | 44 | 45 | 46 | 47 |
|---|---|---|---|---|---|---|---|---|---|---|---|---|---|---|---|---|---|---|---|---|---|---|---|---|---|---|---|---|---|---|---|---|---|---|---|---|---|---|---|---|---|---|
| Maize-Soy | WINTER | | | | | | | | | | | | | | | | | | | | | | | | | | | | | | | | | | | | | | | | | |
| Sunflower | Crop Disturbance | | | | | | | | | | | | | | | | | | | | | | | | | | | | | | | | | | | | | | | | | |
| Vegetables | OPPORTUNITY | | | | | | | | | | | | | | | | | | | | | | | | | | | | | | | | | | | | | | | | | |
| | UNDISTURBED | | | | | | | | | | | | | | 45 | | | | | | | | | | 1 | | | | | | | | | | | | | | 3 | | | |
| | Disturbance-1 | | | | | | | | | | | | | | 216 | | | | | | | | | | | 1 | | | | | | | | | | | | | 1 | | 1 | |
| ITALY: | Disturbance-2 | | | | | | | | 1 | | | | | | 320 | | | | | | | | | | | 1 | | | 2 | | 1 | | | | | | | | | | | |
| LOCAL | Disturbance-3 | | | | | | | | | | | | | | 336 | | | | | | | | | | | 2 | | | 1 | | 2 | | | | 3 | | | | 1 | | | |
| | Disturbance-4 | | | | | | | | | | | | | | 170 | | | | | | | | 8 | | | 1 | | | | | 3 | | | | 4 | | | | 1 | | 1 | |
| | Disturbance-5 | | | | | | | | | | | | 11 | | 279 | | | | | | | | | | | 5 | | | | | 7 | | | | 12 | | | | | | 2 | |
| | UNDISTURBED | | | | | | | | | | | 55 | | | | | | | | | | | | 1 | | | | | | | 1 | | | | 4 | | 2 | | 3 | | | |
| | Disturbance-1 | | | | | | | | | | | 229 | | | | | | | | | | | | 1 | | | | | | | | | | | | | | | 2 | | | |
| DEN-COM | Disturbance-2 | | | | | | | | | | | 303 | | | 2 | | | | | | | | | | | 1 | | | | | | | | | | | 2 | | 3 | | | |
| | Disturbance-3 | | | | | | | | | | | 177 | | | | | | | | | | | | | | | | | | | | | | | | | | | 1 | | | |
| | Disturbance-4 | | | | | | | | | | | 22 | | | | | | | | | | | | | | | | | | | 2 | | | | | | 2 | | 2 | | | |
| | Disturbance-5 | | | | | | | | | | | | | | 66 | | | | | | | | 8 | | 1 | | | | | | | | | | 2 | | 1 | | 1 | | | |
| Seedling Recruitment Cohorts | | | | | | | | | | | | | | | | SPRING-SUMMER | | | | | | | | | | | | | Su | | Su | | | | A | | A | | | AUT | | |
| Potential Seed Productivity | | | | | | | | | 1000x | | | | | | | | | | | | | | 100x | | | | | | 10% | | 1% | | | | 1% | | 1% | | | 1% | | |
| | Un-Disturb: LOCAL | | | | | | | | | | | 6% | | | | | | | | | | | | | 1% | | | | | | | | | | | | | | 1% | | | |
| | Un-Disturb: DEN-COM | | | | | | | | | | | 8% | | | | | | | | | | | | 1% | | | | | | | 1% | | | | 1% | | 1% | | 1% | | | |
| SEEDLING | Disturb: LOCAL | | | | | | | | | | | 26-48% | | | | | | | | | | | | | | | | | 0-1% | | 0-1% | | | | 0-2% | | | | 0-1% | | | |
| INVESTMENT | Disturb: DEN-COM | | | | | | | | | | | 3-44% | | | | | | | | | | | | 0-1% | | | | | | | 0-1% | | | | 0-1% | | | | 1% | | | |
| (%) | Un-Disturb: LOCAL | | | | | | | | | | | 6% | | | | | | | | | | | | | 1% | | | | | | | | | | | | | | 1% | | | |
| | Disturb: LOCAL | | | | | | | | | | | 26-48% | | | | | | | | | | | | | | | | | 0-1% | | 0-1% | | | | 0-2% | | | | 0-1% | | | |
| | Un-Disturb: DEN-COM | | | | | | | | | | | 8% | | | | | | | | | | | | 1% | | | | | | | 1% | | | | 1% | | 1% | | 1% | | | |
| | Disturb: DEN-COM | | | | | | | | | | | 3-44% | | | | | | | | | | | | 0-1% | | | | | | | 0-1% | | | | 0-1% | | | | 1% | | | |

**Calendar 8A**: USA; * disturbance day data not presented (combined emerged, unemerged seedlings); no data collected after June 20, 2006 (JW 26).

| | | | | | | | | | | | | | | | | | | | | | | | | | | | |
|---|---|---|---|---|---|---|---|---|---|---|---|---|---|---|---|---|---|---|---|---|---|---|---|---|---|---|---|
| USA Pana, Illinois | N 40° W 89° | 2006 C.E. | | | | | | YEAR-SEASON-MONTH-JULIAN WEEK | | | | | | | | | | | | | | | | | | | 2006 C.E. |
| | | WINTER | | | | | | | SPRING | | | | | | | | | | SUMMER | | | | | AUTUMN | | | |
| | | MARCH | | | | APRIL | | | MAY | | | | JUNE | | | | | | SEPT | | | | | OCTOBER | | | |
| POPULATION | DISTURBANCE | 10 | 11 | 12 | 13 | 14 | 15 | 16 | 17 | 18 | 19 | 20 | 21 | 22 | 23 | 24 | 25 | | 34 | 35 | 36 | 37 | 38 | 39 | 40 | 41 | 42 |
| | WINTER | ■ | | | | | | | | | | | | | | | | | | | | | | | | | ■ |
| Maize | Seedbed Prep | | | | | | | ■ | ■ | ■ | | | | | | | | | | ■ | ■ | ■ | | | | | |
| Forage | Planting | | | | | | | ■ | ■ | ■ | | | | | | | | | | ■ | ■ | ■ | | | | | |
| | Weed Control | | | | | | | | ■ | ■ | | | | | | | | | | | | | | | | | |
| | Harvest | | | | | | | | | | | | | | | | | | | | | | | | ■ | ■ | |
| | OPPORTUNITY | | ■ | ■ | ■ | ■ | ■ | | | | ■ | ■ | ■ | ■ | ■ | ■ | ■ | ■ | ■ | | | | ■ | ■ | | | |
| | UNDISTURBED | | 1 | 2 | 9 | 10 | 16 | 3 | | | | | | | | | | | | | | | | | | | |
| | Disturbance-1 | | * | | 1 | 4 | 13 | 20 | 4 | | | | | | | | | | | | | | | | | | |
| USA: | Disturbance-2 | | 1 | 2 | * | 1 | 14 | 10 | | | | | | | | | | | | | | | | | | | |
| LOCAL | Disturbance-3 | | | 1 | 10 | 7 | * | 9 | 15 | 2 | | | | | | | | | | | | | | | | | |
| | Disturbance-4 | | 2 | 3 | 9 | 6 | * | 5 | 6 | 1 | 2 | | | | | | | | | | | | | | | | |
| | Disturbance-5 | | 1 | 1 | 9 | 14 | 18 | * | | 7 | 9 | 10 | 2 | 3 | | | | | | | | | | | | | |
| | UNDISTURBED | | 2 | 2 | 22 | 77 | 57 | 54 | 2 | | 1 | | | | | | | | | | | | | | | | |
| | Disturbance-1 | | * | | | 6 | 8 | 12 | 4 | 1 | | 1 | | | | | | | | | | | | | | | |
| USA: | Disturbance-2 | | | 3 | * | 6 | 37 | 38 | 4 | | | | | | | | | | | | | | | | | | |
| DEN-COM | Disturbance-3 | | 2 | 1 | 17 | 79 | * | 95 | 5 | 2 | | | | | | | | | | | | | | | | | |
| | Disturbance-4 | | 2 | 4 | 31 | 81 | * | 14 | 13 | 4 | 2 | | | | | | | | | | | | | | | | |
| | Disturbance-5 | | | 2 | 21 | 65 | 56 | * | 1 | 12 | 11 | 3 | 3 | | | | | | | | | | | | | | |

**Calendar 8B**: USA; * disturbance day data not presented (combined emerged, unemerged seedlings); no data collected after June 20, 2006 (JW 26).

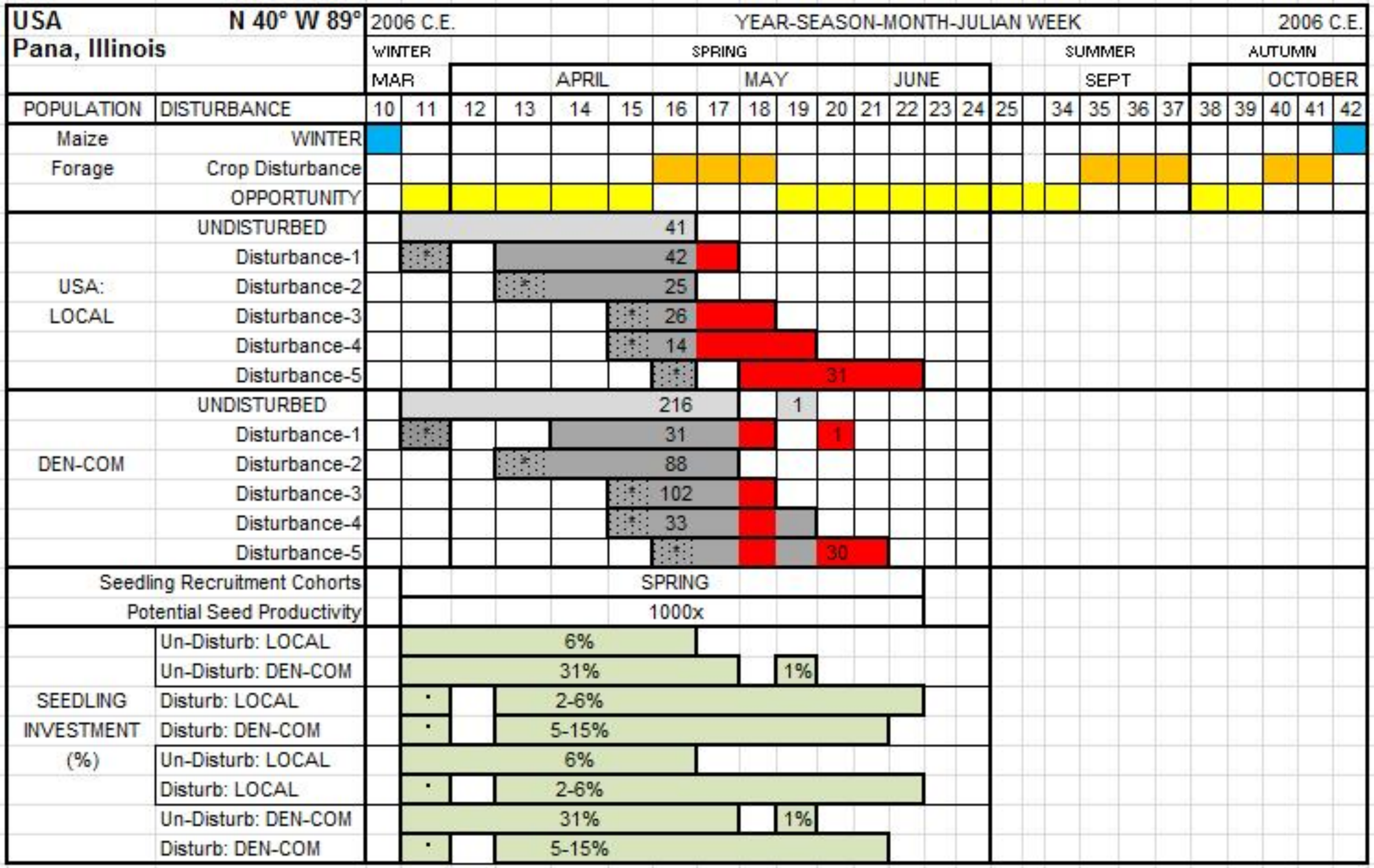

| USA | N 40° W 89° | 2006 C.E. | | | | | | | | | | | YEAR-SEASON-MONTH-JULIAN WEEK | | | | | | | | | | | | | 2006 C.E. | |
|---|---|---|---|---|---|---|---|---|---|---|---|---|---|---|---|---|---|---|---|---|---|---|---|---|---|---|---|
| Pana, Illinois | | WINTER | | | | | | | SPRING | | | | | | | | | | SUMMER | | | | | AUTUMN | | | |
| | | MAR | | | | APRIL | | | | MAY | | | | JUNE | | | | | SEPT | | | | | | OCTOBER | | |
| POPULATION | DISTURBANCE | 10 | 11 | 12 | 13 | 14 | 15 | 16 | 17 | 18 | 19 | 20 | 21 | 22 | 23 | 24 | 25 | | 34 | 35 | 36 | 37 | 38 | 39 | 40 | 41 | 42 |
| Maize | WINTER | | | | | | | | | | | | | | | | | | | | | | | | | | |
| Forage | Crop Disturbance | | | | | | | | | | | | | | | | | | | | | | | | | | |
| | OPPORTUNITY | | | | | | | | | | | | | | | | | | | | | | | | | | |
| | UNDISTURBED | | | | | | | 41 | | | | | | | | | | | | | | | | | | | |
| | Disturbance-1 | | * | | | | | 42 | | | | | | | | | | | | | | | | | | | |
| USA: | Disturbance-2 | | | | * | | | 25 | | | | | | | | | | | | | | | | | | | |
| LOCAL | Disturbance-3 | | | | | | * | 26 | | | | | | | | | | | | | | | | | | | |
| | Disturbance-4 | | | | | | * | 14 | | | | | | | | | | | | | | | | | | | |
| | Disturbance-5 | | | | | | | * | | | | 31 | | | | | | | | | | | | | | | |
| | UNDISTURBED | | | | | | | 216 | | | 1 | | | | | | | | | | | | | | | | |
| | Disturbance-1 | | * | | | | | 31 | | | | 1 | | | | | | | | | | | | | | | |
| DEN-COM | Disturbance-2 | | | | * | | | 88 | | | | | | | | | | | | | | | | | | | |
| | Disturbance-3 | | | | | | * | 102 | | | | | | | | | | | | | | | | | | | |
| | Disturbance-4 | | | | | | * | 33 | | | | | | | | | | | | | | | | | | | |
| | Disturbance-5 | | | | | | | * | | | | 30 | | | | | | | | | | | | | | | |
| Seedling Recruitment Cohorts | | | | | | | | SPRING | | | | | | | | | | | | | | | | | | | |
| Potential Seed Productivity | | | | | | | | 1000x | | | | | | | | | | | | | | | | | | | |
| | Un-Disturb: LOCAL | | | | | 6% | | | | | | | | | | | | | | | | | | | | | |
| | Un-Disturb: DEN-COM | | | | | 31% | | | | | 1% | | | | | | | | | | | | | | | | |
| SEEDLING | Disturb: LOCAL | | * | | | 2-6% | | | | | | | | | | | | | | | | | | | | | |
| INVESTMENT | Disturb: DEN-COM | | * | | | 5-15% | | | | | | | | | | | | | | | | | | | | | |
| (%) | Un-Disturb: LOCAL | | | | | 6% | | | | | | | | | | | | | | | | | | | | | |
| | Disturb: LOCAL | | * | | | 2-6% | | | | | | | | | | | | | | | | | | | | | |
| | Un-Disturb: DEN-COM | | | | | 31% | | | | | 1% | | | | | | | | | | | | | | | | |
| | Disturb: DEN-COM | | * | | | 5-15% | | | | | | | | | | | | | | | | | | | | | |

**Calendar 9A**: UK

| UK | N 52° W 2° | 2005 C.E. | | | | | 2006 C.E. | | | | | | | | | YEAR-SEASON-MONTH-JULIAN WEEK | | | | | | | | | | | | | | | | | | | 2006 C.E. | |
|---|---|---|---|---|---|---|---|---|---|---|---|---|---|---|---|---|---|---|---|---|---|---|---|---|---|---|---|---|---|---|---|---|---|---|---|---|
| | | | | AUTUMN | | | | WINTER | | | | | | | | SPRING | | | | | | | | | | | | | SUMMER | | | | | | | |
| | | NOV | | | DEC | | MARCH | | | | | | APRIL | | | MAY | | | | JUNE | | | | | JULY | | | | | AUGUST | | | SEPT | | | |
| POPULATION | DISTURBANCE | 45 | 46 | 47 | 48 | 49 | 9 | 10 | 11 | 12 | 13 | 14 | 15 | 16 | 17 | 18 | 19 | 20 | 21 | 22 | 23 | 24 | 25 | 26 | 27 | 28 | 29 | 30 | 31 | 32 | 33 | 34 | 35 | 36 | 37 | 38 |
| | WINTER | | | | | | ■ | | | | | | | | | | | | | | | | | | | | | | | | | | | | | |
| | Seedbed Prep | | | | | | | ■ | ■ | ■ | ■ | ■ | ■ | ■ | ■ | ■ | ■ | | | | | | | | | | | | | | | | | | | |
| Vegetables | Planting | | | | | | | ■ | ■ | ■ | ■ | ■ | ■ | ■ | ■ | ■ | ■ | | | | | | | | | | | | | | | | | | | |
| | Weed Control | | | | | | | | | ■ | ■ | ■ | ■ | ■ | ■ | ■ | ■ | ■ | | | | ■ | ■ | ■ | ■ | ■ | ■ | ■ | ■ | ■ | | | | | | |
| | Harvest | | | | | | | | | | | | | | | | | | | | | | | | | | | | | | ■ | ■ | ■ | ■ | ■ | |
| | OPPORTUNITY | ■ | ■ | ■ | ■ | ■ | | | | | | | | | | | | | ■ | ■ | ■ | | | | | | | | | | | | | | | ■ |
| | UNDISTURBED | | | | | | | 1 | | 1 | 2 | 2 | 2 | 3 | | | | | 12 | 8 | | 1 | 1 | | | 2 | | | | | | 2 | | | | |
| | Disturbance-1 | 1 | | | | | | 1 | | 1 | 3 | | 6 | 38 | 15 | 1 | 2 | 6 | 12 | 2 | | | | | | 2 | | 3 | 2 | 1 | | 3 | | | | |
| UK: | Disturbance-2 | | | | | | | | | | 3 | 3 | | 24 | 46 | 1 | | 5 | 28 | 1 | | 4 | 1 | | | 2 | | 2 | | 2 | | 10 | | | | |
| LOCAL | Disturbance-3 | | | | | | | | | | | 3 | 2 | | 48 | 22 | 5 | 14 | 15 | 11 | | 2 | 1 | | | 1 | | 4 | | 2 | | 9 | | | | |
| | Disturbance-4 | | | | | | | | | | 2 | 3 | | 1 | 6 | 96 | 53 | 30 | 15 | 20 | | 2 | | | | | | 2 | | 1 | | 9 | | 1 | | |
| | Disturbance-5 | | | | | | | | | | 1 | 1 | 2 | 2 | 1 | | 59 | 30 | 24 | 11 | | 3 | 1 | | | 1 | | 1 | | 1 | | 6 | | 1 | | |
| | UNDISTURBED | 12 | | 3 | 2 | 1 | | | | | 2 | 17 | 6 | 8 | 4 | 2 | 2 | 3 | 2 | 2 | | | 1 | | | 1 | | | | | | 1 | | | | |
| | Disturbance-1 | 9 | | 1 | 2 | 1 | | 1 | | | 6 | 2 | 20 | 69 | 30 | 3 | 3 | 3 | 2 | 2 | | | | | | 1 | | | | | | | | | | |
| UK: | Disturbance-2 | 10 | | 3 | 2 | | | | | | 1 | 9 | | 49 | 54 | 2 | 2 | 1 | 3 | 5 | | 1 | | | | | | | | | | | | | | |
| DEN-COM | Disturbance-3 | 16 | | 4 | | 1 | | | | | 1 | 14 | 7 | 2 | 125 | 21 | 1 | 5 | 5 | 4 | | | | | | 1 | | | | | | | | | | |
| | Disturbance-4 | 12 | | 5 | | 2 | | | | | | 18 | 5 | 11 | 18 | 226 | 51 | 16 | 6 | 6 | | | | | | | | | | | | | | | | |
| | Disturbance-5 | 14 | | 3 | | 1 | | | | | | 17 | 6 | 8 | 4 | 21 | 143 | 24 | 16 | 7 | | 2 | | | | | | | | | | | | | | |

**Calendar 9B**: UK

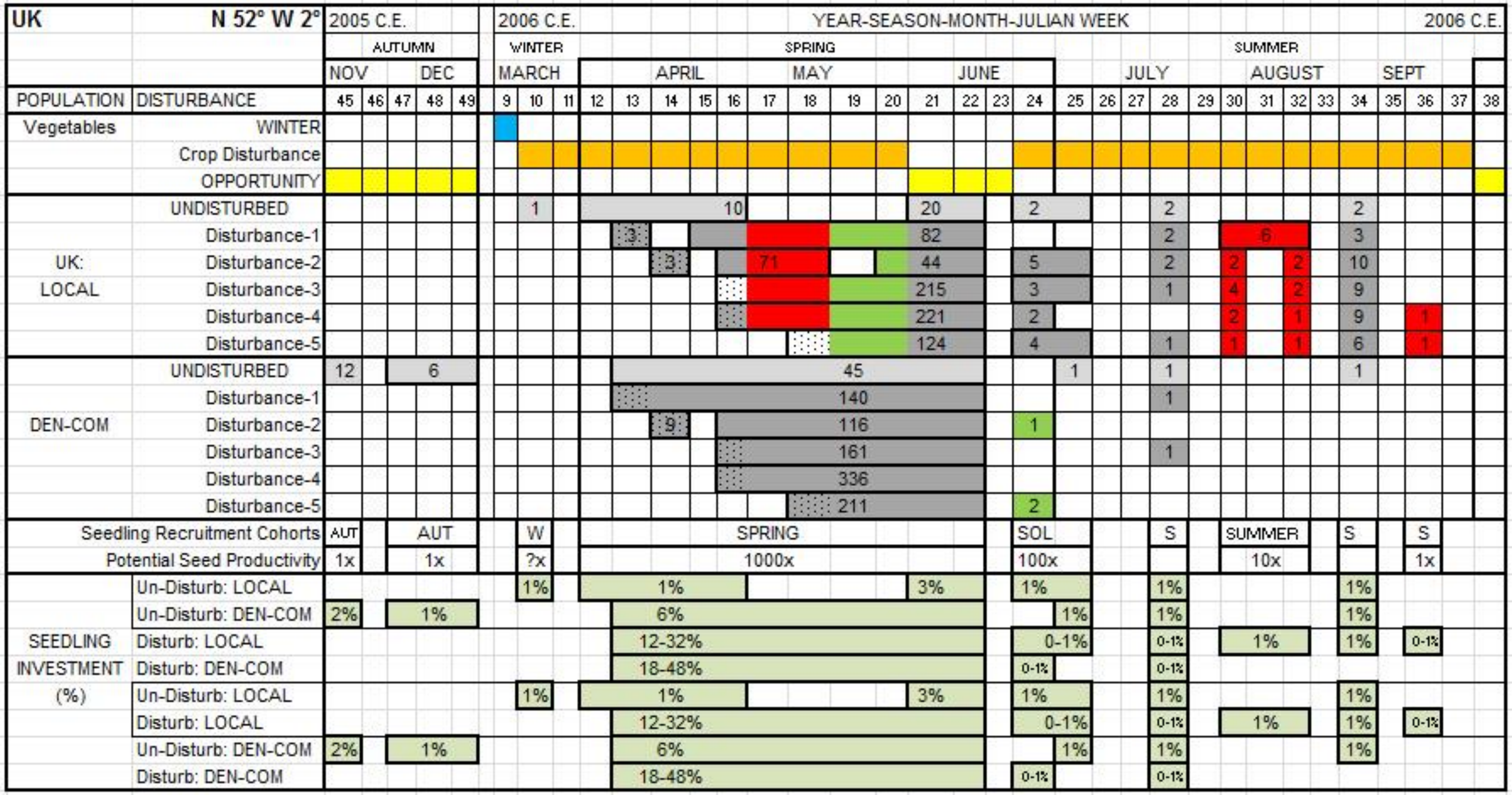

| UK | N 52° W 2° | 2005 C.E. | | | | | 2006 C.E. | | | | | | | | | | | | | YEAR-SEASON-MONTH-JULIAN WEEK | | | | | | | | | | | | | | | 2006 C.E. | |
|---|---|---|---|---|---|---|---|---|---|---|---|---|---|---|---|---|---|---|---|---|---|---|---|---|---|---|---|---|---|---|---|---|---|---|---|---|
| | | | | AUTUMN | | | | WINTER | | | | | | | | SPRING | | | | | | | | | | | | | SUMMER | | | | | | | |
| | | NOV | | | DEC | | | MARCH | | | | APRIL | | | | MAY | | | | JUNE | | | | | JULY | | | | | AUGUST | | | SEPT | | | |
| POPULATION | DISTURBANCE | 45 | 46 | 47 | 48 | 49 | 9 | 10 | 11 | 12 | 13 | 14 | 15 | 16 | 17 | 18 | 19 | 20 | 21 | 22 | 23 | 24 | 25 | 26 | 27 | 28 | 29 | 30 | 31 | 32 | 33 | 34 | 35 | 36 | 37 | 38 |
| Vegetables | WINTER | | | | | | | | | | | | | | | | | | | | | | | | | | | | | | | | | | | |
| | Crop Disturbance | | | | | | | | | | | | | | | | | | | | | | | | | | | | | | | | | | | |
| | OPPORTUNITY | | | | | | | | | | | | | | | | | | | | | | | | | | | | | | | | | | | |
| | UNDISTURBED | | | | | | | 1 | | | | | | 10 | | | | | 20 | | | 2 | | | | 2 | | | | | | 2 | | | | |
| | Disturbance-1 | | | | | | | | | | 3 | | | | | | | | 82 | | | | | | | 2 | | | 6 | | | 3 | | | | |
| UK: | Disturbance-2 | | | | | | | | | | | 3 | | | 71 | | | | 44 | | | 5 | | | | 2 | | 2 | | 2 | | 10 | | | | |
| LOCAL | Disturbance-3 | | | | | | | | | | | | | | | | | | 215 | | | 3 | | | | 1 | | 4 | | 2 | | 9 | | | | |
| | Disturbance-4 | | | | | | | | | | | | | | | | | | 221 | | | 2 | | | | | | 2 | | 1 | | 9 | | 1 | | |
| | Disturbance-5 | | | | | | | | | | | | | | | | | | 124 | | | 4 | | | | 1 | | 1 | | 1 | | 6 | | 1 | | |
| | UNDISTURBED | 12 | | | 6 | | | | | | | | | | | | 45 | | | | | | 1 | | | 1 | | | | | | 1 | | | | |
| | Disturbance-1 | | | | | | | | | | | | | | | | 140 | | | | | | | | | 1 | | | | | | | | | | |
| DEN-COM | Disturbance-2 | | | | | | | | | | | 9 | | | | | 116 | | | | | 1 | | | | | | | | | | | | | | |
| | Disturbance-3 | | | | | | | | | | | | | | | | 161 | | | | | | | | | 1 | | | | | | | | | | |
| | Disturbance-4 | | | | | | | | | | | | | | | | 336 | | | | | | | | | | | | | | | | | | | |
| | Disturbance-5 | | | | | | | | | | | | | | | | 211 | | | | | 2 | | | | | | | | | | | | | | |
| | Seedling Recruitment Cohorts | AUT | | | AUT | | | W | | | | | | | SPRING | | | | | | | SOL | | | | S | | | SUMMER | | | S | | S | | |
| | Potential Seed Productivity | 1x | | | 1x | | | ?x | | | | | | | 1000x | | | | | | | 100x | | | | | | | 10x | | | | | 1x | | |
| | Un-Disturb: LOCAL | | | | | | | 1% | | | | 1% | | | | | | | 3% | | | 1% | | | | 1% | | | | | | 1% | | | | |
| | Un-Disturb: DEN-COM | 2% | | | 1% | | | | | | | 6% | | | | | | | | | | | 1% | | | 1% | | | | | | 1% | | | | |
| SEEDLING | Disturb: LOCAL | | | | | | | | | | | 12-32% | | | | | | | | | | | 0-1% | | | 0-1% | | | 1% | | | 1% | | 0-1% | | |
| INVESTMENT | Disturb: DEN-COM | | | | | | | | | | | 18-48% | | | | | | | | | | 0-1% | | | | 0-1% | | | | | | | | | | |
| (%) | Un-Disturb: LOCAL | | | | | | | 1% | | | | 1% | | | | | | | 3% | | | 1% | | | | 1% | | | | | | 1% | | | | |
| | Disturb: LOCAL | | | | | | | | | | | 12-32% | | | | | | | | | | | 0-1% | | | 0-1% | | | 1% | | | 1% | | 0-1% | | |
| | Un-Disturb: DEN-COM | 2% | | | 1% | | | | | | | 6% | | | | | | | | | | | 1% | | | 1% | | | | | | 1% | | | | |
| | Disturb: DEN-COM | | | | | | | | | | | 18-48% | | | | | | | | | | 0-1% | | | | 0-1% | | | | | | | | | | |

**Calendar 10A**: Portugal

| PORTUGAL | | 2005 C.E. | | | | | 2006 C.E. | | | | | | | | | | | | | |
|---|---|---|---|---|---|---|---|---|---|---|---|---|---|---|---|---|---|---|---|---|
| | N 39° 9°W | AUTUMN | | | | | WINTER | | | | | | | | | SPRING | | | | |
| | | NOV | | | DEC | | JAN | | FEB | | | | MARCH | | | APRIL | | | | |
| POPULATION | DISTURBANCE | 46 | 47 | 48 | 49 | 50 | 2 | 3 | 5 | 6 | 7 | 8 | 9 | 10 | 11 | 12 | 13 | 14 | 15 | 16 |
| Irrigated: | Seedbed Prep | | | | | | | | | | | | | | | | | | | |
| Potato | Planting | | | | | | | | | | | | | | | | | | | |
| Maize | Weed Control | | | | | | | | | | | | | | | | | | | |
| | Harvest | | | | | | | | | | | | | | | | | | | |
| | OPPORTUNITY | | | | | | | | | | | | | | | | | | | |
| | UNDISTURBED | 4 | 3 | 1 | 1 | | 1 | 1 | | | | | 2 | 1 | | | 2 | 2 | | |
| | Disturbance-1 | 1 | 1 | | | | 1 | | | | | | 1 | | | | 2 | 2 | | |
| PORTUGAL: | Disturbance-2 | 3 | 4 | 2 | 2 | | 1 | | | | | | | | | | | 2 | | |
| LOCAL | Disturbance-3 | 2 | 4 | | | | | 1 | | | | | 2 | 1 | | | 2 | | | |
| | Disturbance-4 | 2 | 2 | 2 | | | | | | | | | | | | | | 2 | | |
| | Disturbance-5 | 7 | 3 | 1 | 1 | | | | | | | | 2 | 1 | | | | | | |
| | UNDISTURBED | 9 | 8 | 4 | 2 | 2 | | 1 | 1 | 1 | | | 1 | | | | | 2 | | |
| | Disturbance-1 | 9 | 14 | 6 | | 1 | | 1 | | | | | 1 | | | | | | | |
| PORTUGAL: | Disturbance-2 | 5 | 3 | 3 | | | | | 1 | 1 | | | | | | | | | | |
| DEN-COM | Disturbance-3 | 11 | 12 | 3 | 3 | 2 | | | | | | | 1 | | | | | | | |
| | Disturbance-4 | 6 | 2 | 2 | | | | | | | | | 1 | | | | | 2 | | |
| | Disturbance-5 | 8 | 8 | 3 | 1 | | | 1 | | | | | 2 | | | | | | | |

| PORTUGAL | | YEAR-SEASON-MONTH-JULIAN WEEK | | | | | | | | | | | | | | | | | | | 2006 C.E. | | |
|---|---|---|---|---|---|---|---|---|---|---|---|---|---|---|---|---|---|---|---|---|---|---|---|
| | N 39° 9°W | SPRING | | | | | | | | | | | | | | SUMMER | | | | | AUTUMN | | |
| | | MAY | | | | JUNE | | | | | JULY | | | AUGUST | | | | SEPT | | | | OCTOBER | |
| POPULATION | DISTURBANCE | 17 | 18 | 19 | 20 | 21 | 22 | 23 | 24 | 25 | 26 | 27 | 31 | 32 | 33 | 34 | 35 | 36 | 37 | 38 | 39 | 40 | 41 | 44 |
| Irrigated: | Seedbed Prep | | | | | | | | | | | | | | | | | | | | | | | |
| Potato | Planting | | | | | | | | | | | | | | | | | | | | | | | |
| Maize | Weed Control | | | | | | | | | | | | | | | | | | | | | | | |
| | Harvest | | | | | | | | | | | | | | | | | | | | | | | |
| | OPPORTUNITY | | | | | | | | | | | | | | | | | | | | | | | |
| | UNDISTURBED | 5 | 2 | 2 | 3 | 5 | 3 | 4 | 1 | 2 | | | | 2 | | | | | 9 | 3 | 1 | 3 | | |
| | Disturbance-1 | 124 | 25 | 4 | 2 | 3 | 2 | 5 | 6 | | | | | 1 | | | | | 7 | 2 | | | | |
| PORTUGAL: | Disturbance-2 | 3 | 40 | 65 | 6 | 3 | 2 | 5 | | | | | | | | | | | 3 | | 2 | | | |
| LOCAL | Disturbance-3 | 6 | 9 | 147 | 31 | 8 | 4 | 4 | | | | | | 1 | | | | | 4 | 1 | 2 | | | |
| | Disturbance-4 | 6 | 2 | 2 | 103 | 27 | 2 | 1 | | | | | | | | | | | 6 | 1 | | | | |
| | Disturbance-5 | 4 | | 1 | | 111 | 35 | 1 | | 2 | 2 | | | 1 | | | | | 5 | | | | | |
| | UNDISTURBED | 3 | 1 | 2 | | 1 | 1 | 2 | 1 | 3 | | | | | | | | | 3 | | 2 | 1 | | |
| | Disturbance-1 | 43 | 15 | 1 | | | | 3 | | 2 | | | | | | | | | 1 | | 1 | 1 | | |
| PORTUGAL: | Disturbance-2 | | 18 | 37 | 2 | 3 | | 3 | | | | | | | | | | | 3 | | | 1 | | |
| DEN-COM | Disturbance-3 | 2 | 1 | 130 | 11 | 1 | 1 | 2 | | | | | | | | | | | 2 | | 2 | | | |
| | Disturbance-4 | | | 2 | 39 | 3 | | 1 | 1 | 2 | | | | | | | | | 2 | | 1 | | | |
| | Disturbance-5 | 1 | 1 | | | 17 | 6 | 4 | | 1 | | | | | | | | | 1 | 1 | | | | |

**Calendar 10B**: Portugal

| PORTUGAL | N 39° 9°W | 2005 C.E. | | | | | 2006 C.E. | | | | | | | | | | | | | | | | YEAR-SEASON-MONTH-JULIAN WEEK | | | | | | | | | | | | | | | | | | | | 2006 C.E. |
|---|---|---|---|---|---|---|---|---|---|---|---|---|---|---|---|---|---|---|---|---|---|---|---|---|---|---|---|---|---|---|---|---|---|---|---|---|---|---|---|---|---|---|---|
| | | | | AUTUMN | | | | | | WINTER | | | | | | | | | | | | | SPRING | | | | | | | | | | SUMMER | | | | | | | | AUTUMN | | |
| | | NOV | | | DEC | | JAN | | | FEB | | | MARCH | | | | | APRIL | | | | MAY | | | | JUNE | | | | | JUL | | AUGUST | | | SEPT | | | | | OCT | | |
| POPULATION | DISTURBANCE | 46 | 47 | 48 | 49 | 50 | 2 | 3 | 5 | 6 | 7 | 8 | 9 | 10 | 11 | 12 | 13 | 14 | 15 | 16 | 17 | 18 | 19 | 20 | 21 | 22 | 23 | 24 | 25 | 26 | 27 | 31 | 32 | 33 | 34 | 35 | 36 | 37 | 38 | 39 | 40 | 41 | 44 |
| Irrigated Potato | Crop Disturbance | | | | | | | | | | | | | | | | | | | | | | | | | | | | | | | | | | | | | | | | | | |
| Maize | OPPORTUNITY | | | | | | | | | | | | | | | | | | | | | | | | | | | | | | | | | | | | | | | | | | |
| | UNDISTURBED | | | 9 | | | | 2 | | | | | 3 | | | | | 4 | | | | | | | 27 | | | | | | | | 2 | | | | | 16 | | | | | |
| | Disturbance-1 | | | | | | | | | | | | | | | | | | | | | | | | 171 | | | | | | | | 1 | | | | | 9 | | | | | |
| PORTUGAL: | Disturbance-2 | | | | | | | | | | | | | | | | | | | | | | | | 121 | | | | | | | | | | | | | 3 | | 2 | | | |
| LOCAL | Disturbance-3 | | | | | | | | | | | | | | | | | | | | | | | | 213 | | | | | | | | 1 | | | | | 7 | | | | | |
| | Disturbance-4 | | | | | | | | | | | | | | | | | | | | | | | | 135 | | | | | | | | | | | | | 7 | | | | | |
| | Disturbance-5 | | | | | | | | | | | | | | | | | | | | | | | | 147 | | | | 2 | | | | 1 | | | | | 5 | | | | | |
| | UNDISTURBED | | | 25 | | | | 1 | | 2 | | | 1 | | | | | 2 | | | | | 6 | | | | 8 | | | | | | | | | | | 3 | | 2 | | | |
| | Disturbance-1 | | | | | | | | | | | | | | | | | | | | | | 59 | | | | 3 | | 2 | | | | | | | | | 1 | | 2 | | | |
| DEN-COM | Disturbance-2 | | | | | | | | | | | | | | | | | | | | | | 60 | | | | 3 | | | | | | | | | | | 3 | | | 1 | | |
| | Disturbance-3 | | | | | | | | | | | | | | | | | | | | | | 146 | | | | | | | | | | | | | | | 2 | | 2 | | | |
| | Disturbance-4 | | | | | | | | | | | | | | | | | | | | | | 44 | | | | 4 | | | | | | | | | | | 2 | | 1 | | | |
| | Disturbance-5 | | | | | | | | | | | | | | | | | | | | | | | | | | 27 | | 1 | | | | | | | | | 2 | | | | | |
| | Seedling Recruitment Cohorts | LATE AUT | | | | | EW | | W | | | | LW | | | ES | | | | | | | | | SPRING | | | | | | | | S | | | | | | AUT | | | | |
| | Potential Seed Productivity | | 10x | | | | 10x | | 100x | | | | 1000x | | | 1000x | | | | | | | | | 1000x | | | | | | | | [illegible] | | | | | | 10x | | | | |
| | Un-Disturb: LOCAL | | | 1% | | | | 1% | | | | | 1% | | | | | 1% | | | | | | | 4% | | | | | | | | 1% | | | | | 2% | | | | | |
| | Un-Disturb: DEN-COM | | | 4% | | | | 1% | | 2% | | | 1% | | | | | 2% | | | | | | | 2% | | | | | | | | | | | | | 1% | | 1% | | | |
| SEEDLING | Disturb: LOCAL | | | | | | | | | | | | | | | | | | | | | | | | 17-30% | | | | | | | | 0-1% | | | | | | 1% | | | | |
| INVESTMENT | Disturb: DEN-COM | | | | | | | | | | | | | | | | | | | | | | | | 4-21% | | | | | | | | | | | | | | 1% | | | | |
| (%) | Un-Disturb: LOCAL | | | 1% | | | | 1% | | | | | 1% | | | | | 1% | | | | | | | 4% | | | | | | | | 1% | | | | | 2% | | | | | |
| | Disturb: LOCAL | | | | | | | | | | | | | | | | | | | | | | | | 17-30% | | | | | | | | 0-1% | | | | | | 1% | | | | |
| | Un-Disturb: DEN-COM | | | 4% | | | | 1% | | 2% | | | 1% | | | | | 2% | | | | | | | 2% | | | | | | | | | | | | | 1% | | 1% | | | |
| | Disturb: DEN-COM | | | | | | | | | | | | | | | | | | | | | | | | 4-21% | | | | | | | | | | | | | | 1% | | | | |

**Calendar 11A**: Spain

| SPAIN | N 38° W5° | 2005 C.E. | | | | | | 2006 C.E. | | | | | | | | YEAR-SEASON-MONTH-JULIAN WEEK | | | | | | | | | | | | | | | | | | | | 2006 C.E. |
|---|---|---|---|---|---|---|---|---|---|---|---|---|---|---|---|---|---|---|---|---|---|---|---|---|---|---|---|---|---|---|---|---|---|---|---|---|
| | | AUTUMN | | | | | | WINTER | | | | | | | | | | | | SPRING | | | | | | | SUMMER | | | | | | AUTUMN | | | |
| | | NOV | | | | | DEC | JAN | | | | | | MARCH | | | | APRIL | | | | | MAY | | | | AUG | SEPT | | | | OCTOBER | | NOV | | |
| POPULATION | DISTURBANCE | 47 | 48 | 49 | 50 | 51 | 52 | 2 | 3 | 4 | 5 | 6 | 9 | 10 | 11 | 12 | 13 | 14 | 15 | 16 | 17 | 18 | 19 | 20 | 21 | 30 | 31 | 35 | 36 | 39 | 40 | 42 | 43 | 44 | 45 | 46 |
| Wheat | Seedbed Prep | | | | | | | | | | | | | | | | | | | | | | | | | | | | | | | | | | | |
| Irrigated: | Planting | | | | | | | | | | | | | | | | | | | | | | | | | | | | | | | | | | | |
| Maize | Weed Control | | | | | | | | | | | | | | | | | | | | | | | | | | | | | | | | | | | |
| Cotton | Harvest | | | | | | | | | | | | | | | | | | | | | | | | | | | | | | | | | | | |
| | OPPORTUNITY | | | | | | | | | | | | | | | | | | | | | | | | | | | | | | | | | | | |
| | UNDISTURBED | | | | | | | | | | | | | | | | 1 | | | | | | | | | | | | | | | | | | | |
| | Disturbance-1 | | | | | | | | | | | | | | 37 | | 15 | 2 | | 1 | | | | | | | | | | | | | | | | |
| SPAIN: | Disturbance-2 | | | | | | | | | | | | | 1 | | | | | | | | | | | | | | | | | | | | | | |
| LOCAL | Disturbance-3 | | | | | | | | | | | | | | | | | | | | | | | | | | | | | | | | | | | |
| | Disturbance-4 | | | | | | | | | | | | | | | | | | | | | | | | | | | | | | | | | | | |
| | Disturbance-5 | | | | | | | | | | | | | | | | | | | | | | | | | | | | | | | | | | | |
| | UNDISTURBED | | 15 | 3 | | | | 1 | | | | | | | | | 1 | | | | | | | | | | | | | | | | | | | |
| | Disturbance-1 | | 10 | 24 | 21 | 7 | | 7 | | | | | | | | | 15 | | | | | | | | | | | | | | | | | | | |
| SPAIN: | Disturbance-2 | | 17 | | | | | 1 | | | | | | | | | | | | | | | | | | | | | | | | | | | | |
| DEN-COM | Disturbance-3 | | 17 | | | | | | | | | | | | | | | | | | | | | | | | | | | | | | | | | |
| | Disturbance-4 | | 9 | | | | | | | | | | | | | | | | | | | | | | | | | | | | | | | | | |
| | Disturbance-5 | | 11 | | | | | 1 | | | | | | | | | 1 | | | | | | | | | | | | | | | | | | | |

**Calendar 11B**: Spain

| SPAIN | N 38° W5° | 2005 C.E. | | | | | | 2006 C.E. | | | | | | | YEAR-SEASON-MONTH-JULIAN WEEK | | | | | | | | | | | | | | | | | | | 2006 C.E. |
|---|---|---|---|---|---|---|---|---|---|---|---|---|---|---|---|---|---|---|---|---|---|---|---|---|---|---|---|---|---|---|---|---|---|---|
| | | AUTUMN | | | | | | WINTER | | | | | | | | | | SPRING | | | | | | | SUMMER | | | | | | AUTUMN | | | |
| | | NOV | | | | DEC | | JAN | | | | | MARCH | | | | APRIL | | | | MAY | | | | AUG | SEPT | | | OCTOBER | | | NOV | | |
| POPULATION | DISTURBANCE | 47 | 48 | 49 | 50 | 51 | 52 | 2 | 3 | 4 | 5 | 6 | 9 | 10 | 11 | 12 | 13 | 14 | 15 | 16 | 17 | 18 | 19 | 20 | 21 | 30 | 31 | 35 | 36 | 39 | 40 | 42 | 43 | 44 | 45 | 46 |
| Irrigated Maize, | Crop Disturbance | | | | | | | | | | | | | | | | | | | | | | | | | | | | | | | | | | | |
| Cotton; Wheat | OPPORTUNITY | | | | | | | | | | | | | | | | | | | | | | | | | | | | | | | | | | | |
| SPAIN: LOCAL | UNDISTURBED | | | | | | | | | | | | | | | | 1 | | | | | | | | | | | | | | | | | | | |
| | Disturbance-1 | | | | | | | | | | | | | | 37 | | 17 | | | 1 | | | | | | | | | | | | | | | | |
| | Disturbance-2 | | | | | | | | | | | | | 1 | | | | | | | | | | | | | | | | | | | | | | |
| | Disturbance-3 | | | | | | | | | | | | | | | | | | | | | | | | | | | | | | | | | | | |
| | Disturbance-4 | | | | | | | | | | | | | | | | | | | | | | | | | | | | | | | | | | | |
| | Disturbance-5 | | | | | | | | | | | | | | | | | | | | | | | | | | | | | | | | | | | |
| DEN-COM | UNDISTURBED | | 18 | | | | | 1 | | | | | | | | | 1 | | | | | | | | | | | | | | | | | | | |
| | Disturbance-1 | | 62 | | | | | 7 | | | | | | | | | 15 | | | | | | | | | | | | | | | | | | | |
| | Disturbance-2 | | | | | | | | | | | | | | | | | | | | | | | | | | | | | | | | | | | |
| | Disturbance-3 | | | | | | | | | | | | | | | | | | | | | | | | | | | | | | | | | | | |
| | Disturbance-4 | | | | | | | | | | | | | | | | | | | | | | | | | | | | | | | | | | | |
| | Disturbance-5 | | | | | | | | | | | | | | | | | | | | | | | | | | | | | | | | | | | |
| Seedling Recruitment Cohorts | | | LATE AUT | | | | | W | | | | | | WINT | | | SPR | | | S | | | | | | | | | | | | | | | | |
| Potential Seed Productivity | | | 10x | | | | | 1x | | | | | | 1000x | | | 1000x | | | [illegible] | | | | | | | | | | | | | | | | |
| SEEDLING INVESTMENT (%) | Un-Disturb: LOCAL | | | | | | | | | | | | | | | | | 1% | | | | | | | | | | | | | | | | | | |
| | Un-Disturb: DEN-COM | | 2% | | | | | 1% | | | | | | | | | 1% | | | | | | | | | | | | | | | | | | | |
| | Disturb: LOCAL | | | | | | | | | | | | | 0-5% | | | 0-2% | | | 0-1% | | | | | | | | | | | | | | | | |
| | Disturb: DEN-COM | | 0-9% | | | | | 0-1% | | | | | | | | | 0-2% | | | | | | | | | | | | | | | | | | | |
| | Un-Disturb: LOCAL | | | | | | | | | | | | | | | | | 1% | | | | | | | | | | | | | | | | | | |
| | Disturb: LOCAL | | | | | | | | | | | | | 0-5% | | | 0-2% | | | 0-1% | | | | | | | | | | | | | | | | |
| | Un-Disturb: DEN-COM | | 2% | | | | | 1% | | | | | | | | | 1% | | | | | | | | | | | | | | | | | | | |
| | Disturb: DEN-COM | | 0-9% | | | | | 0-1% | | | | | | | | | 0-2% | | | | | | | | | | | | | | | | | | | |

**Calendar 12**. Calendar of local *Chenopodium album* population seedling emergence magnitude (mean numbers per JW of 700 possible buried seeds) with time (2005, 2006) in undisturbed soil at 11 nursery locations. Comparisons of local population emergence times with those of DEN-COM at Denmark: earlier (green), later (red).

| LOCAL | | 2005 C.E. | | | | 2006 C.E. | | | | | | | | | | YEAR-SEASON-MONTH-JULIAN WEEK | | | | | | | | | | | | | | | | | | | | | 2006 C.E. | |
|---|---|---|---|---|---|---|---|---|---|---|---|---|---|---|---|---|---|---|---|---|---|---|---|---|---|---|---|---|---|---|---|---|---|---|---|---|---|---|
| Population: | | | AUTUMN | | | | | WINTER | | | | | | | | | SPRING | | | | | | | | | | | | | SUMMER | | | | | | AUTUMN | | |
| Un-Disturbed | | NOV | | DEC | | JAN | | MARCH | | | | | APRIL | | | | MAY | | | | JUNE | | | | | JULY | | | | AUGUST | | | | SEPT | | OCT | NOV | |
| LAT | LOCAL POP | 46 | 47 | 48 | 49 | 2 | 3 | 9 | 10 | 11 | 12 | 13 | 14 | 15 | 16 | 17 | 18 | 19 | 20 | 21 | 22 | 23 | 24 | 25 | 26 | 27 | 28 | 30 | 31 | 32 | 33 | 34 | 37 | 38 | 39 | 40 | 44 | 45 |
| N 61° | Finland | ● | | ● | | | | | | | | | | | ● | ● | ● | ● | | | ● | 27 | 2 | | | 🟥 1 | | | | | | | | | | | | |
| N 60° | Norway | ● | | ● | | | | | | | | | | | ● | 2 | 7 | 8 | | 🟥 1 | 8 | 7 | ● | | 🟥 6 | | | | | | | | | | | | | |
| N 58° | Sweden | ● | | ● | | | | | | | | | | | 11 | 36 | 79 | 102 | 🟥 5 | 🟥 7 | ● | ● | ● | | | | | | | 🟥 12 | 🟥 8 | | | | | | | |
| N 55° | DEN-LOCAL | ● | | 2 | | | | | | | | | | | 1 | 1 | 1 | 1 | | | 2 | ● | ● | | | | | | | 🟥 1 | | | | | | | | |
| | DEN-COM | 1 | | 1 | | | | | | | | | | | 1 | 4 | 3 | 4 | | | 9 | 4 | 1 | | | | | | | | | | | | | | | |
| N 49° | Czech | ● | | ● | | | | | | | | | | 🟩 1 | 11 | 7 | 7 | 2 | | 🟥 1 | 1 | ● | ● | | | | 🟥 3 | | | 🟥 2 | | | | | | | | |
| N 45° | Canada | ● | | ● | | | | | | | | | 🟩 1 | 🟩 1 | 5 | 10 | 2 | 3 | 🟥 3 | 🟥 2 | 2 | ● | ● | | | | | 🟥 2 | 🟥 1 | | | | | | | | | |
| N 43° | Italy | ● | | ● | | | | | | | 🟩 1 | 🟩 1 | 🟩 1 | 🟩 3 | 2 | 8 | 17 | 8 | 🟥 4 | | ● | ● | ● | | | | | 🟥 2 | | | | | | | | | 🟥 1 | 🟥 2 |
| N 40° | USA | ● | | ● | | | | | | 🟥 1 | 🟩 2 | 🟩 9 | 🟩 10 | 🟩 16 | 3 | ● | ● | ● | | | ● | ● | ● | | | | | | | | | | | | | | | |
| N 52° | UK | ● | | ● | | | | | 🟥 1 | | 🟩 1 | 🟩 2 | 🟩 2 | 🟩 2 | 3 | ● | ● | ● | | 🟥 12 | 8 | ● | 1 | 🟥 1 | | | 🟥 2 | | | | | 🟥 2 | | | | | | |
| N 39° | Portugal | 4 | 🟥 3 | 1 | 🟥 1 | 🟥 1 | 🟥 1 | 🟥 2 | 🟥 1 | | | 🟩 2 | 🟩 2 | | ● | 5 | 2 | 2 | 🟥 3 | 🟥 5 | 3 | 4 | 1 | 🟥 2 | | | | | | 🟥 2 | | | 🟥 9 | 🟥 3 | 🟥 1 | 🟥 3 | | |
| N 37° | Spain | ● | | ● | | | | | | | | 🟩 1 | | | ● | ● | ● | ● | | | ● | ● | ● | | | | | | | | | | | | | | | |

| Key: | | |
|---|---|---|
| | 🟩 | Local population EARLIER than DEN-COM in Denmark |
| | 🟥 | Local population LATER than DEN-COM in Denmark |
| | ● | Local population emergence ABSENT compared to DEN-COM in Denmark |

**Calendar 13**. Calendar of a common Denmark (DEN-COM) *Chenopodium album* population seedling emergence magnitude (mean numbers per JW of 700 possible buried seeds) with time (2005, 2006) in undisturbed soil at 11 nursery locations. Comparisons of DEN-COM population emergence times at various locations with those of DEN-COM at Denmark: earlier (green), later (red).

| DEN-COM | | 2005 C.E. | | | | | | 2006 C.E. | | | | | | | | | | | YEAR-SEASON-MONTH-JULIAN WEEK | | | | | | | | | | | | | | | | | | | | | | | 2006 C.E. |
|---|---|---|---|---|---|---|---|---|---|---|---|---|---|---|---|---|---|---|---|---|---|---|---|---|---|---|---|---|---|---|---|---|---|---|---|---|---|---|---|---|---|---|
| Population: | | | | AUTUMN | | | | | | WINTER | | | | | | | | | | SPRING | | | | | | | | | | | | SUMMER | | | | | | | AUTUMN | | | |
| Un-Disturbed | | NOV | | | DEC | | | JAN | | FEB | | MAR | | | | APRIL | | | | MAY | | | | JUNE | | | | JULY | | | | AUGUST | | | SEPT | | | OCT | | | NOV | |
| LAT | NURSERY | 45 | 46 | 47 | 48 | 49 | 50 | 2 | 3 | 5 | 6 | 9 | 11 | 12 | 13 | 14 | 15 | 16 | 17 | 18 | 19 | 20 | 21 | 22 | 23 | 24 | 25 | 26 | 27 | 28 | 29 | 32 | 33 | 34 | 35 | 36 | 37 | 39 | 40 | 42 | 44 | 46 |
| N 61° | Finland | | ● | | ● | | | | | | | | | | | | | ● | ● | ● | ● | | | ● | 96 | 8 | | | | | | | | 2 | 3 | | | | | | | |
| N 60° | Norway | | ● | | ● | | | | | | | | | | | | | ● | 3 | 28 | 58 | | 2 | 10 | ● | ● | 2 | 14 | 2 | | | | | | | | | | | | | 3 |
| N 58° | Sweden | | ● | | ● | | | | | | | | | | | | 2 | 13 | 26 | 69 | 84 | 5 | 7 | ● | ● | ● | | | | | | 5 | 10 | | | | | | | | | |
| N 55° | DEN-COM | | 1 | | 1 | | | | | | | | | | | | | 1 | 4 | 3 | 4 | | | 9 | 4 | 1 | | | | | | | | | | | | | | | | |
| N 49° | Czech | | ● | | ● | | | | | | | | | | | | 6 | 77 | 17 | 22 | 4 | 2 | 2 | 1 | ● | ● | 1 | | 1 | 1 | | | | | | | | | | | | |
| N 45° | Canada | | 1 | | ● | | | | | | | | | | | 3 | 6 | 42 | 25 | 11 | 22 | 5 | | ● | ● | ● | | | | | | | | | | | | | | | | |
| N 43° | Italy | | ● | | ● | | | | | | | | | 6 | 10 | 10 | 14 | 6 | 5 | 4 | ● | | | ● | ● | ● | | | | | 1 | | | | | 1 | | | 4 | 2 | 3 | |
| N 40° | USA | | ● | | ● | | | | | | | | 2 | 2 | 22 | 77 | 57 | 54 | 2 | ● | 1 | | | ● | ● | ● | | | | | | | | | | | | | | | | |
| N 52° | UK | 12 | ● | 3 | 2 | 1 | | | | | | | | | 2 | 17 | 6 | 8 | 4 | 2 | 2 | 3 | 2 | 2 | ● | ● | 1 | | | 1 | | | | 1 | | | | | | | | |
| N 39° | Portugal | | 9 | 8 | 4 | 2 | 2 | | 1 | 1 | 1 | 1 | | | | 2 | | ● | 3 | 1 | 2 | | 1 | 1 | 2 | 1 | 3 | | | | | | | | | | 3 | 2 | 1 | | | |
| N 37° | Spain | | ● | | 15 | 3 | | 1 | | | | | | | 1 | | | ● | ● | ● | ● | | | ● | ● | ● | | | | | | | | | | | | | | | | |

| Key: | | |
|---|---|---|
| | 🟩 | Local population EARLIER than DEN-COM in Denmark |
| | 🟥 | Local population LATER than DEN-COM in Denmark |
| | ● | Local population emergence ABSENT compared to DEN-COM Denmark |

**Calendar 14**. Calendar of local and a common Denmark (DEN-COM) *Chenopodium album* population seedling emergence magnitude (mean numbers per JW of 700 possible buried seeds) with time (2005, 2006) in undisturbed soil at 11 nursery locations. Comparisons of local population emergence times with those of DEN-COM at the same nursery location: earlier (green), later (red).

| Un-Disturbed: LOCAL and DEN-COM | | 2005 C.E. | | | | | | 2006 C.E. | | | | | | | YEAR-SEASON-MONTH-JULIAN WEEK | | | | | | | | | | | | | | | | | | | | | | | | | | | | | | 2006 C.E. | | |
|---|---|---|---|---|---|---|---|---|---|---|---|---|---|---|---|---|---|---|---|---|---|---|---|---|---|---|---|---|---|---|---|---|---|---|---|---|---|---|---|---|---|---|---|---|---|---|---|
| | | AUTUMN | | | | | | WINTER | | | | | | | SPRING | | | | | | | | | | | | | SUMMER | | | | | | | | | | | | | AUTUMN | | | | | | |
| | | NOV | | | DEC | | | JAN | | | FEB | MARCH | | | | | | APRIL | | | MAY | | | | JUNE | | | | | JULY | | | | AUGUST | | | | SEPT | | | | | OCT | | NOV | | |
| SITE | POP | 45 | 46 | 47 | 48 | 49 | 50 | 2 | 3 | 5 | 6 | 9 | 10 | 11 | 12 | 13 | 14 | 15 | 16 | 17 | 18 | 19 | 20 | 21 | 22 | 23 | 24 | 25 | 26 | 27 | 28 | 29 | 30 | 31 | 32 | 33 | 34 | 35 | 36 | 37 | 38 | 39 | 40 | 42 | 44 | 45 | 46 |
| Finland | LOCAL | | | | | | | | | | | | | | | | | | | | | | | | | 27 | 2 | | | 1 (red) | | | | | | | • | • | | | | | | | | | |
| N 61° | DEN-COM | | | | | | | | | | | | | | | | | | | | | | | | | 96 | 8 | | | | | | | | | | 2 | 3 | | | | | | | | | |
| Norway | LOCAL | | | | | | | | | | | | | | | | | | | 2 | 7 | 8 | | 1 | 8 | 7 (red) | | • | 6 | • | | | | | | | | | | | | | | | | | • |
| N 60° | DEN-COM | | | | | | | | | | | | | | | | | | | 3 | 28 | 58 | | 2 | 10 | | | 2 | 14 | 2 | | | | | | | | | | | | | | | | | 3 |
| Sweden | LOCAL | | | | | | | | | | | | | | | | | • | 11 | 36 | 79 | 102 | 5 | 7 | | | | | | | | | | | 12 | 8 | | | | | | | | | | | |
| N 58° | DEN-COM | | | | | | | | | | | | | | | | | 2 | 13 | 26 | 69 | 84 | 5 | 7 | | | | | | | | | | | 5 | 10 | | | | | | | | | | | |
| Denmark | LOCAL | | • | | 2 | | | | | | | | | | | | | | 1 | 1 | 1 | 1 | | | 2 | • | • | | | | | | | | 1 (red) | | | | | | | | | | | | |
| N 55° | DEN-COM | | 1 | | 1 | | | | | | | | | | | | | | 1 | 4 | 3 | 4 | | | 9 | 4 | 1 | | | | | | | | | | | | | | | | | | | | |
| Czech | LOCAL | | | | | | | | | | | | | | | | | 1 | 11 | 7 | 7 | 2 | | 1 | 1 | | | • | | • | 3 | | | | 2 (red) | | | | | | | | | | | | |
| N 49° | DEN-COM | | | | | | | | | | | | | | | | | 6 | 77 | 17 | 22 | 4 | 2 | 2 | 1 | | | 1 | | 1 | 1 | | | | | | | | | | | | | | | | |
| Canada | LOCAL | | • | | | | | | | | | | | | | | 1 | 1 | 5 | 10 | 2 | 3 | 3 | 2 (red) | 2 (red) | | | | | | | | 2 (red) | 1 (red) | | | | | | | | | | | | | |
| N 45° | DEN-COM | | 1 | | | | | | | | | | | | | | 3 | 6 | 42 | 25 | 11 | 22 | 5 | | | | | | | | | | | | | | | | | | | | | | | | |
| Italy | LOCAL | | | | | | | | | | | | | | 1 | 1 | 1 | 3 | 2 | 8 | 17 | 8 (red) | 4 (red) | | | | | | | | | • | 2 (red) | | | | | | • | | | | • | • | 1 | 2 (red) | |
| N 43° | DEN-COM | | | | | | | | | | | | | | 6 | 10 | 10 | 14 | 6 | 5 | 4 | | | | | | | | | | | 1 | | | | | | | 1 | | | | 4 | 2 | 3 | | |
| USA | LOCAL | | | | | | | | | | | | | 1 | 2 | 9 | 10 | 16 | 3 | | | | | | | | | | | | | | | | | | | | | | | | | | | | |
| N 40° | DEN-COM | | | | | | | | | | | | | 2 | 2 | 22 | 77 | 57 | 54 | 2 | | 1 | | | | | | | | | | | | | | | | | | | | | | | | | |
| UK | LOCAL | • | | • | • | • | | | | | | | 1 (green) | | 1 (green) | 2 | 2 | 2 | 3 | • | • | • | | 12 | 8 | | 1 (green) | 1 | | | 2 | | | | | | 2 | | | | | | | | | | |
| N 52° | DEN-COM | 12 | | 3 | 2 | 1 | | | | | | | | | | 2 | 17 | 6 | 8 | 4 | 2 | 2 | 3 | 2 | 2 | | | 1 | | | 1 | | | | | | 1 | | | | | | | | | | |
| Portugal | LOCAL | | 4 | 3 | 1 | 1 | | 1 (green) | 1 | • | • | 2 | 1 (red) | | | 2 (green) | 2 | | | 5 | 2 | 2 | 3 | 5 | 3 | 4 | 1 | 2 | | | | | | | 2 (red) | | | | | 9 | 3 | 1 | 3 | | | | |
| N 39° | DEN-COM | | 9 | 8 | 4 | 2 | 2 | | 1 | 1 | 1 | 1 | | | | | 2 | | | 3 | 1 | 2 | | 1 | 1 | 2 | 1 | 3 | | | | | | | | | | | | 3 | | 2 | 1 | | | | |
| Spain | LOCAL | | | | • | • | | • | | | | | | | | 1 | | | | | | | | | | | | | | | | | | | | | | | | | | | | | | | |
| N 37° | DEN-COM | | | | 15 | 3 | | 1 | | | | | | | | 1 | | | | | | | | | | | | | | | | | | | | | | | | | | | | | | | |

| Key: | | |
|---|---|---|
| | (green) | Local population EARLIER than DEN-COM in same nursery location |
| | (red) | Local population LATER than DEN-COM in same nursery location |
| | • | Local population emergence ABSENT compared to DEN-COM in the same nursery location |

**Calendar 15**. Calendar of local *Chenopodium album* population seedling emergence magnitude (mean numbers per JW of 700 possible buried seeds) with time (2005, 2006) in undisturbed and disturbed soil at 11 nursery locations.

**Calendar 16**. Calendar of local and a common Denmark (DEN-COM) *Chenopodium album* population seedling emergence magnitude (mean numbers per JW of 700 possible buried seeds) with time (2005, 2006) in undisturbed and disturbed soil at 11 nursery locations.

| | | |
|---|---|---|
| KEY: | | Undisturbed soil seedling emergence time |
| | | Disturbed soil seedling emergence EARLIER than undisturbed |
| | | Disturbed soil seedling emergence LATER than undisturbed |
| | • | Disturbed soil seedling emergence ABSENT relative to undisturbed |

**Calendar 15**. Undisturbed and disturbed soil; local population.

LOCAL Population: Disturbed — 2006 C.E. — YEAR-SEASON-MONTH-JULIAN WEEK — 2006 C.E.

AUTUMN | WINTER | SPRING | SUMMER | AUTUMN

DEC | MARCH | APRIL | MAY | JUNE | JULY | AUGUST | SEP | OCT | NOV

| NURSERY | DIST | 47 | 48 | 49 | 9 | 10 | 11 | 12 | 13 | 14 | 15 | 16 | 17 | 18 | 19 | 20 | 21 | 22 | 23 | 24 | 25 | 26 | 27 | 28 | 29 | 30 | 31 | 32 | 33 | 34 | 35 | 36 | 37 | 38 | 39 | 40 | 44 | 45 | 46 |
|---|---|---|---|---|---|---|---|---|---|---|---|---|---|---|---|---|---|---|---|---|---|---|---|---|---|---|---|---|---|---|---|---|---|---|---|---|---|---|---|
| Finland | Un-D | | | | | | | | | | | | | | | | | | 27 | 2 | | | 1 | | | | | | | | | | | | | | | | |
| | D-1 | | | | | | | | | | | | | | | | | 5 | 9 | 9 | | | 1 | | | | | | | | | | | | | | | | |
| | D-2 | | | | | | | | | | | | | | | | | 22 | 1 | 21 | 1 | 1 | 3 | | | | | | | | 1 | | | | | | | | |
| | D-3 | | | | | | | | | | | | | | | | | | :21: | ● | 4 | 1 | 12 | | | | | | | 1 | | | | | | | | | |
| | D-4 | | | | | | | | | | | | | | | | | | | :4: | | | 7 | | | | | | | 1 | | | | | | | | | |
| | D-5 | | | | | | | | | | | | | | | | | | | ● | ::: | | 4 | | | | | | | 1 | | | | | | | | | |
| Norway | Un-D | | | | | | | | | | | | 2 | 7 | 8 | | 1 | 8 | 7 | | | 6 | | | | | | | | | | | | | | | | | |
| | D-1 | | | | | | | | | | | | ::: | 3 | 4 | | 1 | 5 | 1 | | | 1 | | | | | | | | | | | | | | | | | |
| | D-2 | | | | | | | | | | | | | | :8: | | 20 | 183 | 12 | 2 | | 21 | 1 | | 1 | | | | | | | | | | | | | | |
| | D-3 | | | | | | | | | | | | | | :6: | | 1 | 83 | 63 | 1 | 2 | 23 | | | | | | | | | | | | | | | | | |
| | D-4 | | | | | | | | | | | | | | | | | :3: | 12 | 6 | 3 | 40 | | 1 | | | | | | | | | | | | | | | |
| | D-5 | | | | | | | | | | | | | | | | | | ::: | 2 | 2 | 62 | 13 | | | | | | | | | | | | | | | | |
| Sweden | Un-D | | | | | | | | | | | 11 | 36 | 79 | 102 | 5 | 7 | | | | | | | | | | | 12 | 8 | | | | | | | | | | |
| | D-1 | | | | | | | | | | ::: | 37 | 60 | 76 | 103 | 6 | 10 | 2 | | | | | | | | | | 10 | 9 | | | | | | | | | | |
| | D-2 | | | | | | | | | | | | | :76: | 42 | 6 | 8 | 1 | 2 | | | | | | | | | 10 | 6 | | | | | | | | | | |
| | D-3 | | | | | | | | | | | | | | :80: | 4 | 14 | 1 | | | | | | | | | | 11 | 9 | | | | | | | | | | |
| | D-4 | | | | | | | | | | | | | | :86: | 14 | 14 | 6 | | | | | | | | | | 8 | 8 | | | | | | | | | | |
| | D-5 | | | | | | | | | | | | | | | | :38: | 13 | 5 | | | | | | | | | 10 | 8 | | | | | | | | | | |
| Denmark: Common | Un-D | | | | | | | | | | | 1 | 4 | 3 | 4 | | | 9 | 4 | 1 | | | | | | | | | | | | | | | | | | | |
| | D-1 | | | | | | | | | | | ::: | 2 | 27 | 63 | 1 | 2 | 12 | 2 | ● | | | | | | | | | | | | | | | | | | | |
| | D-2 | | | | | | | | | | | | :4: | ● | 14 | 6 | 19 | 24 | 5 | ● | | | | | | | | | | | | | | | | | | | |
| | D-3 | | | | | | | | | | | | | | :2: | 26 | 19 | 100 | 4 | ● | | | | | | | | | | | | | | | | | | | |
| | D-4 | | | | | | | | | | | | | | | ::: | 2 | 105 | 8 | ● | | | | | | | | | | | | | | | | | | | |
| | D-5 | | | | | | | | | | | | | | | | | | :8: | ● | 14 | 36 | 5 | | 1 | | | | | | | | | | | | | | |
| Denmark: Local | Un-D | | | | | | | | | | | 1 | 1 | 1 | 1 | | | 2 | | | | | | | | | | 1 | | | | | | | | | | | |
| | D-1 | | | | | | | | | | | ::: | 1 | 1 | 9 | | 1 | 1 | 2 | | | | | | | | | 1 | | | | | | | | | | | |
| | D-2 | | | | | | | | | | | | :1: | ● | ● | | 16 | 6 | | | | | | | | | | 1 | | | | | | | | | | | |
| | D-3 | | | | | | | | | | | | | | ::: | | 3 | 33 | 7 | | | | | | | | | ● | | | 1 | | | | | | | | |
| | D-4 | | | | | | | | | | | | | | | ::: | | 34 | 7 | | | | | | | | | ● | 1 | | | | | | | | | | |
| | D-5 | | | | | | | | | | | | | | | | | ● | 1 | | 5 | 20 | 3 | 1 | 3 | | | ● | | | | | | | | 1 | | | |
| Czech | Un-D | | | | | | | | | | 1 | 11 | 7 | 7 | 2 | | 1 | 1 | | | | | | 3 | | | | 2 | | | | | | | | | | | |
| | D-1 | | | | | | | | | | | :3: | ● | 49 | 9 | 5 | 2 | 4 | 4 | | | | 1 | ● | 1 | | | ● | | | | | | | | | | | |
| | D-2 | | | | | | | | | | | | :3: | 30 | 83 | 4 | ● | 5 | 2 | | | | 1 | ● | 2 | | | ● | | | 2 | | | | | | | | |
| | D-3 | | | | | | | | | | | | | :5: | 10 | 12 | 8 | 31 | 34 | 3 | | | 11 | 3 | | | | ● | 1 | | 1 | | | | | | | | |
| | D-4 | | | | | | | | | | | | | | ::: | 6 | 11 | 11 | 12 | 3 | | | 8 | 3 | | 1 | | 1 | | 1 | | | | | | | | | |
| | D-5 | | | | | | | | | | | | | | ::: | | 5 | 4 | 21 | | | | 4 | 2 | 1 | | | ● | 1 | | | | | | | | | | |
| Canada | Un-D | | | | | | | | | 1 | 1 | 5 | 10 | 2 | 3 | 3 | 2 | 2 | | | | | | | | 2 | 1 | | | | | | | | | | | | |
| | D-1 | | | | | | | | | :1: | | 1 | 10 | 5 | 3 | 8 | 2 | 1 | 1 | 2 | 1 | | | | | ● | ● | | | | | | | | | | | | |
| | D-2 | | | | | | | | | | | :5: | ● | 2 | 121 | 23 | ● | ● | 1 | 2 | | | | | | ● | ● | | | | | | | | | | | | |
| | D-3 | | | | | | | | | | | | | :2: | 6 | 143 | 4 | ● | | | | | | | | ● | ● | | | | | | | | | | | | |
| | D-4 | | | | | | | | | | | | | | :8: | 309 | 58 | 1 | | | | | | | | ● | ● | | | | | | | | | | | | |
| | D-5 | | | | | | | | | | | | | | | :5: | 71 | 23 | 4 | | | | 1 | | | ● | ● | | | | | | | | | | | | |
| Italy | Un-D | | | | | | | 1 | 1 | 1 | 3 | 2 | 8 | 17 | 8 | 4 | | | | | | | | | | 2 | | | | | | | | | | | 1 | 2 | |
| | D-1 | | | | | | ::: | | 24 | 44 | 7 | 14 | 102 | 17 | 4 | 4 | | | | | | | | | | ● | 1 | | | | | | | | | | 1 | ● | 1 |
| | D-2 | | | | | | | | | :2: | ● | 100 | 102 | 80 | 15 | 20 | 3 | | | | | | | | | ● | 1 | | | 2 | | 1 | | | | | ● | ● | |
| | D-3 | | | | | | | | | | ::: | 3 | 101 | 172 | 39 | 19 | 2 | | | | | | | | | ● | 2 | | | 1 | | 2 | | | | 3 | 1 | ● | |
| | D-4 | | | | | | | | | | | | :3: | 60 | 13 | 85 | 3 | | | | | | | 1 | 8 | ● | 1 | | | | | 3 | | | | 4 | 1 | ● | 1 |
| | D-5 | | | | | | | | | | | | | :11: | ● | 232 | 36 | 1 | 1 | 1 | 1 | 1 | 1 | 5 | | 1 | 4 | | | | | 7 | | | | 12 | ● | 1 | 1 |
| USA | Un-D | | | | | | 1 | 2 | 9 | 10 | 16 | 3 | | | | | | | | | | | | | | | | | | | | | | | | | | | |
| | D-1 | | | | | | ::: | ● | 1 | 4 | 13 | 20 | 4 | | | | | | | | | | | | | | | | | | | | | | | | | | |
| | D-2 | | | | | | | | ::: | 1 | 14 | 10 | | | | | | | | | | | | | | | | | | | | | | | | | | | |
| | D-3 | | | | | | | | | | ::: | 9 | 15 | 2 | | | | | | | | | | | | | | | | | | | | | | | | | |
| | D-4 | | | | | | | | | | ::: | 5 | 6 | 1 | 2 | | | | | | | | | | | | | | | | | | | | | | | | |
| | D-5 | | | | | | | | | | | ::: | | 7 | 9 | 10 | 2 | 3 | | | | | | | | | | | | | | | | | | | | | |
| UK | Un-D | | | | | 1 | | 1 | 2 | 2 | 2 | 3 | | | | | 12 | 8 | | 1 | 1 | | | 2 | | | | | | 2 | | | | | | | | | |
| | D-1 | | | | | | | | :3: | ● | 6 | 38 | 15 | 1 | 2 | 6 | 12 | 2 | | ● | ● | | | 2 | | 3 | 2 | 1 | | 3 | | | | | | | | | |
| | D-2 | | | | | | | | | :3: | ● | 24 | 46 | 1 | | 5 | 28 | 1 | | 4 | 1 | | | 2 | | 2 | | 2 | | 10 | | | | | | | | | |
| | D-3 | | | | | | | | | | | ::: | 48 | 22 | 5 | 14 | 15 | 11 | | 2 | 1 | | | 1 | | 4 | | 2 | | 9 | | | | | | | | | |
| | D-4 | | | | | | | | | | | :1: | 6 | 36 | 53 | 30 | 15 | 20 | | 2 | ● | | | ● | | 2 | | 1 | | 9 | | 1 | | | | | | | |
| | D-5 | | | | | | | | | | | | | | 59 | 30 | 24 | 11 | | 3 | 1 | | | 1 | | 1 | | 1 | | 6 | | 1 | | | | | | | |
| Portugal | Un-D | | | | | | | | | | | | 5 | 2 | 2 | 3 | 5 | 3 | 4 | 1 | 2 | | | | | | | 2 | | | | | 9 | 3 | 1 | 3 | | | |
| | D-1 | | | | | | | | | | | | :124: | 25 | 4 | 2 | 3 | 2 | 5 | 6 | ● | | | | | | | 1 | | | | | 7 | 2 | ● | ● | | | |
| | D-2 | | | | | | | | | | | | | :40: | 65 | 6 | 3 | 2 | 5 | ● | ● | | | | | | | ● | | | | | 3 | ● | 2 | ● | | | |
| | D-3 | | | | | | | | | | | | | :8: | 147 | 31 | 8 | 4 | 4 | ● | ● | | | | | | | 1 | | | | | 4 | 1 | 2 | ● | | | |
| | D-4 | | | | | | | | | | | | | | :2: | 103 | 27 | 2 | 1 | ● | ● | | | | | | | ● | | | | | 6 | 1 | ● | ● | | | |
| | D-5 | | | | | | | | | | | | | | | ::: | 111 | 35 | 1 | ● | 2 | 2 | | | | | | 1 | | | | | 5 | ● | ● | ● | | | |
| Spain | Un-D | | | | | | | | 1 | | | | | | | | | | | | | | | | | | | | | | | | | | | | | | |
| | D-1 | | ::: | | | | 37 | | 15 | 2 | | 1 | | | | | | | | | | | | | | | | | | | | | | | | | | | |
| | D-2 | | | | | 1 | | | ● | | | | | | | | | | | | | | | | | | | | | | | | | | | | | | |
| | D-3 | | | | | | | | ● | | | | | | ::: | | | | | | | | | | | | | | | | | | | | | | | | |
| | D-4 | | | | | | | | ● | | | | | | | ::: | | | | | | | | | | | | | | | | | | | | | | | |
| | D-5 | | | | | | | | ● | | | | | | | | ::: | | | | | | | | | | | | | | | | | | | | | | |
| NURSERY | DIST | 47 | 48 | 49 | 9 | 10 | 11 | 12 | 13 | 14 | 15 | 16 | 17 | 18 | 19 | 20 | 21 | 22 | 23 | 24 | 25 | 26 | 27 | 28 | 29 | 30 | 31 | 32 | 33 | 34 | 35 | 36 | 37 | 38 | 39 | 40 | 44 | 45 | 46 |

**Calendar 16**. Undisturbed and disturbed soil; DEN-COM population.

| DEN-COM Population: Disturbed | | 10 | 11 | 12 | 13 | 14 | 15 | 16 | 17 | 18 | 19 | 20 | 21 | 22 | 23 | 24 | 25 | 26 | 27 | 28 | 29 | 30 | 31 | 32 | 33 | 34 | 35 | 36 | 37 | 38 | 39 | 40 | 41 | 42 | 43 | 44 | 45 | 46 |
|---|---|---|---|---|---|---|---|---|---|---|---|---|---|---|---|---|---|---|---|---|---|---|---|---|---|---|---|---|---|---|---|---|---|---|---|---|---|---|
| | | 2006 C.E. | | | | | | | | | YEAR-SEASON-MONTH-JULIAN WEEK | | | | | | | | | | | | | | | | | | | | | | | | | 2006 C.E. | | |
| | | | | | | | | | | SPRING | | | | | | | | | | | | SUMMER | | | | | | | | | | | AUTUMN | | | | | |
| | | | | | | | APRIL | | | MAY | | | | JUNE | | | | | JULY | | | | | AUGUST | | | SEPT | | | | | OCTOBER | | | NOV | | |
| NURSERY | DIST | 10 | 11 | 12 | 13 | 14 | 15 | 16 | 17 | 18 | 19 | 20 | 21 | 22 | 23 | 24 | 25 | 26 | 27 | 28 | 29 | 30 | 31 | 32 | 33 | 34 | 35 | 36 | 37 | 38 | 39 | 40 | 41 | 42 | 43 | 44 | 45 | 46 |
| Finland | Un-D | | | | | | | | | | | | | | 96 | 8 | | | | | | | | | | 2 | 3 | | | | | | | | | | | |
| | D-1 | | | | | | | | | | | | | 13 | 14 | 37 | | | 1 | | | | | | | 2 | • | 1 | | | | | | | | | | |
| | D-2 | | | | | | | | | | | | | | 7 | 111 | 22 | | 14 | | | | | | | • | 1 | 1 | | | | | | | | | | |
| | D-3 | | | | | | | | | | | | | | 96 | • | 13 | | 79 | | 3 | | | | | 5 | 8 | 2 | 4 | | | | 1 | | | | | |
| | D-4 | | | | | | | | | | | | | | | 14 | | | 43 | | 8 | | | | | 11 | 16 | 1 | 3 | 1 | | | 3 | | | | | |
| | D-5 | | | | | | | | | | | | | | | | :: | | 22 | 2 | 7 | | | | | 14 | 11 | 1 | 1 | | | | 1 | | | | | |
| Norway | Un-D | | | | | | | | 3 | 28 | 58 | | 2 | 10 | | | 2 | 14 | 2 | | | | | | | | | | | | | | | | | | | 3 |
| | D-1 | | | | | | | | :2: | 6 | 26 | | 3 | 9 | 1 | | • | • | • | | | | | | | | | | | | | | | | | | | • |
| | D-2 | | | | | | | | | | :58: | | 22 | 112 | 18 | 1 | 2 | 12 | 1 | | | | | | | | | | | | | | | | | | | • |
| | D-3 | | | | | | | | | | :56: | | • | 66 | 34 | 1 | • | 7 | 2 | | | | | | | | | | | | | | | | | | | • |
| | D-4 | | | | | | | | | | | | | 14 | 2 | | 2 | 19 | • | | | | | | | | | | | | | | | | | | | • |
| | D-5 | | | | | | | | | | | | | | :: | | 1 | 29 | 5 | | | | | | | | | | | | | | | | | | | • |
| Sweden | Un-D | | | | | | 2 | 13 | 26 | 69 | 84 | 5 | 7 | | | | | | | | | | | 5 | 10 | | | | | | | | | | | | | |
| | D-1 | | | | | | :1: | 41 | 61 | 80 | 82 | 5 | 11 | | | | | | | | | | | 5 | 8 | | | | | | | | | | | | | |
| | D-2 | | | | | | | | | :64: | 32 | 3 | 9 | | | | | | | | | | | 9 | 16 | | | | | | | | | | | | | |
| | D-3 | | | | | | | | | | :72: | 3 | 11 | | | | | | | | | | | 6 | 7 | | | | | | | | | | | | | |
| | D-4 | | | | | | | | | | :86: | 15 | 13 | 6 | | | | | | | | | | 6 | 9 | | | | | | | | | | | | | |
| | D-5 | | | | | | | | | | | | :31: | 14 | 3 | 2 | 1 | | | | | | | 7 | 7 | | | | | | | | | | | | | |
| Denmark | Un-D | | | | | | | 1 | 4 | 3 | 4 | | | 9 | 4 | 1 | | | | | | | | | | | | | | | | | | | | | | |
| DEN-COM | D-1 | | | | | | | :: | 2 | 27 | 63 | 1 | 2 | 12 | 2 | • | | | | | | | | | | | | | | | | | | | | | | |
| | D-2 | | | | | | | | :4: | • | 14 | 6 | 19 | 24 | 5 | • | | | | | | | | | | | | | | | | | | | | | | |
| | D-3 | | | | | | | | | | :2: | 26 | 19 | 100 | 4 | • | | | | | | | | | | | | | | | | | | | | | | |
| | D-4 | | | | | | | | | | | :: | 2 | 105 | 8 | • | | | | | | | | | | | | | | | | | | | | | | |
| | D-5 | | | | | | | | | | | | | | 8 | • | 14 | 36 | 5 | | 1 | | | | | | | | | | | | | | | | | |
| Czech | Un-D | | | | | | 6 | 77 | 17 | 22 | 4 | 2 | 2 | 1 | | | 1 | | 1 | 1 | | | | | | | | | | | | | | | | | | |
| | D-1 | | | | | | | :25: | 19 | 84 | 10 | 3 | 2 | 1 | | | • | | | | | | | | | | | | | | | | | | | | | |
| | D-2 | | | | | | | | :16: | 110 | 60 | 4 | 1 | 3 | 3 | | • | | | | | | | | | | | | | | | | | | | | | |
| | D-3 | | | | | | | | | :14: | 29 | 22 | 23 | 12 | 7 | | • | | 1 | | | | | | | | | | | | | | | | | | | |
| | D-4 | | | | | | | | | | :2: | 5 | 8 | 8 | 15 | 5 | • | | 6 | 1 | 1 | | | | | | | | | | | | | | | | | |
| | D-5 | | | | | | | | | | :4: | • | 1 | 8 | 4 | 2 | • | | 4 | | | | | | | | | | | | | | | | | | | |
| Canada | Un-D | | | | | 3 | 4 | 42 | 37 | 11 | 17 | 5 | | | | | | | | | | | | | | | | | | | | | | | | | | |
| | D-1 | | | | | :: | • | 14 | 38 | 8 | 13 | 5 | | | 1 | | | | | | | | | | | | | | | | | | | | | | | |
| | D-2 | | | | | | | :32: | 6 | 35 | 110 | 7 | 2 | | 1 | | | | | | | | | | | | | | | | | | | | | | | |
| | D-3 | | | | | | | | | :10: | 36 | 131 | | 1 | | | | | | | | | | | | | | | | | | | | | | | | |
| | D-4 | | | | | | | | | | :9: | 147 | 8 | 8 | | 1 | | | | | | | | | | | | | | | | | | | | | | |
| | D-5 | | | | | | | | | | | :15: | 57 | 12 | | | | | | | | | | | | | | | | | | | | | | | | |
| Italy | Un-D | | | 6 | 10 | 10 | 14 | 6 | 5 | 4 | | | | | | | | | | | 1 | | | | | | | 1 | | | | 4 | | 2 | | 3 | | |
| | D-1 | | :: | 1 | 76 | 104 | 22 | 16 | 8 | 3 | 1 | | | | | | | | | | 1 | | | | | | | • | | | | • | | • | | 1 | 1 | |
| | D-2 | | | | | :17: | 9 | 172 | 80 | 22 | | 2 | | | | | | | | | • | | 1 | | | | | • | | | | • | | 2 | | 1 | 2 | |
| | D-3 | | | | | | :13: | 15 | 98 | 42 | 5 | 4 | | | | | | | | | • | | | | | | | • | | | | • | | • | | 1 | | |
| | D-4 | | | | | | | | :4: | 10 | 2 | 4 | 2 | | | | | | | | • | | | | | | | 2 | | | | • | | 2 | | 2 | | |
| | D-5 | | | | | | | | | :: | | 62 | 4 | | | | | | | 8 | • | 1 | | | | | | • | | | | 2 | | 1 | | 1 | | |
| USA | Un-D | | 2 | 2 | 22 | 77 | 57 | 54 | 2 | | 1 | | | | | | | | | | | | | | | | | | | | | | | | | | | |
| | D-1 | | :: | • | • | 6 | 8 | 12 | 4 | 1 | • | 1 | | | | | | | | | | | | | | | | | | | | | | | | | | |
| | D-2 | | | | :: | 6 | 37 | 38 | 4 | | • | | | | | | | | | | | | | | | | | | | | | | | | | | | |
| | D-3 | | | | | | :: | 95 | 5 | 2 | • | | | | | | | | | | | | | | | | | | | | | | | | | | | |
| | D-4 | | | | | | :: | 14 | 13 | 4 | 2 | | | | | | | | | | | | | | | | | | | | | | | | | | | |
| | D-5 | | | | | | | :: | 1 | 12 | 11 | 3 | 3 | | | | | | | | | | | | | | | | | | | | | | | | | |
| UK | Un-D | | | | 2 | 17 | 6 | 8 | 4 | 2 | 2 | 3 | 2 | 2 | | | 1 | | | 1 | | | | | | 1 | | | | | | | | | | | | |
| | D-1 | | | | :6: | 2 | 20 | 69 | 30 | 3 | 3 | 3 | 2 | 2 | | | • | | | 1 | | | | | | | | | | | | | | | | | | |
| | D-2 | | | | | :3: | | 49 | 54 | 2 | 2 | 1 | 3 | 5 | | 1 | • | | | • | | | | | | | | | | | | | | | | | | |
| | D-3 | | | | | | | :2: | 125 | 21 | 1 | 5 | 5 | 4 | | | • | | | 1 | | | | | | | | | | | | | | | | | | |
| | D-4 | | | | | | | :11: | 18 | 226 | 51 | 16 | 6 | 6 | | | • | | | • | | | | | | | | | | | | | | | | | | |
| | D-5 | | | | | | | | | :21: | 143 | 24 | 16 | 7 | | 2 | • | | | • | | | | | | | | | | | | | | | | | | |
| Portugal | Un-D | | | | | | | | 3 | 1 | 2 | | 1 | 1 | 2 | 1 | 3 | | | | | | | | | | | | 3 | | 2 | 1 | | | | | | |
| | D-1 | | | | | | | | :43: | 15 | 1 | | • | • | 3 | • | 2 | | | | | | | | | | | | 1 | | 1 | 1 | | | | | | |
| | D-2 | | | | | | | | | :18: | 37 | 2 | 3 | | 3 | • | • | | | | | | | | | | | | 3 | | • | 1 | | | | | | |
| | D-3 | | | | | | | | | :1: | 130 | 11 | 1 | 1 | 2 | • | • | | | | | | | | | | | | 2 | | 2 | • | | | | | | |
| | D-4 | | | | | | | | | | :2: | 39 | 3 | • | 1 | 1 | 2 | | | | | | | | | | | | 2 | | 1 | • | | | | | | |
| | D-5 | | | | | | | | | | | :: | 17 | 6 | 4 | • | 1 | | | | | | | | | | | | 1 | 1 | • | • | | | | | | |
| NURSERY | DIST | 10 | 11 | 12 | 13 | 14 | 15 | 16 | 17 | 18 | 19 | 20 | 21 | 22 | 23 | 24 | 25 | 26 | 27 | 28 | 29 | 30 | 31 | 32 | 33 | 34 | 35 | 36 | 37 | 38 | 39 | 40 | 41 | 42 | 43 | 44 | 45 | 46 |

| NURSERY | DIST | 48 | 49 | 50 | 51 | | 2 | | 10 | 11 | 12 | 13 | 14 | 15 | 16 | 17 | 18 | 19 | 20 | 21 | 22 | 23 | 24 | 25 | 26 | 27 | 28 | 29 | 30 | 31 | 32 | 33 | 34 | 35 | 36 | 37 | 38 | 39 |
|---|---|---|---|---|---|---|---|---|---|---|---|---|---|---|---|---|---|---|---|---|---|---|---|---|---|---|---|---|---|---|---|---|---|---|---|---|---|---|
| Spain | Un-D | 15 | 3 | | | | 1 | | | | | 1 | | | | | | | | | | | | | | | | | | | | | | | | | | |
| | D-1 | :10: | 24 | 21 | 7 | | 7 | | | | | 15 | | | | | | | | | | | | | | | | | | | | | | | | | | |
| | D-2 | | | | | | | | :: | | | | | | | | | | | | | | | | | | | | | | | | | | | | | |
| | D-3 | | | | | | | | | | | | | | | | | :: | | | | | | | | | | | | | | | | | | | | |
| | D-4 | | | | | | | | | | | | | | | | | | :: | | | | | | | | | | | | | | | | | | | |
| | D-5 | | | | | | | | | | | | | | | | | | | :: | | | | | | | | | | | | | | | | | | |
| NURSERY | DIST | 48 | 49 | 50 | 51 | | 2 | | 10 | 11 | 12 | 13 | 14 | 15 | 16 | 17 | 18 | 19 | 20 | 21 | 22 | 23 | 24 | 25 | 26 | 27 | 28 | 29 | 30 | 31 | 32 | 33 | 34 | 35 | 36 | 37 | 38 | 39 |

**Calendar 17**. Local and common Denmark (DEN-COM) Chenopodium album population seedling emergence cohort structure and pattern in undisturbed and (+) disturbed soil (see Table 3).

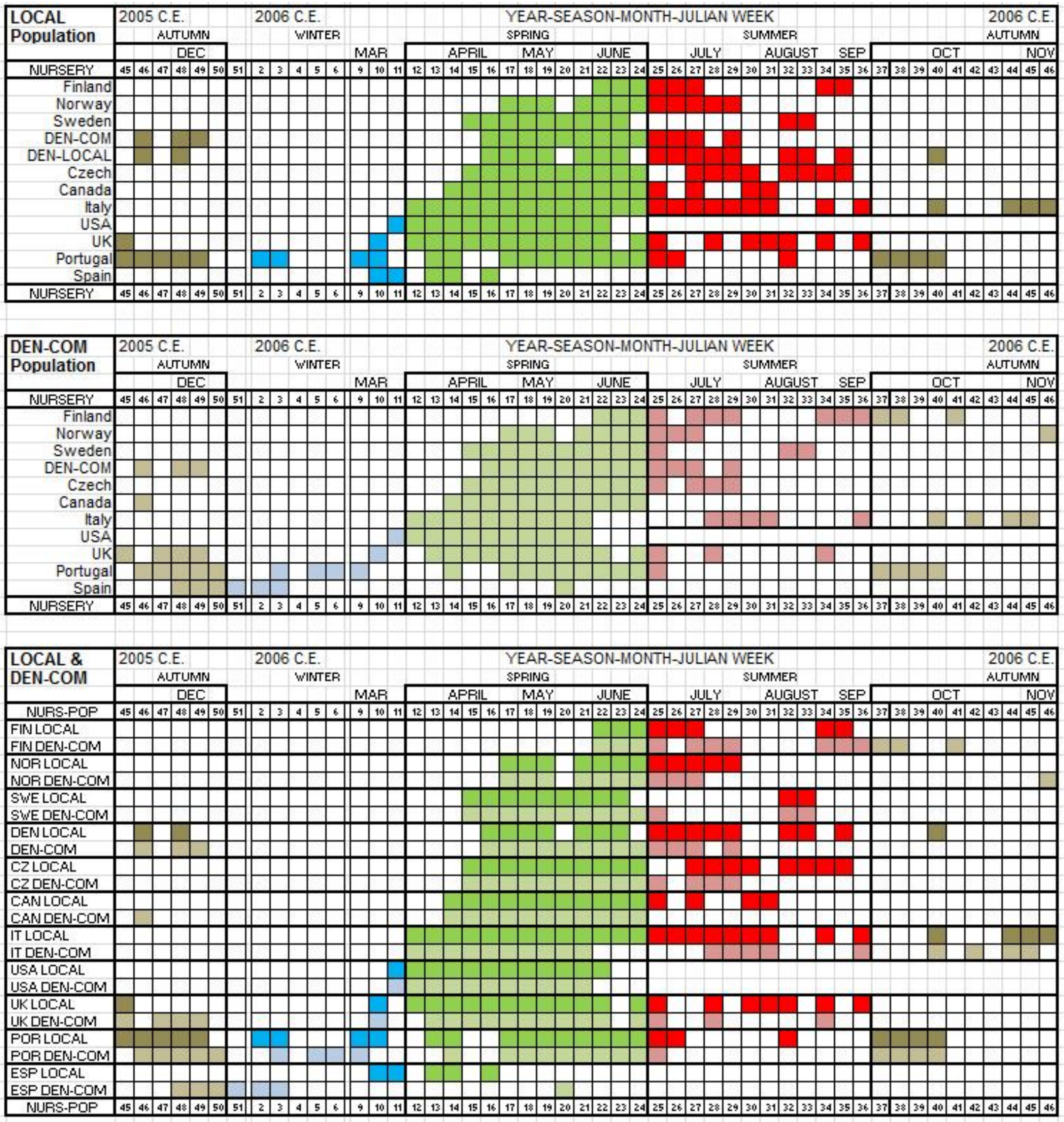

**Calendar 18**. *Chenopodium album* behavior with time (JW, season): aggregate seedling emergence over population, location, disturbance.

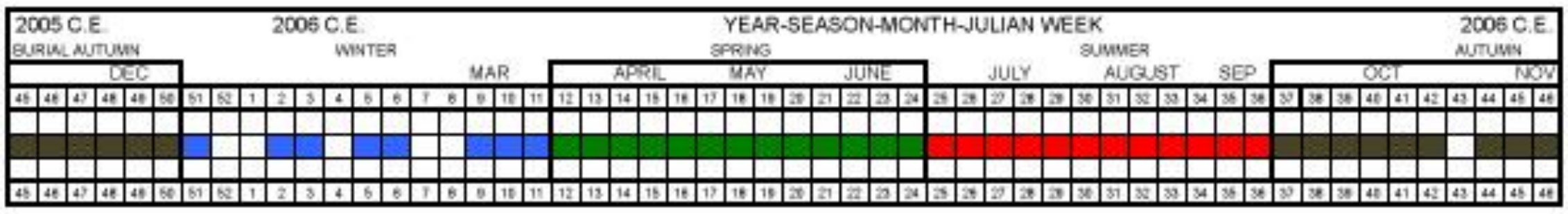

**Calendar 19**. Aggregate location, population

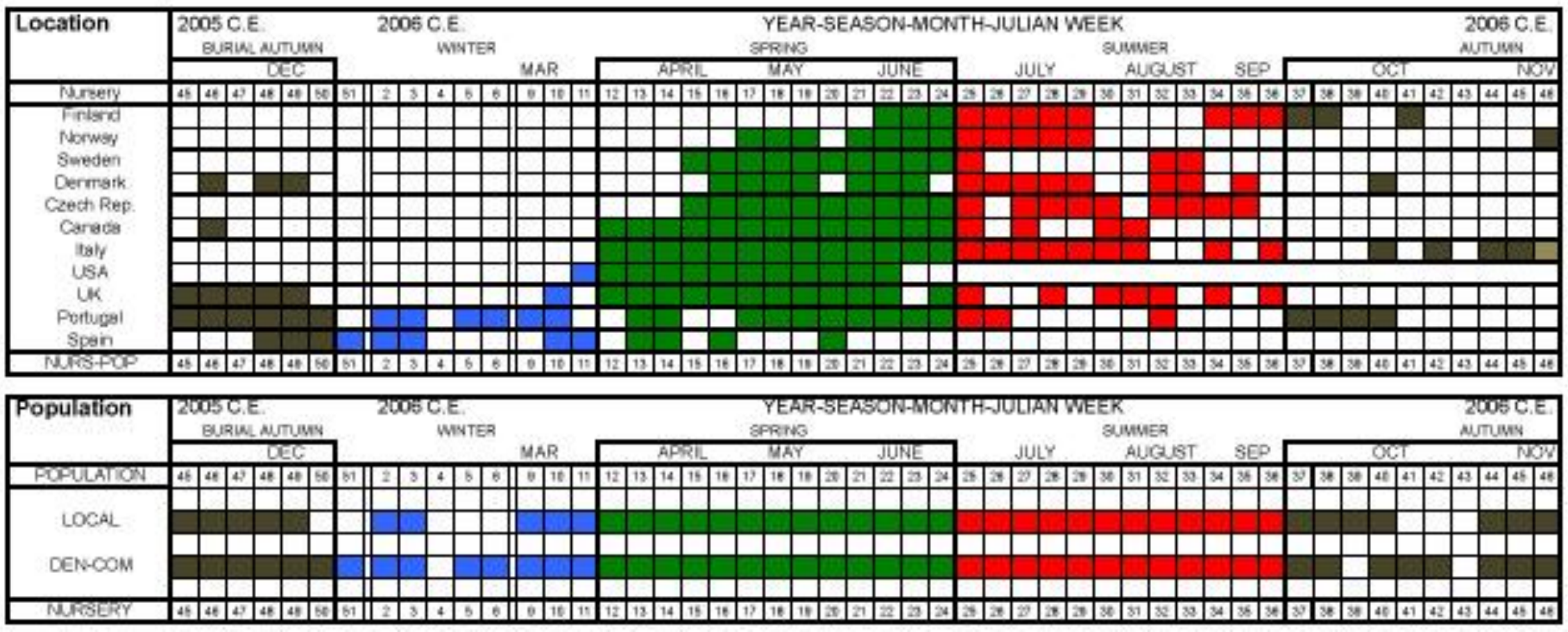

**Calendar 23**. Aggregate undisturbed and disturbed soil

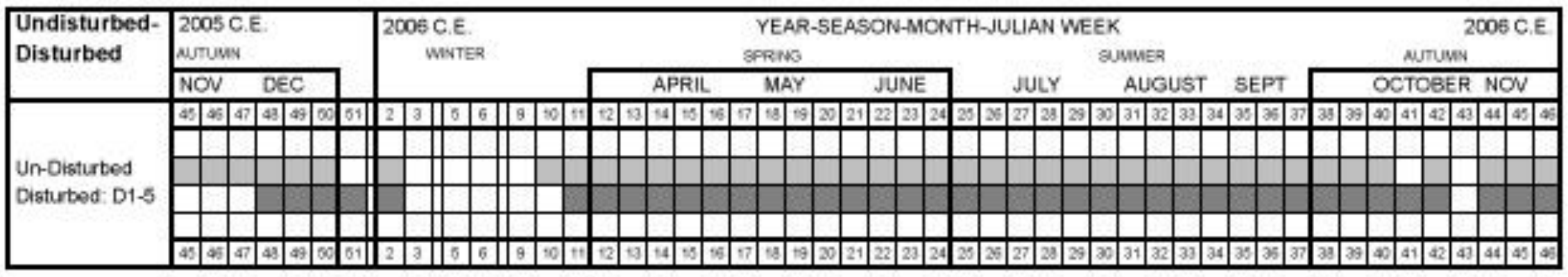

**Calendar 22**. Aggregate undisturbed and disturbed soil by location

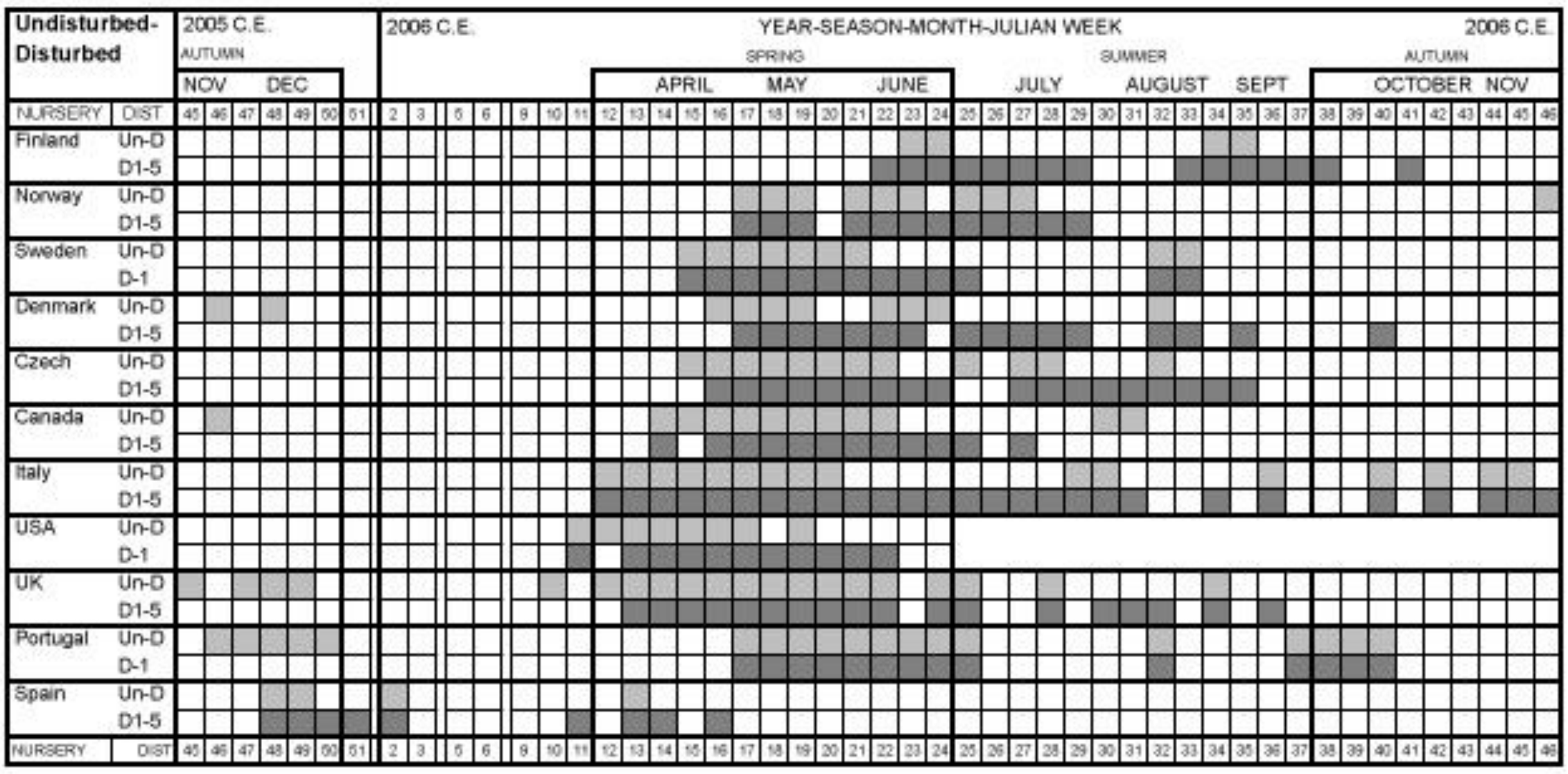

**Calendar 20**. Local, undisturbed and disturbed, aggregated over population and disturbance regimes (D1-5).

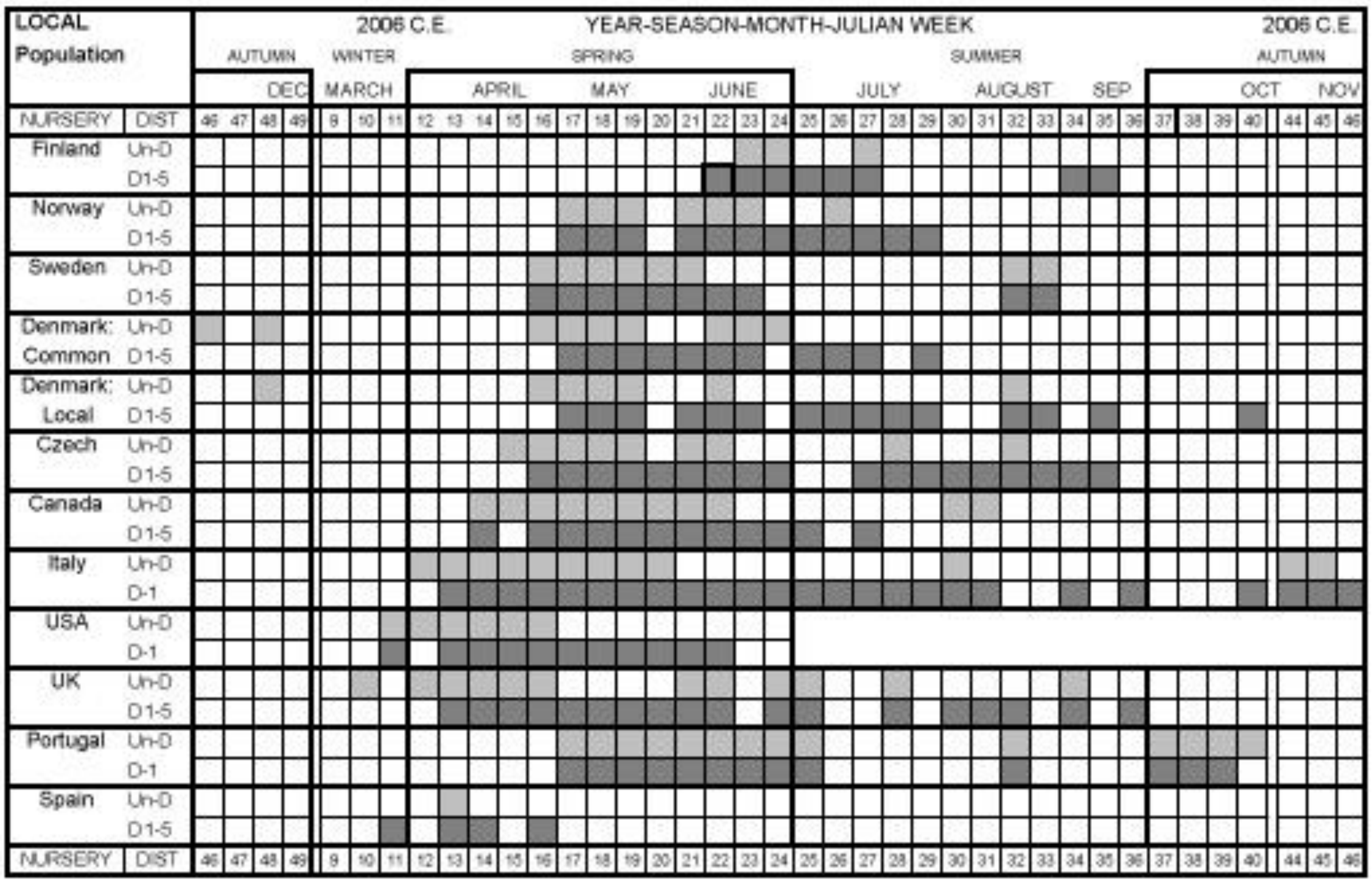

**Calendar 21**. DEN-COM, undisturbed and disturbed, aggregated over population and disturbance regimes (D1-5).

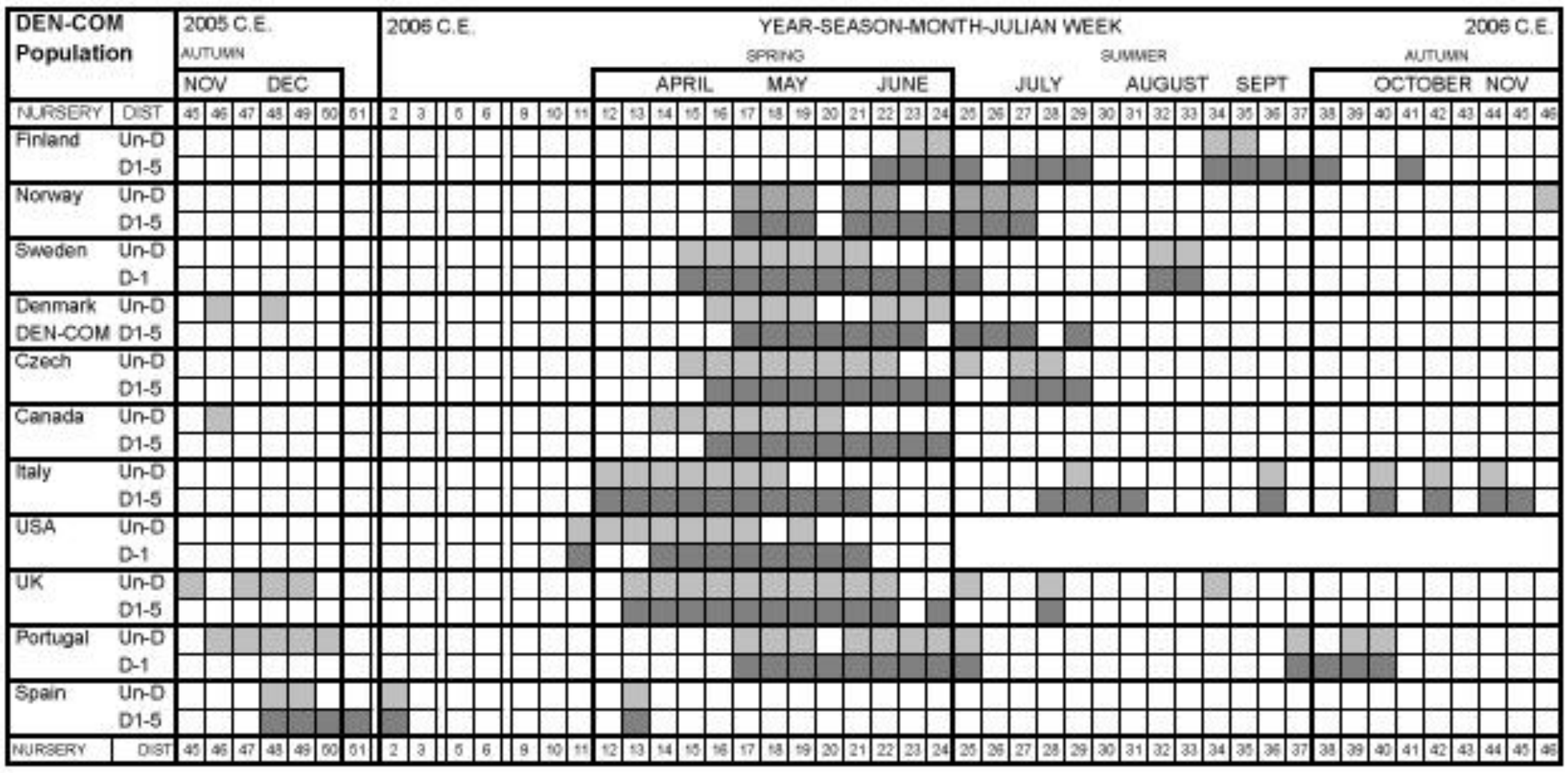
DEN-COM Population
2005 C.E.
AUTUMN
NOV DEC
2006 C.E.
YEAR-SEASON-MONTH-JULIAN WEEK
2006 C.E.
SPRING
SUMMER
AUTUMN
APRIL MAY JUNE
JULY AUGUST SEPT
OCTOBER NOV
NURSERY DIST
45 46 47 48 49 50 51 2 3 5 6 9 10 11 12 13 14 15 16 17 18 19 20 21 22 23 24 25 26 27 28 29 30 31 32 33 34 35 36 37 38 39 40 41 42 43 44 45 46
Finland Un-D
D1-5
Norway Un-D
D1-5
Sweden Un-D
D-1
Denmark Un-D
DEN-COM D1-5
Czech Un-D
D1-5
Canada Un-D
D1-5
Italy Un-D
D1-5
USA Un-D
D-1
UK Un-D
D1-5
Portugal Un-D
D-1
Spain Un-D
D1-5
NURSERY DIST

## 6 SEEDLING EMERGENCE TABLES

**Table 5**. [T6E]. *Chenopodium album* seedling emergence season magnitude (total number) in undisturbed and disturbed (D; total, D1-5) soil: location (nursery), population (LOCAL, DEN-COM). '*', disturbance with the greatest number of emerged seedlings within the location and population.

| Location | Population | **Seedling Recruitment Season Magnitude** (Total number) | | | | | | |
|---|---|---|---|---|---|---|---|---|
| | | **UN-DISTURBED** | **DISTURBED** | | | | | |
| | | | **Total** | **D-1** | **D-2** | **D-3** | **D-4** | **D-5** |
| **Finland** | LOCAL | 28 | 5-50 | 24 | *50** | 39 | 12 | 5 |
| | DEN-COM | 109 | 59-226 | 68 | *226** | 201 | 100 | 59 |
| **Norway** | LOCAL | 39 | 15-247 | 15 | *247** | 179 | 71 | 79 |
| | DEN-COM | 122 | 35-227 | 47 | *227** | 166 | 39 | 35 |
| **Sweden** | LOCAL | 260 | 74-316 | *315** | 151 | 129 | 146 | 74 |
| | DEN-COM | 221 | 65-295 | *296** | 133 | 99 | 136 | 65 |
| **Denmark** | LOCAL | 9 | 16-44 | 16 | 24 | *44** | 42 | 34 |
| | DEN-COM | 28 | 64-151 | 109 | 72 | *151** | 115 | 64 |
| **Czech R.** | LOCAL | 35 | 40-131 | 78 | *131** | 119 | 58 | 40 |
| | DEN-COM | 134 | 23-197 | 144 | *197** | 108 | 51 | 23 |
| **Canada** | LOCAL | 32 | 35-376 | 35 | 154 | 155 | *376** | 104 |
| | DEN-COM | 120 | 79-193 | 79 | *193** | 178 | 174 | 84 |
| **Italy** | LOCAL | 49 | 188-345 | 219 | 325 | *345** | 188 | 316 |
| | DEN-COM | 66 | 28-311 | 232 | *311** | 178 | 28 | 79 |
| **UK** | LOCAL | 37 | 96-236 | 96 | 139 | 234 | *236** | 138 |
| | DEN-COM | 66 | 126-336 | 141 | 126 | 162 | *336** | 213 |
| **Portugal** | LOCAL | 63 | 126-221 | 181 | 126 | *221** | 142 | 155 |
| | DEN-COM | 50 | 30-150 | 67 | 67 | *150** | 51 | 30 |
| **Spain** | LOCAL | 1 | 0-55 | *55** | 1 | 0 | 0 | 0 |
| | DEN-COM | 20 | 0-84 | *84** | 0 | 0 | 0 | 0 |
| **Key to Between POPULATION Comparisons: Within LOCATION and DISTURBANCE** | | | | | | | | |
| DEN-COM magnitude GREATER than LOCAL | | | | | | | | |
| DEN-COM magnitude LESSER than LOCAL | | | | | | | | |

**Table 6**. [T6G]. *Chenopodium album* seedling emergence season magnitude (total number) in undisturbed soil during five seasonal cohorts (burial autumn, winter, spring, summer, autumn): location (nursery), population (LOCAL, DEN-COM).

| Population | Location | Seedling Recruitment Magnitude in Un-Disturbed Soil: by Season | | | | |
|---|---|---|---|---|---|---|
| | | Bur Aut | Winter | Spring | Summer | Autumn |
| Finland | LOCAL | | | 29 | 1 | |
| | DEN-COM | | | 104 | 5 | |
| Norway | LOCAL | | | 33 | 6 | 3 |
| | DEN-COM | | | 101 | 18 | |
| Sweden | LOCAL | | | 240 | 20 | |
| | DEN-COM | | | 206 | 15 | |
| Denmark | LOCAL | 2 | | 6 | 1 | |
| | DEN-COM | 2 | | 26 | | |
| Czech R. | LOCAL | | | 30 | 5 | |
| | DEN-COM | | | 131 | 3 | |
| Canada | LOCAL | | | 29 | 3 | |
| | DEN-COM | 1 | | 119 | | |
| Italy | LOCAL | | | 45 | 1 | 3 |
| | DEN-COM | | | 55 | 2 | 9 |
| USA | LOCAL | | 1 | 40 | | |
| | DEN-COM | | 2 | 215 | | |
| UK | LOCAL | | 1 | 31 | 5 | |
| | DEN-COM | 18 | | 45 | 3 | |
| Portugal | LOCAL | 9 | 5 | 29 | 13 | 7 |
| | DEN-COM | 25 | 4 | 13 | 6 | 3 |
| Spain | LOCAL | | | 1 | | |
| | DEN-COM | 18 | 1 | 1 | | |
| DEN-COM magnitude GREATER than LOCAL | | | | | | |
| DEN-COM magnitude LESS than LOCAL | | | | | | |

**Table 7**. [T6H]. *Chenopodium album* seedling emergence season magnitude (total number) in disturbed (D; D1-5) soil during five seasonal cohorts (burial autumn, winter, spring, summer, autumn): location (nursery), population (LOCAL, DEN-COM).

| Pop | Disturb | Seedling Recruitment Magnitude in Disturbed Soil: by Season and Population | | | | | |
|---|---|---|---|---|---|---|---|
| | | Spring | | Summer | | Autumn | |
| | | LOCAL | DEN-COM | LOCAL | DEN-COM | LOCAL | DEN-COM |
| Finland | D-1 | 23 | 64 | 1 | 4 | | |
| | D-2 | 44 | 118 | 6 | 38 | | |
| | D-3 | 21 | 96 | 18 | 114 | | 1 |
| | D-4 | 4 | 14 | 8 | 82 | | 4 |
| | D-5 | | | 5 | 58 | | 1 |
| Norway | D-1 | 14 | 47 | 1 | | | |
| | D-2 | 225 | 212 | 22 | 15 | | |
| | D-3 | 154 | 157 | 25 | 10 | | |
| | D-4 | 21 | 16 | 50 | 21 | | |
| | D-5 | 2 | | 77 | 35 | | |
| Sweden | D-1 | 294 | 282 | 19 | 13 | | |
| | D-2 | 135 | 108 | 16 | 25 | | |
| | D-3 | 109 | 86 | 20 | 13 | | |
| | D-4 | 130 | 120 | 16 | 16 | | |
| | D-5 | 56 | 51 | 18 | 14 | | |
| Denmark | D-1 | 15 | 109 | 1 | | | |
| | D-2 | 23 | 72 | 1 | | | |
| | D-3 | 43 | 151 | 1 | | | |
| | D-4 | 41 | 115 | 1 | | | |
| | D-5 | 1 | 8 | 32 | 56 | 1 | |
| Czech R. | D-1 | 76 | 144 | 2 | | | |
| | D-2 | 127 | 197 | 4 | | | |
| | D-3 | 103 | 107 | 16 | 1 | | |
| | D-4 | 44 | 43 | 14 | 8 | | |
| | D-5 | 30 | 19 | 10 | 4 | | |
| Canada | D-1 | 34 | 79 | 1 | | | |
| | D-2 | 154 | 193 | | | | |
| | D-3 | 155 | 177 | | | | |
| | D-4 | 376 | 174 | | | | |
| | D-5 | 103 | 84 | 1 | | | |
| Italy | D-1 | 216 | 229 | 1 | 1 | 2 | 2 |
| | D-2 | 321 | 305 | 4 | 1 | | 5 |
| | D-3 | 336 | 177 | 5 | | 4 | 1 |
| | D-4 | 170 | 22 | 12 | 2 | 6 | 4 |
| | D-5 | 282 | 66 | 20 | 9 | 14 | 6 |
| UK | D-1 | 85 | 140 | 11 | 1 | | |
| | D-2 | 122 | 126 | 17 | | | |
| | D-3 | 217 | 161 | 17 | 1 | | |
| | D-4 | 223 | 336 | 13 | | | |
| | D-5 | 127 | 213 | 11 | | | |
| Portugal | D-1 | 171 | 62 | 8 | 3 | 2 | 2 |
| | D-2 | 121 | 63 | 3 | 3 | 2 | 1 |
| | D-3 | 213 | 146 | 5 | 2 | 3 | 2 |
| | D-4 | 135 | 46 | 6 | 4 | 1 | 1 |
| | D-5 | 147 | 27 | 10 | 2 | | 1 |
| DEN-COM magnitude GREATER than LOCAL | | | | | | | |
| DEN-COM magnitude LESS than LOCAL | | | | | | | |

**Table 8**. [T7C]. Cohort number for *Chenopodium album* seedling emergence cohort number for local and common Denmark (DEN-COM) populations in undisturbed and disturbed (D; D1-5) soil.

| Location | Population | **Seedling Emergence Cohort Number** | | | | | | |
|---|---|---|---|---|---|---|---|---|
| | | **UN-DISTURBED** | **DISTURBED** | | | | | |
| | | | Total | D-1 | D-2 | D-3 | D-4 | D-5 |
| **Finland** | LOCAL | 2 | 2-3 | 2 | 2 | 3 | 3 | 2 |
| | DEN-COM | 2 | 3-6 | 🟩 4 | 🟩 3 | 🟩 6 | 🟩 5 | 🟩 3 |
| **Norway** | LOCAL | 3 | 1-4 | 3 | 4 | 2 | 2 | 1 |
| | DEN-COM | 🟩 4 | 1-3 | 🟥 2 | 🟥 2 | 🟩 3 | 2 | 1 |
| **Sweden** | LOCAL | 2 | 2 | 2 | 2 | 2 | 2 | 2 |
| | DEN-COM | 2 | 2 | 2 | 2 | 2 | 2 | 2 |
| **Denmark** | LOCAL | 4 | 2-3 | 3 | 3 | 2 | 2 | 3 |
| | DEN-COM | 4 | 1-3 | 🟥 1 | 🟥 2 | 🟥 1 | 🟥 1 | 3 |
| **Czech Rep.** | LOCAL | 4 | 3-5 | 4 | 5 | 4 | 5 | 3 |
| | DEN-COM | 🟥 3 | 1-3 | 🟥 1 | 🟥 1 | 🟥 2 | 🟥 2 | 3 |
| **Canada** | LOCAL | 2 | 1-3 | 2 | 3 | 1 | 1 | 2 |
| | DEN-COM | 2 | 1-2 | 2 | 🟥 2 | 🟩 2 | 🟩 2 | 🟥 1 |
| **Italy** | LOCAL | 3 | 4-7 | 4 | 5 | 6 | 7 | 6 |
| | DEN-COM | 🟩 6 | 3-6 | 🟥 3 | 5 | 🟥 2 | 🟥 4 | 6 |
| **UK** | LOCAL | 6 | 5-8 | 5 | 8 | 6 | 6 | 7 |
| | DEN-COM | 6 | 🟥 1-3 | 🟥 2 | 🟥 3 | 🟥 2 | 🟥 1 | 🟥 2 |
| **Portugal** | LOCAL | 7 | 2-4 | 3 | 3 | 3 | 2 | 4 |
| | DEN-COM | 🟩 9 | 3-5 | 🟩 5 | 🟩 4 | 3 | 🟩 4 | 🟥 3 |
| **Spain** | LOCAL | 1 | 0-3 | 3 | 1 | 0 | 0 | 0 |
| | DEN-COM | 🟩 3 | 0-3 | 3 | 🟥 0 | 0 | 0 | 0 |
| **Key to Between POPULATION Comparisons: Within LOCATION and DISTURBANCE** | | | | | | | | |
| DEN-COM cohort number GREATER than LOCAL | | 🟩 | 🟩 | 🟩 | 🟩 | 🟩 | 🟩 | 🟩 |
| DEN-COM cohort number LESSER than LOCAL | | 🟥 | 🟥 | 🟥 | 🟥 | 🟥 | 🟥 | 🟥 |

**Table 9**. [6D]. *Chenopodium album* seedling emergence season duration (JW period; total JW): comparisons between populations (local, DEN-COM) within location (nursery), for each disturbance (undisturbed, disturbed); comparison criteria (longer, shorter): greater than 1 week difference. Period: Julian weeks period within which emergence occurred, including intervening weeks of no emergence between cohorts. Time: number of weeks in which some emergence occurred, not including intervening weeks with no emergence between cohorts.

| Location | Population | Seedling Recruitment Season Duration in Un-Disturbed Soil: Period and Time | |
|---|---|---|---|
| | | Duration Period (JW) | Duration Time (weeks) |
| **Finland** | LOCAL | 23-27 | 3 |
| | DEN-COM | 23-35 | 4 |
| **Norway** | LOCAL | 17-26 | 7 |
| | DEN-COM | 17-27 | 8 |
| **Sweden** | LOCAL | 16-33 | 8 |
| | DEN-COM | 15-33 | 9 |
| **Denmark** | LOCAL | 46-32 | 7 |
| | DEN-COM | 46-24 | 9 |
| **Czech R.** | LOCAL | 15-32 | 9 |
| | DEN-COM | 15-28 | 11 |
| **Canada** | LOCAL | 14-31 | 11 |
| | DEN-COM | 46-20 | 8 |
| **Italy** | LOCAL | 12-45 | 12 |
| | DEN-COM | 12-44 | 12 |
| **UK** | LOCAL | 45-34 | 12 |
| | DEN-COM | 46-34 | 17 |
| **Portugal** | LOCAL | 46-40 | 24 |
| | DEN-COM | 46-40 | 21 |
| **Spain** | LOCAL | 13 | 1 |
| | DEN-COM | 48-13 | 4 |
| **Key to Between POPULATION Comparisons: Within LOCATION** | | | |
| DEN-COM period or time LONGER than LOCAL | | 🟩 | 🟩 |
| DEN-COM period or time SHORTER than LOCAL | | 🟥 | 🟥 |

**Table 10**. [6F]. *Chenopodium album* seedling emergence season duration (JW period; total JW): comparisons between populations (local, DEN-COM) within location (nursery), for each disturbance (undisturbed, disturbed); comparison criteria (longer, shorter): greater than 1 week difference. Period: Julian weeks period within which emergence occurred; commencing with week of disturbance; including intervening weeks of no emergence between cohorts. Time: number of weeks in which some emergence occurred, not including intervening weeks with no emergence between cohorts. '*', longest duration seedling emergence among disturbed (D1-5).

| Location | Population | Seedling Recruitment Season Duration in Disturbed Soil: Period and Time | | | | | | | | | | | |
|---|---|---|---|---|---|---|---|---|---|---|---|---|---|
| | | DURATION PERIOD (JW) | | | | | | DURATION TIME (weeks) | | | | | |
| | | Total | D-1 | D-2 | D-3 | D-4 | D-5 | Total | D-1 | D-2 | D-3 | D-4 | D-5 |
| **Finland** | LOCAL | 22-35 | 22-27 | 22-35* | 23-34 | 24-34 | 25-34 | 2-7 | 4 | 7* | 5 | 3 | 2 |
| | DEN-COM | 22-41 | 22-36 | 23-36 | 23-41* | 24-41 | 25-41 | 6-9 | 6 | 6 | 9* | 9* | 8 |
| **Norway** | LOCAL | 17-29 | 17-26 | 19-29* | 19-26 | 22-28 | 23-27 | 4-8 | 6 | 8* | 7 | 6 | 4 |
| | DEN-COM | 17-27 | 17-23 | 19-27* | 19-27* | 22-26 | 23-27 | 3-8 | 6 | 8* | 6 | 4 | 3 |
| **Sweden** | LOCAL | 15-33 | 15-33* | 18-33 | 19-33 | 19-33 | 21-33 | 5-9 | 9* | 8 | 6 | 6 | 5 |
| | DEN-COM | 15-33 | 15-33* | 18-33 | 19-33 | 19-33 | 21-33 | 5-9 | 9* | 6 | 5 | 6 | 7 |
| **Denmark** | LOCAL | 16-40 | 16-32 | 17-32 | 19-35 | 20-33 | 23-40* | 3-7 | 7* | 4 | 4 | 3 | 7* |
| | DEN-COM | 16-29 | 16-23* | 17-23 | 19-23 | 20-23 | 23-29 | 3-7 | 7* | 6 | 5 | 3 | 5 |
| **Czech R.** | LOCAL | 16-35 | 16-29 | 17-35* | 18-35 | 19-34 | 19-33 | 7-11 | 9 | 11* | 11* | 10 | 7 |
| | DEN-COM | 16-29 | 16-22 | 17-23 | 18-27 | 19-29* | 19-27 | 6-9 | 7 | 7 | 7 | 9* | 6 |
| **Canada** | LOCAL | 14-27 | 14-25* | 16-24 | 18-21 | 19-22 | 20-27 | 4-11 | 11* | 6 | 4 | 4 | 5 |
| | DEN-COM | 14-24 | 14-23* | 16-23 | 18-22 | 19-24 | 20-22 | 3-7 | 6 | 7* | 4 | 5 | 3 |
| **Italy** | LOCAL | 11-46 | 11-46* | 14-36 | 15-44 | 17-46 | 18-46 | 11-17 | 11 | 10 | 11 | 12 | 17* |
| | DEN-COM | 11-45 | 11-45* | 14-45 | 15-44 | 17-44 | 18-44 | 7-11 | 11* | 10 | 7 | 8 | 7 |
| **UK** | LOCAL | 13-36 | 13-34* | 14-34 | 16-34 | 16-36 | 18-36 | 11-14 | 14* | 13 | 12 | 12 | 11 |
| | DEN-COM | 13-28 | 13-28* | 14-24 | 16-28 | 16-22 | 18-24 | 6-11 | 11* | 9 | 8 | 7 | 6 |
| **Portugal** | LOCAL | 17-39 | 17-38* | 18-39* | 18-39* | 19-38 | 20-37 | 7-11 | 11* | 8 | 10 | 7 | 7 |
| | DEN-COM | 17-40 | 17-40* | 18-40 | 18-39 | 19-39 | 20-38 | 7-8 | 8* | 7 | 8* | 8* | 7 |
| **Spain** | LOCAL | 48-21 | 48-16* | 10 | 19 | 20 | 21 | 0-4 | 4* | 1 | 0 | 0 | 0 |
| | DEN-COM | 48-21 | 48-13* | 10 | 19 | 20 | 21 | 0-6 | 6* | 0 | 0 | 0 | 0 |
| **Key to Between POPULATION Comparisons: Within LOCATION and DISTURBANCE** | | | | | | | | | | | | | |
| DEN-COM period or time LONGER than LOCAL | | | | | | | | | | | | | |
| DEN-COM period or time SHORTER than LOCAL | | | | | | | | | | | | | |

**Table 11**. Pattern summary of population differences in *Chenopodium album* seedling emergence cohort (burial autumn, spring, summer, autumn) timing (early, late) comparisons for all sites: soil disturbance (undisturbed, disturbed) and population (local, DEN-COM).
Note: Earlier is also presence when the other population is absent.

| Seedling Emergence Pattern: Population Seasonal Cohorts and Disturbance | | | | | | | | |
|---|---|---|---|---|---|---|---|---|
| | Un-Disturbed Soil | | | | Disturbed Soil | | | |
| Location | BUR AUT | SPR | SUM | AUT | BUR AUT | SPR | SUM | AUT |
| Finland | | | LOC | | | | D-C | D-C |
| Norway | | LOC | D-C D-C | D-C | | D-C | LOC | |
| Sweden | | D-C | | | | D-C D-C | | |
| Denmark | D-C | D-C | LOC | | | | LOC | LOC |
| Czech Rep. | | | D-C | | | | LOC | |
| Canada | D-C | LOC | LOC | | | LOC | LOC | |
| Italy | | LOC | D-C D-C | D-C | | D-C LOC | LOC LOC | LOC |
| UK | D-C | LOC LOC | | | | | LOC LOC | |
| Portugal | D-C | LOC | | LOC | | | | D-C |
| Spain | D-C | | | | D-C | LOC | | |
| Key to Between Cohort and POPULATION Comparisons: Within LOCATION and DISTURBANCE | | | | | | | | |
| Cohort starts EARLIER | | | | | | | | |
| Cohort ends LATER | | | | | | | | |
| No Cohort | | | | | | | | |

**Table 12**. [calendars 15-16]

| | | LOCAL | | | DEN-COM | | |
|---|---|---|---|---|---|---|---|
| | | Seedling Cohort Season | | | Seedling Cohort Season | | |
| Location | DIST | Spring | Summer | Autumn | Spring | Summer | Autumn |
| Finland | D1 | | | | | | |
| | D2 | | | | | | |
| | D3 | • | | | • | | |
| | D4 | | | | | | |
| | D5 | • | | | | | |
| Norway | D1 | | | | | • | • |
| | D2 | | | | | | • |
| | D3 | | | | | • | • |
| | D4 | | | | | • | • |
| | D5 | | | | | | • |
| Sweden | D1 | | | | | | |
| | D2 | | | | | | |
| | D3 | | | | | | |
| | D4 | | | | | | |
| | D5 | | | | | | |
| Denmark LOCAL | D1 | | | | | | |
| | D2 | • | | | | | |
| | D3 | | • | | | | |
| | D4 | | • | | | | |
| | D5 | | • | | | | |

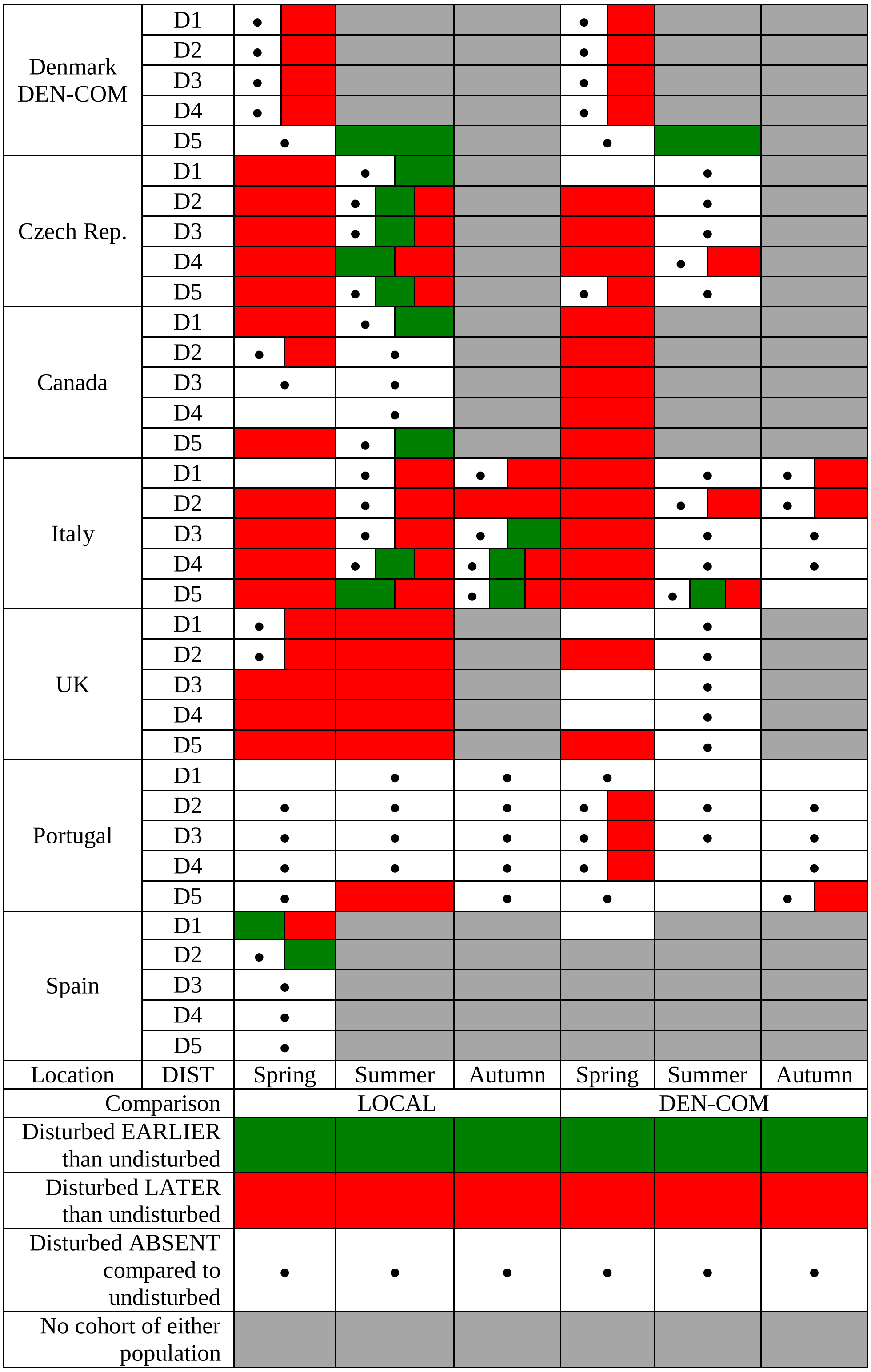
Denmark DEN-COM
Czech Rep.
Canada
Italy
UK
Portugal
Spain
D1
D2
D3
D4
D5
Location
DIST
Spring
Summer
Autumn
Spring
Summer
Autumn
Comparison
LOCAL
DEN-COM
Disturbed EARLIER than undisturbed
Disturbed LATER than undisturbed
Disturbed ABSENT compared to undisturbed
No cohort of either population